\documentclass[aps, prb, preprint,showpacs,superscriptaddress,groupedaddress]{revtex4-1}

\usepackage{amsmath}
\usepackage{amssymb}
\usepackage{braket}
\usepackage{bm}

\usepackage{graphicx}
\usepackage{subcaption}

\begin{document}

\title{Near-exact treatment of seniority-zero ground and excited states with a Richardson-Gaudin mean-field}
\author{Charles-Émile Fecteau, Samuel Cloutier, Jean-David Moisset, J\'{e}r\'{e}my Boulay}
 \affiliation{D\'{e}partement de chimie, Universit\'{e} Laval, Qu\'{e}bec, Qu\'{e}bec, G1V 0A6, Canada}

\author{Patrick Bultinck}
 \affiliation{Department of Chemistry, Ghent University, Gent, Belgium}

\author{Alexandre Faribault}
\affiliation{Universit\'{e} de Lorraine, CNRS, LPCT, F-54000 Nancy, France}

\author{Paul A. Johnson}
\email{paul.johnson@chm.ulaval.ca}
 \affiliation{D\'{e}partement de chimie, Universit\'{e} Laval, Qu\'{e}bec, Qu\'{e}bec, G1V 0A6, Canada}

\date{\today}

\begin{abstract}
Eigenvectors of the reduced Bardeen-Cooper-Schrieffer Hamiltonian, Richardson-Gaudin (RG) states,  are used as a variational wavefunction Ansatz for strongly-correlated electronic systems. These states are geminal products whose coefficients are solutions of non-linear equations. Previous results showed un-physical behaviour but in this contribution it is shown that with only the variational solution for the ground state, all the seniority-zero states are quite well approximated. The difficulty is in choosing the correct RG state. The systems studied showed a clear choice and we expect it should always be possible to reason physically which state to choose.
\end{abstract}

\maketitle

\section{Introduction}
Most systems in quantum chemistry are weakly-correlated and thus well-described in the orbital picture. Qualitatively, the wavefunction is one Slater determinant, with small contributions from single- and double-excitations. This picture is well-understood as a mean-field of electrons, so methods built from such a mean-field, in particular Kohn-Sham density function theory (DFT) and coupled-cluster (CC) with singles and doubles, perform quite well.\cite{helgaker_book}

This picture fails for strongly-correlated systems. The physical wavefunction has important contributions from many Slater determinants, and all must be included to arrive at a qualitatively-correct result. If there are a few important Slater determinants, then the complete active space self-consistent field (CASSCF) or complete active space configuration interaction (CASCI) solve the problem. As the number of important Slater determinants grows, these methods become much more expensive and the correct Slater determinants to include are not always obvious. State of the art methods are effective, but often require experts to use.\cite{huron:1973,chan:2002,chan:2004,thom:2005,booth:2010,chan:2011,booth:2013,sharma:2017,holmes:2017,li:2018,yao:2021} 

Strongly-correlated systems are broadly-defined as any system that is not weakly-correlated. To make measurable progress it is necessary to subdivide these systems into different types based on what they \emph{are}, rather than what they are not. It has been shown\cite{bytautas:2011,bytautas:2015} that classifying systems based on their number of unpaired electrons, their \emph{seniority}, is a productive way to subdivide strongly-correlated systems. The first step is thus to solve the problem for systems with no unpaired electrons, otherwise stated, seniority-zero systems. Our goal is to show definitively that seniority-zero systems are effectively treated as weakly-correlated pairs of electrons. 

The idea of using pairs, or geminals, is an old idea.\cite{fock:1950,mcweeny:1959,mcweeny:1960,mcweeny:1963,nicely:1971,siems:1976} The most general seniority-zero geminal wavefunction, the antisymmetrized product of interacting geminals (APIG),\cite{silver:1969,silver:1970a,silver:1970b,silver:1970c} is intractable for numerical computation, both variationally and by projection. Many of APIG's degenerate cases are however feasible, though each have drawbacks. The antisymmetrized geminal power (AGP)\cite{coleman:1965,ortiz:1981,sarma:1989,coleman:1997} has a long history in quantum chemistry, and is under current study.\cite{henderson:2019,khamoshi:2019,henderson:2020,khamoshi:2020,dutta:2020,khamoshi:2021,dutta:2021} Variationally it is feasible, though unfortunately AGP is not size-consistent which is a problem for molecular dissociations. Neuscamman has shown that size-consistency may be restored at the cost of introducing Jastrow factors.\cite{neuscamman:2012,neuscamman:2013,neuscamman:2016} The antisymmetrized product of strongly-orthogonal geminals (APSG)\cite{hurley:1953,kutzelnigg:1964} is variationally feasible and dissociates molecules correctly. It is however difficult to assign the orbitals correctly into disjoint spaces. The antisymmetrized product of 1-reference orbital geminals (AP1roG),\cite{limacher:2013} equivalently pair-coupled-cluster doubles (pCCD),\cite{stein:2014} treats the ground state of seniority-zero systems quite well.\cite{limacher:2014a,limacher:2014b,henderson:2014a,henderson:2014b,boguslawski:2014a,boguslawski:2014b,boguslawski:2014c,tecmer:2014,boguslawski:2015} However, it is a state-specific method that must be solved by projection. AP1roG / pCCD is at present the method to beat. 

The final degenerate case of APIG is the eigenvectors of the reduced Bardeen-Cooper-Schrieffer (BCS)\cite{bardeen:1957a,bardeen:1957b,schrieffer_book} Hamiltonian, which we refer to as Richardson\cite{richardson:1963,richardson:1964,richardson:1965}-Gaudin\cite{gaudin:1976} (RG) states. These states are geminal product wavefunctions whose geminal coefficients are solutions of a set of non-linear equations. Scalar products and reduced density matrices (RDM) of RG states are easily evaluated, and thus variational optimizations with RG states are feasible. Recently,\cite{johnson:2020} we performed variational calculations for Hydrogen chain dissociations using RG \emph{ground} states. We observed un-physical behaviour which was quite disappointing. To judge whether this was the fault of the RG wavefunction form, we performed variational calculations with RG states that were not eigenvectors of a reduced BCS Hamiltonian, so-called off-shell RG states.\cite{moisset:2022} Off-shell RG states have much more variational freedom, but as they are not eigenvectors of a physical model, they are not computationally feasible. Our results showed that for H$_4$ and H$_6$ the off-shell RG results matched APIG to at least 0.01 milliHartree. For H$_8$, we were forced to relax our convergence criterion, but the agreement was quite clear in any case. These results show that the form of the RG wavefunction was not the problem, but perhaps the specific RG state used. In this contribution, we find the correct RG states to use for Hydrogen chains, as well as N$_2$. The agreement with the best seniority-zero wavefunction possible, doubly-occupied configuration interaction (DOCI),\cite{weinhold:1967a,weinhold:1967b,cook:1975} is excellent everywhere though deviates linearly in the weakly-correlated limit.

In section \ref{sec:RG_states} we very briefly outline the structure of RG states, the approach for numerically obtaining them, and the energy functional to be evaluated. In section \ref{sec:results} variational calculations are performed with RG states. In particular, for H$_4$ a reduced BCS Hamiltonian is optimized for each possible RG state. The correct RG state is found, and with that reduced BCS Hamiltonian, all the RG states are computed and found to match each of the DOCI states. For H$_6$ and H$_8$ similar results are obtained, and for N$_2$ the optimal RG state matches quite well with DOCI.

\section{Richardson-Gaudin States} \label{sec:RG_states}
RG states are geminal product wavefunctions with a particular structure. They describe $M$ pairs among $N$ spatial orbitals built from the pair representation of su(2)
\begin{align} \label{eq:su2_objects}
S^+_i = a^{\dagger}_{i\uparrow} a^{\dagger}_{i\downarrow}, \quad S^-_i = a_{i\downarrow} a_{i\uparrow}, \quad
S^z_i = \frac{1}{2} \left( a^{\dagger}_{i\uparrow}a_{i\uparrow} + a^{\dagger}_{i\downarrow}a_{i\downarrow} -1 \right)
\end{align}
where $a^{\dagger}_{i\uparrow}$ ($a_{i\downarrow}$) creates (removes) an up-spin (down-spin) electron in spatial orbital $i$. The pair operators satisfy the structure
\begin{subequations} \label{eq:su2_structure}
\begin{align} 
[S^+_i , S^-_j] &= 2 \delta_{ij} S^z_i \\
[S^z_i , S^{\pm}_j] &= \pm \delta_{ij} S^{\pm}_i.
\end{align}
\end{subequations}
With the geminals,
\begin{align}
S^{\pm}(v) = \sum_i \frac{S^+_i}{v - \varepsilon_i},
\end{align}
the RG states
\begin{align} \label{eq:rg_state}
\ket{\{v\}} = S^+(v_1) S^+(v_2) \dots S^+(v_M) \ket{\theta}
\end{align}
are the eigenvectors of the reduced BCS Hamiltonian
\begin{align} \label{eq:hbcs}
\hat{H}_{BCS} = \frac{1}{2} \sum_i \varepsilon_i \hat{n}_i - \frac{g}{2} \sum_{ij} S^+_i S^-_j,
\end{align}
provided that the set of complex numbers $\{u\}$, the \emph{rapidities}, are solutions of Richardson's equations
\begin{align} \label{eq:rich_eq}
\frac{2}{g} + \sum_i \frac{1}{u_a - \varepsilon_i} + \sum_{b(\neq a)} \frac{2}{u_b - u_a} = 0.
\end{align}
This construction is a particular example of the algebraic Bethe Ansatz (ABA).\cite{bethe:1931,faddeev:1980,korepin_book} An introduction in terms of electrons is presented in refs.\cite{carrier:2020,moisset:2021} Many algorithms exist to solve Richardson's equations,\cite{rombouts:2004,guan:2012,pogosov:2012,debaerdemacker:2012,claeys:2015} though most struggle near the critical points, where two rapidities are exactly equal to one of the single-particle energies $\varepsilon$. The fastest and most robust approach\cite{faribault:2011,elaraby:2012} involves changing to so-called \emph{eigenvalue-based variables} (EBV) 
\begin{align}
U_i = \sum_a \frac{g}{ \varepsilon_i - u_a}
\end{align}
which satisfy the coupled non-linear equations 
\begin{align} \label{eq:ebv_equations}
U^2_i - 2U_i - g \sum_{k \neq i} \frac{U_k - U_i}{\varepsilon_k - \varepsilon_i} = 0, \quad \forall i =1,\dots,N.
\end{align}
These equations are much easier to solve numerically, as there are no longer the divergences in the denominators for the rapidities. Taking the sum of Richardson's equations, one can see that the sum of the EBV is a constant, in particular, it is twice the number of pairs
\begin{align} \label{eq:sum_ebv}
\sum_i U_i = 2M.
\end{align}
It is essential to enforce \eqref{eq:sum_ebv} as an additional constraint so that there are $N+1$ equations defining the $N$ variables $\{U\}$. Otherwise, the solutions \emph{will} cross into different sectors of $M$.

When $g=0$, the equations \eqref{eq:ebv_equations} decouple, and have a clear solution
\begin{align} \label{eq:ebv_zero_g}
U_i \left( U_i -2 \right) = 0,
\end{align}
so that each EBV is either zero or two. Since the sum of the EBV is $2M$, $M$ of the EBV are two, and the remaining $N-M$ are zero. The complete set of $\binom{N}{M}$ solutions are obtained from the possible choices of assigning the values of zero or two to the EBV, and the states are closed-shell Slater determinants. By convention\cite{faribault:2008,faribault:2010,faribault:2011,elaraby:2012} the states are labelled as a string of 1's and 0's based on the choice of occupations at $g=0$. We refer to such a choice of 1's and 0's as a \emph{distribution}. The equations \eqref{eq:ebv_equations} are solved by choosing a solution at $g=0$ and evolving $g$ slowly to its final value. Remarkably, each state evolves continuously, so it is unambiguous to label an RG state at arbitrary $g$ based on its occupations at $g=0$. Individual states can, and do, cross though the ground state is always the state labelled $1\dots 10\dots 0$, while the highest excited state is always $0\dots 01\dots 1$. For example, a 4-site system with 2 pairs has the states 1100 (which is the ground state at all values of $g$), 1010, 1001, 0110, 0101, and 0011 (which is the highest energy state at all values of $g$). The labelling of the states suggests an aufbau principle, though as curves cross, the ordering of states does not respect the principle at all couplings. As we have shown,\cite{johnson:2021} the dominant couplings between two RG states occur for ``single-pair'' excitations, with double-pair excitations having non-zero contributions. Past doubles, the couplings go to zero quite quickly. A similar conclusion has been drawn in ref.\cite{faribault:2010} The results in section \ref{sec:results} will show that the ground state of the Coulomb Hamiltonian \eqref{eq:C_ham} is \emph{not} an RG ground state, but a particular RG state is near identical to the DOCI ground state. 

\subsection{Solving Richardson's equations}
We will briefly summarize the approach\cite{faribault:2011,elaraby:2012} while highlighting details that we have found to be important. The solver has two components: a solver for the EBV equations \eqref{eq:ebv_equations} and an optimized root-finding procedure to extract the rapidities $\{u\}$. Unfortunately, the expressions for RDM elements require the rapidities, so substantial effort will be devoted to obtaining them. Scalar products and 1-RDM expressions are known in terms of the EBV,\cite{claeys:2017b} but 2-RDM expressions at present are not.

Solving the EBV equations begins by choosing a distribution and assigning the corresponding solution at $g=0$ based on \eqref{eq:ebv_zero_g}. With a set of EBV, a step $\delta g$ is taken by first expanding the EBV as a Taylor series in $g$ to a particular order, then polishing the solution with a Newton-Raphson procedure. Of course the solution could be evolved to the final $g$ by only solving with Newton-Raphson at each step, but the Taylor series allows one to take much larger steps while maintaining the accuracy. The terms in the Taylor series are easily calculated. They are solutions of sets of over-determined linear equations who all share the same matrix. In particular, the first-derivatives of the EBV with respect to $g$, subject to the constraints \eqref{eq:sum_ebv} are the solutions of the linear equations
\begin{align}
A \frac{\partial \textbf{U}}{\partial g} = \textbf{r}_1
\end{align}
where the $(N+1) \times N$ matrix $A$ has elements
\begin{align} \label{eq:ebv_jac}
A_{ij} &= \begin{cases}
2U_i - 2 + \sum_{k \neq i} \frac{g}{\varepsilon_k - \varepsilon_i}, &\quad i = j \\
-\frac{g}{\varepsilon_j - \varepsilon_i}, &\quad i \neq j
\end{cases}, \quad  &1 \leq i,j \leq N, \nonumber \\
A_{N+1,j} &= 1
\end{align}
and the elements of the first-order right-hand-side (RHS) are
\begin{align}
(\textbf{r}_1)_i &= \sum_{k \neq i} \frac{U_k - U_i}{\varepsilon_k - \varepsilon_i}, \quad 1 \leq i \leq N \nonumber \\
(\textbf{r}_1)_{N+1} &= 0. 
\end{align}
It can be shown by induction that the $p$th derivatives $\frac{\partial^p \textbf{U}}{\partial g^p}$ satisfy the set of linear equations
\begin{align}
A \frac{\partial^p \textbf{U}}{\partial g^p} = \textbf{r}_p
\end{align}
with the elements of the $p$th RHS being
\begin{align}
(\textbf{r}_p)_i &= p \sum_{k \neq i} \frac{\frac{\partial^{p-1}U_k}{\partial g^{p-1}}  - \frac{\partial^{p-1}U_i}{\partial g^{p-1}}}{\varepsilon_k - \varepsilon_i}
- \sum^{p-1}_{m=1} \binom{p}{m} \frac{\partial^m U_i }{\partial g}\frac{\partial^{p-m}U_i}{\partial g^{p-m}}, \quad 1 \leq i \leq N \nonumber \\
(\textbf{r}_p)_{N+1} &= 0. 
\end{align}
Each set of linear equations is solved by QR-factorizing the matrix $A=QR$, and solving the linear equations 
\begin{align}
R\frac{\partial^p \textbf{U}}{\partial g^p} = Q^{\dagger} \textbf{r}_p
\end{align}
by backward substitution. This gives the least-squares solution to each set of linear equations, though physically we know there is always an ``exact'' solution. Since the matrix $A$ is common, only one QR-factorization is required to compute the Taylor series. How many terms are required? While it may not be optimal, we have seen that 4th-order is more than sufficient.

With a Taylor-series approximation to the EBV at $g+\delta g$, a Newton-Raphson solution quickly converges the EBV to any desired accuracy. The Jacobian of the EBV equations is the same matrix $A$ \eqref{eq:ebv_jac}, and is once-again factored as $A=QR$, to iteratively solve the solutions of equations
\begin{align}
R  (\textbf{U}_{n+1} - \textbf{U}_n) = - Q^{\dagger} \textbf{f}(\textbf{U}_n)
\end{align}
for the change $\textbf{U}_{n+1} - \textbf{U}_n$. Unfortunately, the Jacobian, and hence its QR-factorization, must be computed at each iteration. The RHS is the EBV equations \eqref{eq:ebv_equations} along with the sum rule \eqref{eq:sum_ebv} evaluated at the $n$th iteration
\begin{align}
f_i (\textbf{U}_n) &= 
\begin{cases}
U^2_{i,n} - 2U_{i,n} -g \sum_{k \neq i} \frac{U_{k,n} - U_{i,n}}{\varepsilon_k - \varepsilon_i} , &1 \leq i \leq N  \\
\sum_k U_{k,n} - 2M , &i = N+1
\end{cases}.
\end{align}

\begin{figure}[ht!] 
	\includegraphics[width=0.475\textwidth]{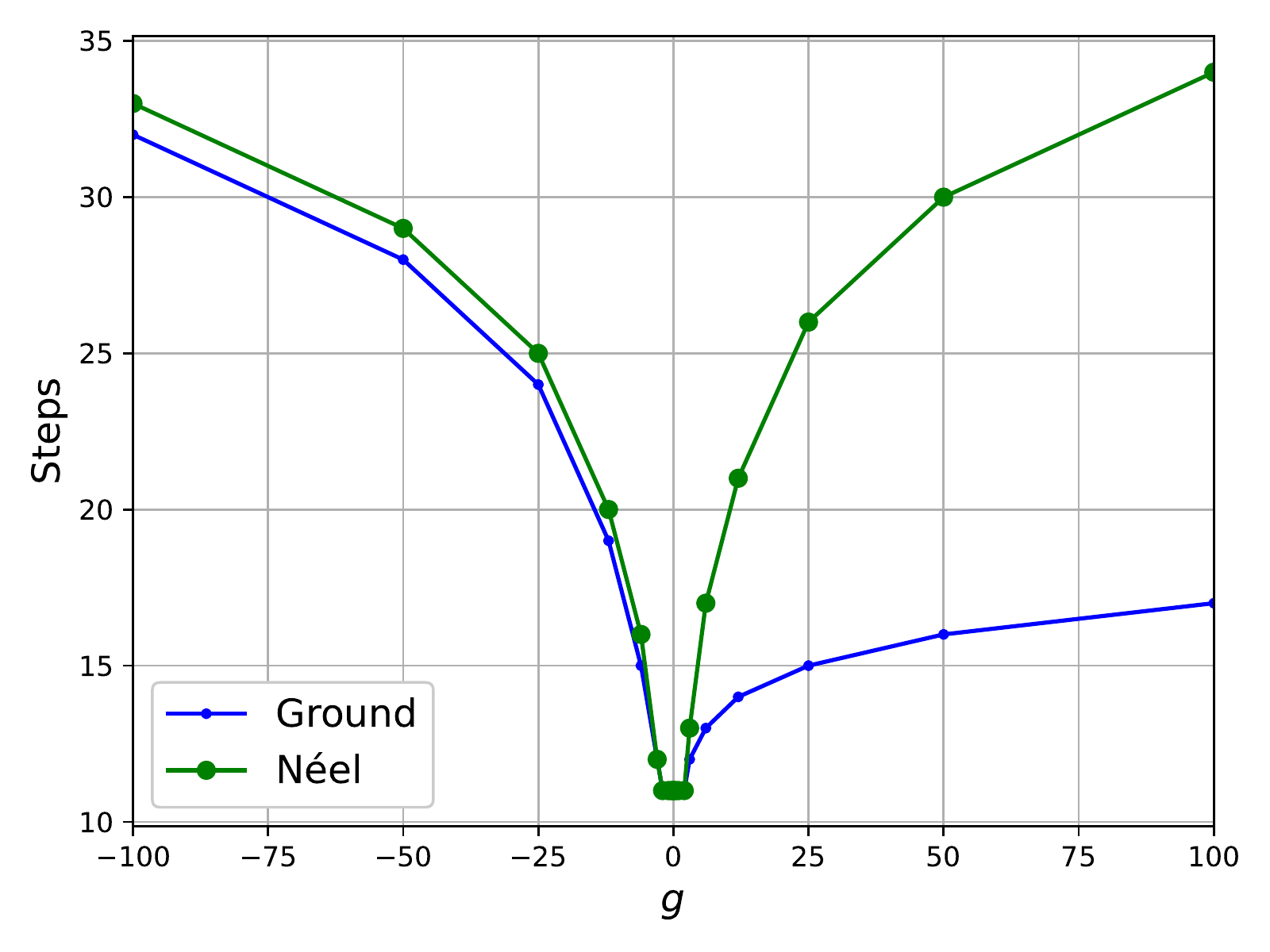} \hfill
	\caption{Number of steps required to solve EBV equations, for the ground-state and the N\'{e}el state, with dynamic step-size algorithm for a reduced BCS Hamiltonian of 40 equally spaced single-particle energies and 20 pairs.}
	\label{fig:log}
\end{figure}

Thus, each step from $g=0$ to $g$ consists of a Taylor series approximation followed by a Newton-Raphson solution. This solution works quite well, and its computational cost scales with the number of steps required, which is in itself a problem. To ensure that the solutions are correctly and unambiguously followed, the initial step from $g=0$ must be smaller than the smallest difference in single particle energies $\Delta \varepsilon$. If there are two $\varepsilon$ that are close in energy, this means a \emph{very large} number of steps will be required. We have adopted a dynamic step size algorithm to minimize the number of steps: a step $\delta g$ is attempted, but is rejected if either the terms computed in the Taylor series grow in size or if after the Newton-Raphson solution the change in the EBV with respect to the previous step is too large. (While we haven't performed exhaustive optimization, a 25\% change in the norm of the EBV is judged to be too large.) A rejected step is reattempted with $\frac{1}{2} \delta g$. With the dynamic step-size procedure, the number of steps appears to scale logarithmically with $g$: the growth in the number of steps required for a half-filled 40-site reduced BCS Hamiltonians with equally-spaced levels is shown in Figure \ref{fig:log}. In section \ref{sec:results}, it will be pertinent to employ the state with alternating occupations $1010\dots 10$ which we will call the N\'{e}el state. In the pair representation of su(2), this state (at zero coupling) corresponds to alternating pair-empty occupations of the spatial orbitals. In a $\frac{1}{2}$-spin representation of su(2) this state is the up-down N\'{e}el ordering. 

With a solution for the EBV, the rapidities are extracted by the procedure of ref.\cite{elaraby:2012} Consider a polynomial $P(z)$, whose roots are the rapidities
\begin{align}
P(z) = \prod^M_{a=1} (z-u_a).
\end{align}
It is straightforward to show that the logarithmic derivative of $P(z)$ evaluated at the single-particle energies gives the EBV scaled by $g$
\begin{align}
\frac{1}{g} U_i = \frac{P'(\varepsilon_i)}{P(\varepsilon_i)}.
\end{align}
Further, Richardson's equations \eqref{eq:rich_eq} are equivalent to a 2nd-order ordinary differential equation (ODE) for $P(z)$:
\begin{align} \label{eq:ODE}
P''(z) - F(z)P'(z) + G(z)P(z) = 0,
\end{align}
with
\begin{align}
F(z) &= \frac{2}{g} + \sum^N_{i=1} \frac{1}{(z-\varepsilon_i)} \\
G(z) &= \frac{1}{g} \sum^n_{i=1} \frac{U_i}{(z-\varepsilon_i)}.
\end{align}
Now it should be clear why the rapidities are special: they are the roots of an orthogonal polynomial. To obtain them, the polynomial $P(z)$ is Lagrange-interpolated on a grid of $M+1$ points $\{z_a\}$ which reduces the ODE \eqref{eq:ODE} to a set of linear equations for the weights of $P(z)$ in the Lagrange basis. This yields a representation of $P(z)$ which solves \eqref{eq:ODE}, whose roots are extracted one at a time by Laguerre's method with deflation. This procedure works very well provided that the chosen grid is a reasonable guess for the rapidities. This means that an initial grid chosen at $g=0$ can be evolved by solving for the rapidities at intermediate steps in the EBV solution. It is not necessary to solve for the rapidities at each step. The grid evolves well enough by solving at every other step of the EBV solution. In refs.\cite{faribault:2010,elaraby:2012} the authors point out that if the rapidities are only required at a large $g$, then a physical argument can be made to estimate the grid-points, eliminating intermediate calculations. Rather than Lagrange interpolation, we considered using Chebyshev polynomials as for real-valued functions Chebyshev interpolation seems to be the unambiguously correct approach.\cite{trefethen_book} We abandoned this idea rather quickly as the rapidities generally occur in complex-conjugate pairs, for which Chebyshev interpolation is much less efficient.

The polynomial $P(z)$ is written in the basis of Lagrange polynomials with the grid $\{z_a\}$ as
\begin{align}
P(z) = l(z) \sum^{M+1}_{a=1} \frac{w_a}{z-z_a}
\end{align}
with $\{w\}$ the weights, and 
\begin{align}
l(z) = \prod^{M+1}_{a=1} (z-z_a).
\end{align}
The first derivative of $P(z)$ is easily evaluated 
\begin{align}
P'(z) = l'(z) \sum^{M+1}_{a=1} \frac{w_a}{z-z_a} - l(z) \sum^{M+1}_{a=1} \frac{w_a}{(z-z_a)^2},
\end{align}
and simplifies when divided by $l(z)$
\begin{align}
\frac{P'(z)}{l(z)} = \sum_{a\neq b} \frac{w_a}{(z-z_a)(z-z_b)}.
\end{align}
Likewise, the second derivative scaled by $l(z)$ is simple
\begin{align}
\frac{P''(z)}{l(z)} = \sum_{a\neq b \neq c} \frac{w_a}{(z-z_a)(z-z_b)(z-z_c)}.
\end{align}
Substituting these expressions into the ODE \eqref{eq:ODE}, and evaluating at the grid-points gives a set of linear equations for the weights $\{w\}$
\begin{align}
\sum_{b\neq c (\neq a)} \frac{w_a + 2w_b}{(z_a-z_b)(z_a-z_c)} - F(z_a) \left( \frac{w_a +w_b}{z_a-z_b} \right) + G(z_a) w_a = 0, \quad 1 \leq a \leq M+1,
\end{align}
along with a normalization on the weights
\begin{align}
\sum^{M+1}_{a=1} w_a = 1.
\end{align}
Taken together this is a system of over-determined linear equations 
\begin{align}
B \textbf{w} = \textbf{b}
\end{align}
for the weights which is again solved by QR factorization, with $B=QR$ as
\begin{align} \label{eq:linear_weight}
R \textbf{w} = Q^{\dagger} \textbf{b}.
\end{align}
The $(M+2) \times (M+1)$ matrix $B$ has elements
\begin{align}
B_{ab} &= \begin{cases}
\sum_{c \neq d (\neq a)} \frac{1}{(z_a-z_c)(z_a-z_d)} - \sum_{c(\neq a)} \frac{F(z_a)}{z_a-z_c}  +G(z_a), &\quad a = b \\
\sum_{c (\neq a,b)} \frac{2}{(z_a-z_b)(z_a-z_c)} - \frac{F(z_a)}{z_a-z_b}  , &\quad a \neq b
\end{cases}, \quad  &1 \leq a,b \leq M+1, \nonumber \\
B_{M+2,b} &= 1
\end{align}
and the RHS $\textbf{b}$ is $M+1$ zeros followed by a single entry 1. Solving the linear equations \eqref{eq:linear_weight} gives a representation of the polynomial $P(z)$ which satisfies the ODE \eqref{eq:ODE}. A single root $\alpha$ may be found with Laguerre's method, and the polynomial is deflated by removing the grid point $z_{\alpha}$ closest to the root. In the Lagrange polynomial basis, this amounts to removing the weight $w_{\alpha}$ corresponding to the root, and causes the other weights to be modified
\begin{align}
w_b \rightarrow w_b \frac{z_b - z_{\alpha}}{z_b - \alpha}.
\end{align}
The new weights represent a polynomial of the original rank minus one, and the procedure may be repeated until all roots are found. Provided the choice of grid is ``close enough'' to the roots, this method works quite well.

The grid consists of $M+1$ points. At $g=0$, we assign $M$ of them to be the single-particle energies of the occupied levels (corresponding to which EBV are non-zero) plus some small random noise, while the last grid-point is chosen as 10 times the largest single-particle energy. At each iteration the grid is updated to the rapidities from the previous iteration, unless the difference between a grid point and a previous rapidity is less than a threshold. As $g$ increases, many rapidities will stabilize and this threshold is imposed to prevent the matrix $B$ from having zero denominators. The final grid-point is updated to be 10 times the absolute value of the largest rapidity.

Finally, with a set of rapidities at the desired $g$, a Newton-Raphson routine applied directly to Richardson's equations \eqref{eq:rich_eq} can be used to ``polish'' the rapidities to higher precision. The whole construction could thus be viewed as a way of pre-conditioning a Newton-Raphson solver at the final $g$. With this approach we've been able to obtain rapidities to double precision.

\subsection{Energy functional}
We want to solve the Coulomb Hamiltonian for molecules
\begin{align} \label{eq:C_ham}
\hat{H}_C = \sum_{ij} h_{ij} \sum_{\sigma} a^{\dagger}_{i \sigma} a_{j \sigma} + \frac{1}{2} \sum_{ijkl} V_{ijkl} \sum_{\sigma \tau} a^{\dagger}_{i \sigma} a^{\dagger}_{j \tau} a_{l \tau} a_{k \sigma}
\end{align}
in which the 1- and 2-electron integrals are computed in a basis $\{\phi\}$
\begin{align}
h_{ij} &= \int d\mathbf{r} \phi^*_i (\mathbf{r}) \left( - \frac{1}{2} \nabla^2 - \sum_I \frac{Z_I}{| \mathbf{r} - \mathbf{R}_I |} \right) \phi_j (\mathbf{r}) \\
V_{ijkl} &= \int d\mathbf{r}_1 d\mathbf{r}_2 \frac{\phi^*_i(\mathbf{r}_1)  \phi^*_j(\mathbf{r}_2)  \phi_k(\mathbf{r}_1)  \phi_l(\mathbf{r}_2)  }{| \mathbf{r}_1 - \mathbf{r}_2|}.
\end{align}
RG states have zero seniority, meaning that the energy of \eqref{eq:C_ham} computed with an RG state $\ket{\{u\}}$
\begin{align} \label{eq:sz_energy}
E [\{\varepsilon\},g] = 2 \sum_i h_{ii} \gamma_i + \sum_{i \neq j} (2V_{ijij} - V_{ijji})D_{ij} + \sum_{ij} V_{iijj} P_{ij}
\end{align}
depends only on the  1-RDM elements $\gamma_i$ along with the non-zero elements of the 2-RDM, the diagonal-correlation function $D_{ij}$ and the pair-correlation function $P_{ij}$. Explicitly, these elements are
\begin{subequations} \label{eq:sz_dm}
\begin{align}
\gamma_i &= \frac{1}{2} \frac{\braket{ \{u\} | \hat{n}_i | \{u\} }}{\braket{ \{u\} | \{u\} }} \\
D_{ij} &= \frac{1}{4}   \frac{\braket{ \{u\} | \hat{n}_i \hat{n}_j | \{u\} }} {\braket{ \{u\} | \{u\} }} \\
P_{ij} &= \frac{\braket{\{u\} | S^+_k S^-_l | \{u\}}} {\braket{\{u\} | \{u\} }}.
\end{align}
\end{subequations}
It is important to note that as written, $D_{ii}$ and $P_{ii}$ correspond to the same element of the 2-RDM. As a convention, we assign this element to $P_{ii}=\gamma_i$ while taking $D_{ii}=0$. The energy \eqref{eq:sz_energy} will be minimized with respect to the single-particle energies $\{\varepsilon\}$ and the pairing strength $g$ as they are the variables we consider. A choice of $\{\varepsilon\}$ and $g$ dictates the values of the rapidities $\{u\}$ through Richardson's equations \eqref{eq:rich_eq}.

The RDM elements for RG states have been written many times.\cite{amico:2002,faribault:2008,faribault:2010,gorohovsky:2011,fecteau:2020} RDM elements of any order can be evaluated with the derivatives of the rapidities with respect to the single-particle energies, which are obtained as solutions of the linear equations
\begin{align} \label{eq:gmat_lin}
G \frac{\partial \textbf{u}}{\partial \varepsilon_k} = b_k
\end{align}
with $G$ the Gaudin matrix
\begin{align} \label{eq:gmat}
G_{ab} = 
\begin{cases}
\sum_i \frac{1}{(u_a - \varepsilon_i)^2} -\sum_{c \neq a} \frac{2}{( u_a - u_c )^2} , &\quad a = b \\
\frac{2}{(u_a - u_b )^2}, &\quad a \neq b.
\end{cases}
\end{align}
In particular, the normalized 1-RDM elements are
\begin{align}
\gamma_k = \sum_a \frac{\partial u_a}{\partial \varepsilon_k}
\end{align}
while the 2-RDM elements are, for $i \neq j$,
\begin{align}
D_{ij} &= \sum_{a<b} \frac{(u_a - \varepsilon_i)(u_b-\varepsilon_j) + (u_a - \varepsilon_j)(u_b-\varepsilon_i)}{(\varepsilon_i - \varepsilon_j) (u_b-u_a)} \left( \frac{\partial u_a}{\partial \varepsilon_i}  \frac{\partial u_b}{\partial \varepsilon_j} -  \frac{\partial u_a}{\partial \varepsilon_j} \frac{\partial u_b}{\partial \varepsilon_i} \right) \\
P_{ij} &= \sum_a \frac{(u_a - \varepsilon_i)}{(u_a - \varepsilon_j)} \frac{\partial u_a}{\partial \varepsilon_i}
-2 \sum_{a<b} \frac{(u_b - \varepsilon_i)(u_a-\varepsilon_i) }{(\varepsilon_i - \varepsilon_j) (u_b-u_a)}
\left( \frac{\partial u_a}{\partial \varepsilon_i}  \frac{\partial u_b}{\partial \varepsilon_j} -  \frac{\partial u_a}{\partial \varepsilon_j} \frac{\partial u_b}{\partial \varepsilon_i} \right).
\end{align}
Thus to compute the 2-RDM elements all that is required is to solve the $N$ sets of linear equations \eqref{eq:gmat_lin}. As $G$ is square, it is cheaper to use LU-factorization.\cite{trefethen_book_2} Each set of linear equations can be solved with cost $\mathcal{O}(M^2)$, so that the $N$ sets of linear equations are thus solvable with a cost of $\mathcal{O}(NM^2)$. The scaling bottleneck comes from evaluating the 2-RDM: there are $\mathcal{O}(N^2)$ elements, and each requires a summation over $M^2$ objects. The energy \eqref{eq:sz_energy} can thus be evaluated with $\mathcal{O}(N^2M^2)$ scaling. Solving Richardson's equations should be of similar or lower scaling. Each step requires a few linear algebra operations, thus should scale like $\mathcal{O}(N^3)$ and the number of steps grows logarithmically with the pairing strength for the dynamic step-size approach. 

\section{Numerical Results} \label{sec:results}
Dissociation curves were computed for the symmetric dissociation of hydrogen chains as well as for molecular nitrogen dissociation in the minimal basis STO-6G. In a given set of orbitals, RG is strictly a variational approximation to DOCI. The best possible case for RG is orbital-optimized (OO)-DOCI, and thus all presented data are in the OO-DOCI orbitals originally computed for ref.\cite{johnson:2020} using a multi-configurational self-consistent field (MCSCF) in the complete doubly-occupied space with GAMESS (US)\cite{barca:2020}. Full configuration interaction (FCI) results, also from ref.\cite{johnson:2020} computed with psi4,\cite{sherill:1999,parrish:2017} demonstrate that for the hydrogen chain dissociations OO-DOCI is close to exact while for molecular nitrogen it is not. Restricted Hartree-Fock (RHF) results were computed with Gaussian 16.\cite{gaussian_16}

The variational optimization has room for improvement. At present the variables $\{\varepsilon\}$ and $g$ are pre-conditioned with the covariance matrix adaptation evolution strategy (CMA-ES)\cite{hansen:2001} and optimized with the Nelder-Mead simplex algorithm.\cite{nelder:1965} This is more or less a brute force approach which is acceptable at present as these results are proof-of-principle. 

It is necessary to impose a consistency check on the RDM expressions. The energy of the reduced BCS Hamiltonian \eqref{eq:hbcs} can be computed in several different manners. It is the sum of the rapidities
\begin{align} \label{eq:rap_bcs}
E_{BCS} = \sum^M_{a=1} u_a,
\end{align}
it has an expression in terms of the EBV
\begin{align} \label{eq:ebv_bcs}
E_{BCS} = \frac{g}{2} M (M-N-1) + \frac{1}{2} \sum^N_{i=1} \varepsilon_i U_i,
\end{align}
and it can evaluated directly
\begin{align} \label{eq:rdm_bcs}
E_{BCS} = \sum^N_{i=1} \varepsilon_i \gamma_i - \frac{g}{2} \sum_{ij} P_{ij}.
\end{align}
The conditions \eqref{eq:rap_bcs} and \eqref{eq:ebv_bcs} are verified in the solver for Richardson's equations. At each energy evaluation in the variational optimization, the reduced BCS energy is computed with both \eqref{eq:rap_bcs} and \eqref{eq:rdm_bcs}. If the difference is larger than $1\times 10^{-6}$ the point is rejected. This loss in precision occurs when two single-particle energies come too close to one another, and could be avoided using the degenerate version of the EBV equations solver.\cite{elaraby:2012} This will be done in a future iteration, as at present the results converge quite well to the correct answer.

\begin{figure}[ht!] 
	\includegraphics[width=0.475\textwidth]{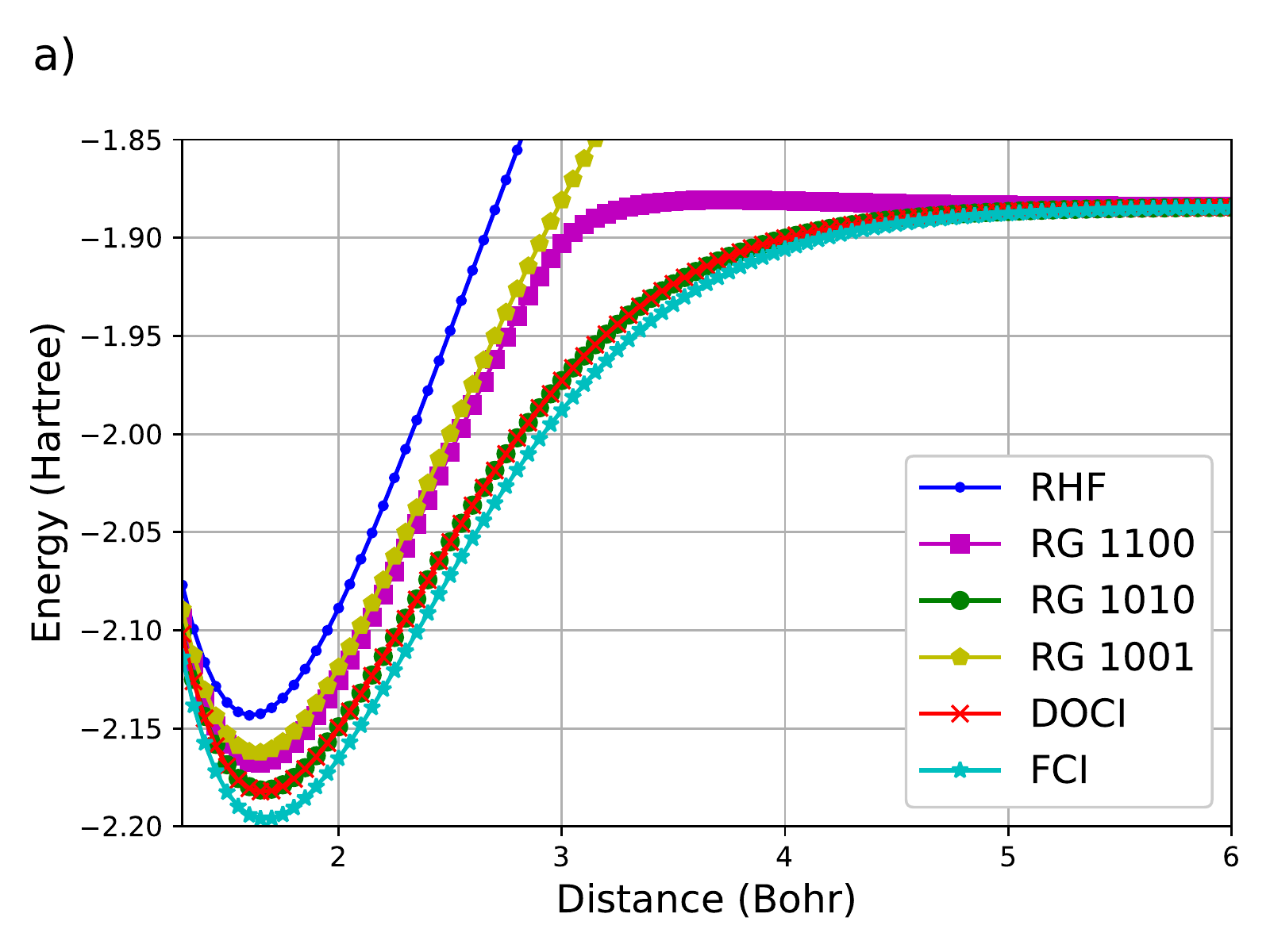} \hfill
	\includegraphics[width=0.475\textwidth]{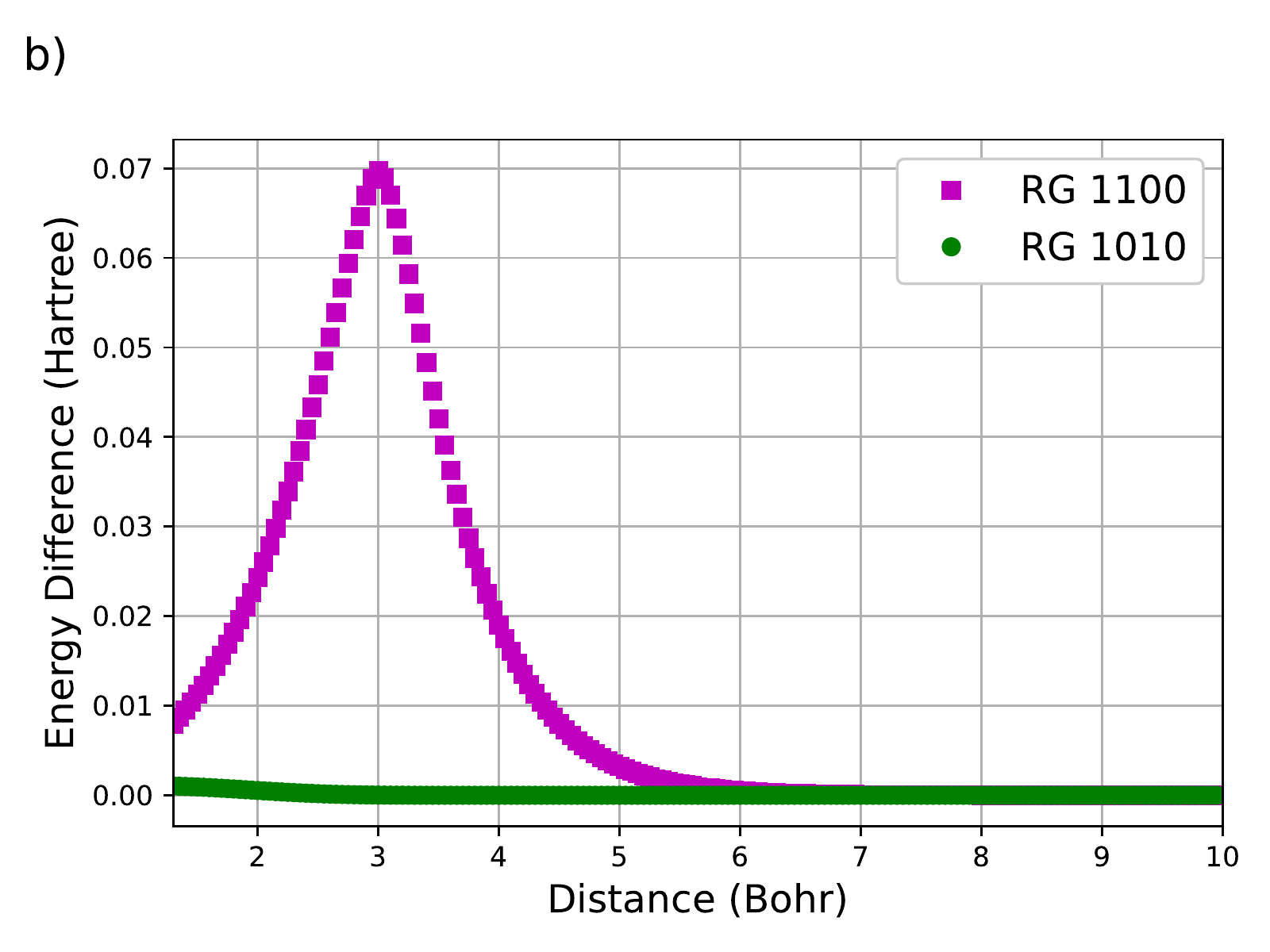}
	\caption{(a) Bond dissociation curves for H$_4$. (b) Deviations from DOCI for the 1100 and the 1010 RG states. All results computed with the STO-6G basis set in the basis of OO-DOCI orbitals.}
	\label{fig:H4_gs_all}
\end{figure}

In ref.\cite{johnson:2020} we observed that the RG ground state does not describe hydrogen chain dissociation correctly at intermediate distances. Near the minimum, the treatment is acceptable, and at dissociation the treatment is correct, but in the middle there is an unphysical local maximum. Similar behaviour for H$_8$ was observed in ref.\cite{bytautas:2011} We have recently shown\cite{moisset:2022} that this was not a failure of the RG wavefunction form, but must be due to the choice of using only RG ground states. We therefore look at all possible RG states to choose the best. For H$_4$ we consider the $\binom{4}{2}$ possible states and variationally optimize parameters for each. The results are presented in figure \ref{fig:H4_gs_all}.
First, as we are choosing a distribution (state) and optimizing the parameters $\{\varepsilon\}$ and $g$, not all distributions are unique: the ground state of a reduced BCS Hamiltonian with a negative (repulsive) interaction is equivalent to the highest excited state of a different reduced BCS Hamiltonian with a positive (attractive) interaction. As a result, there are only three distinct choices since with this equivalence $1100 = 0011$, $1010 = 0101$, and $1001 = 0110$. Variational calculations were performed for all six choices, but as the results are doubly-degenerate, only one set are plotted.

\begin{figure}[ht!]
	\begin{subfigure}{\textwidth}
		\includegraphics[width=0.475\textwidth]{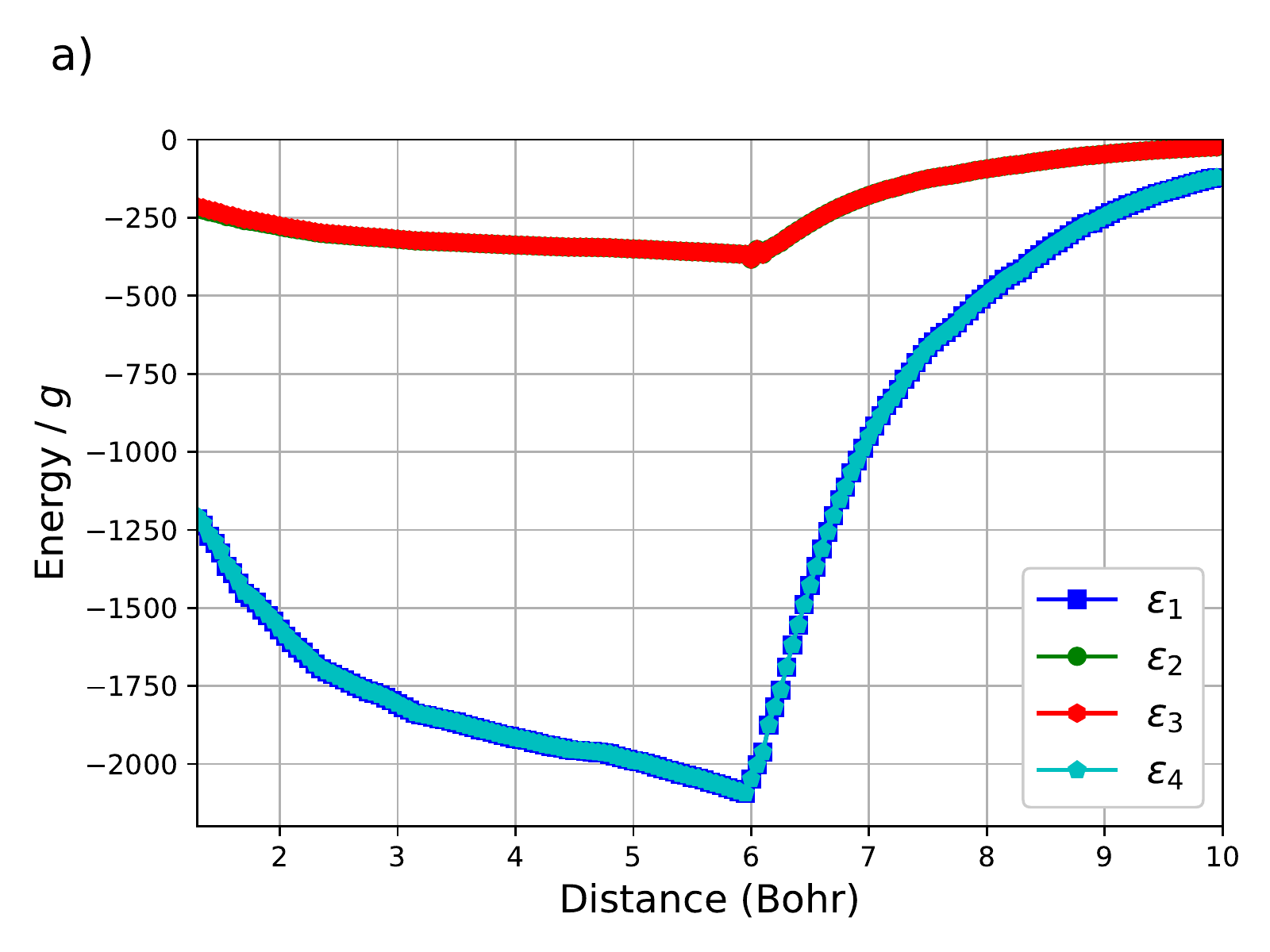} \hfill
		\includegraphics[width=0.475\textwidth]{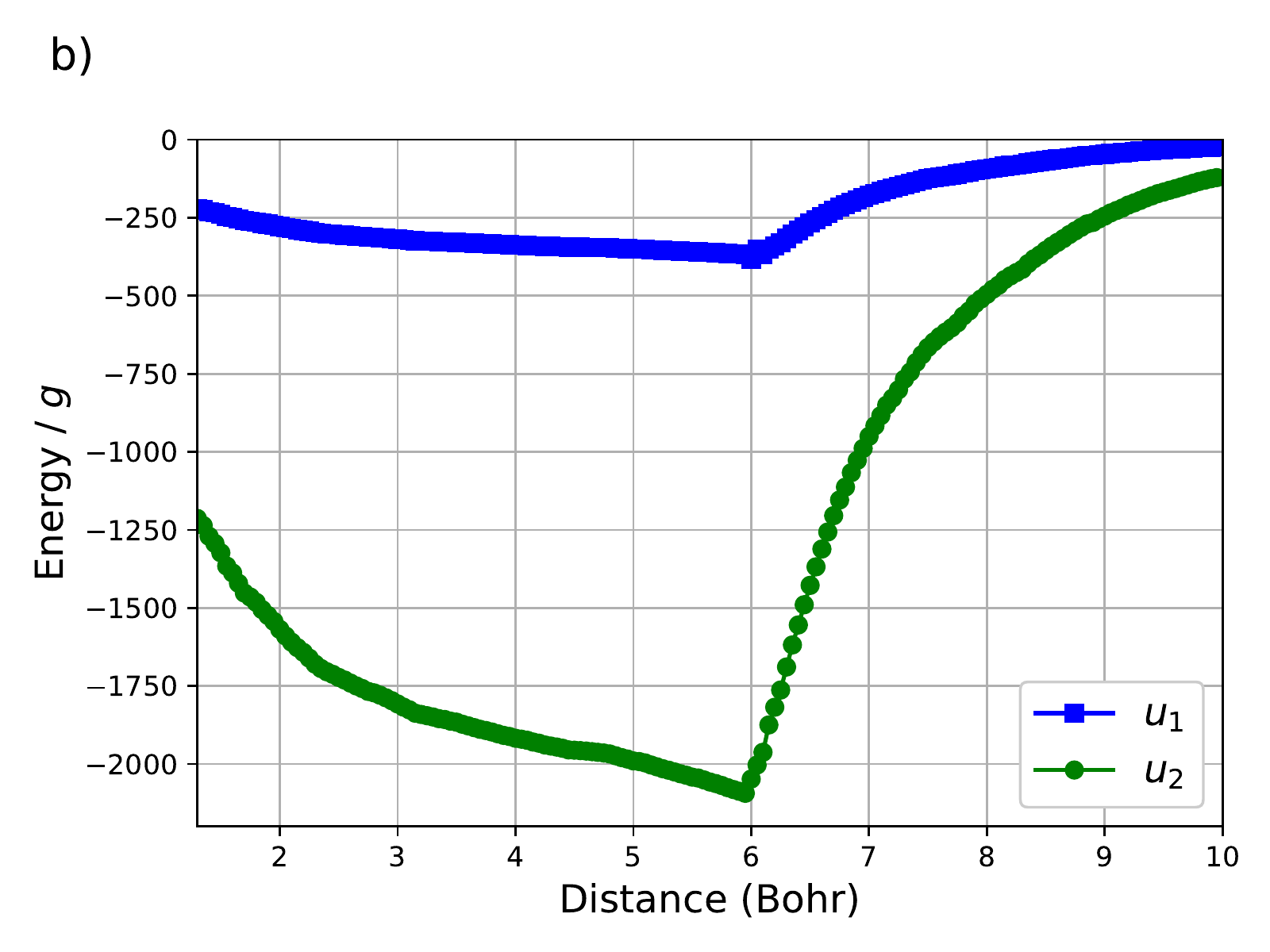}
	\end{subfigure}
		\caption{(a) Optimal reduced BCS parameters $\{\varepsilon\}$ scaled by $|g|$ for the 1010 RG state. (b) Rapidities of the 1010 RG state.}
		\label{fig:H4_BCS_params}
\end{figure}

\begin{figure}[ht!]
	\begin{subfigure}{\textwidth}
		\includegraphics[width=0.3\textwidth]{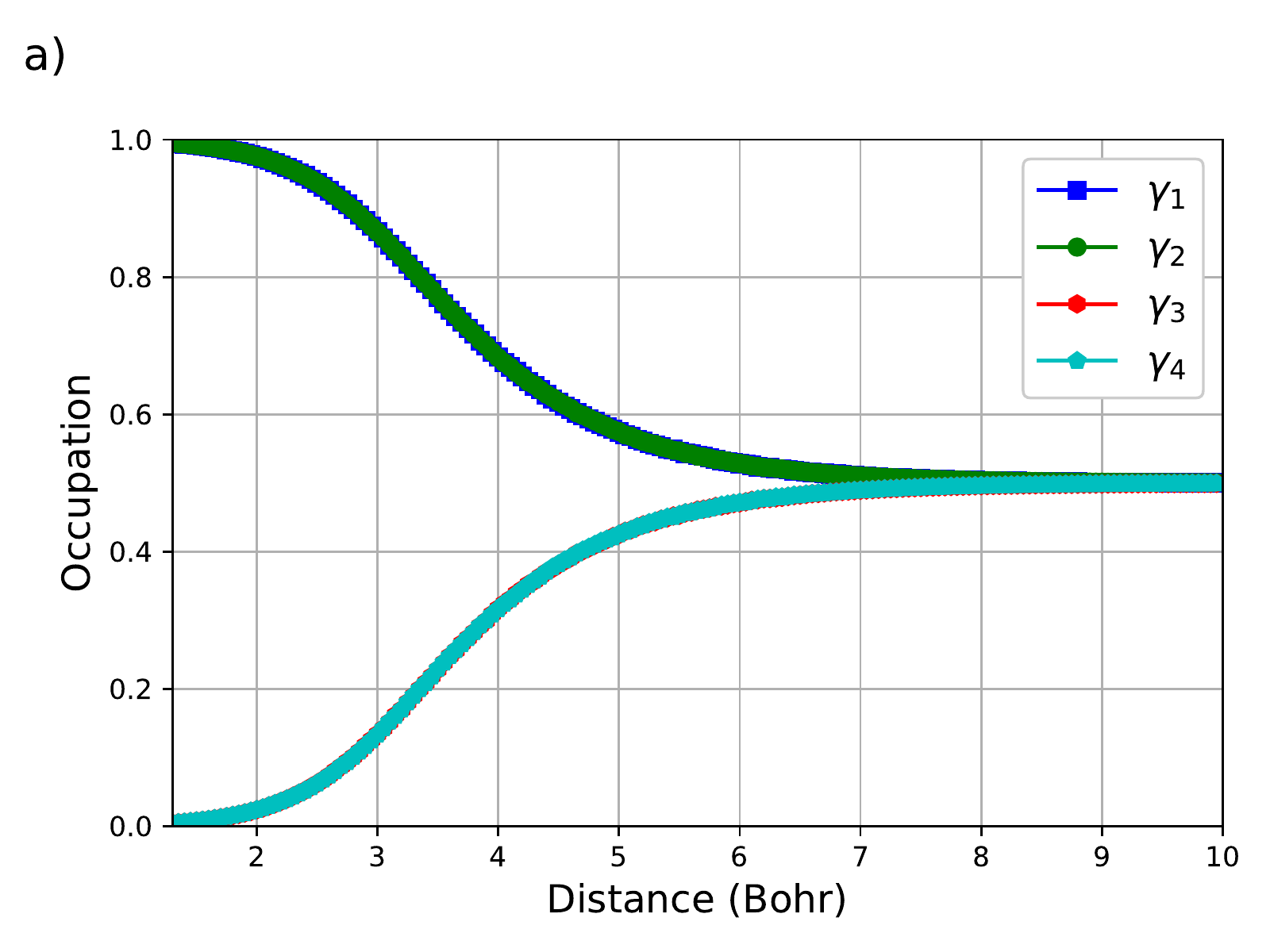} \hfill
		\includegraphics[width=0.3\textwidth]{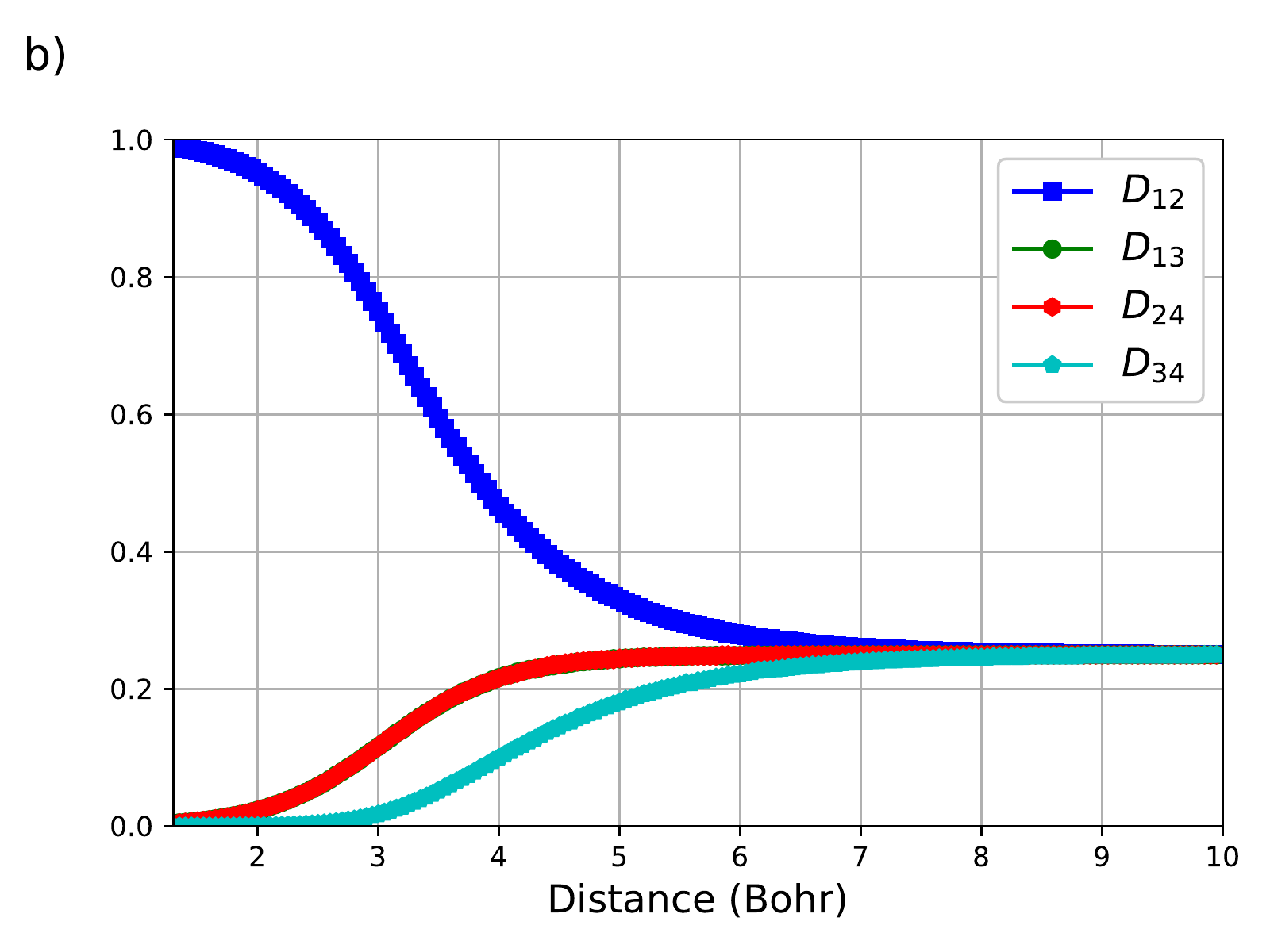} \hfill
		\includegraphics[width=0.3\textwidth]{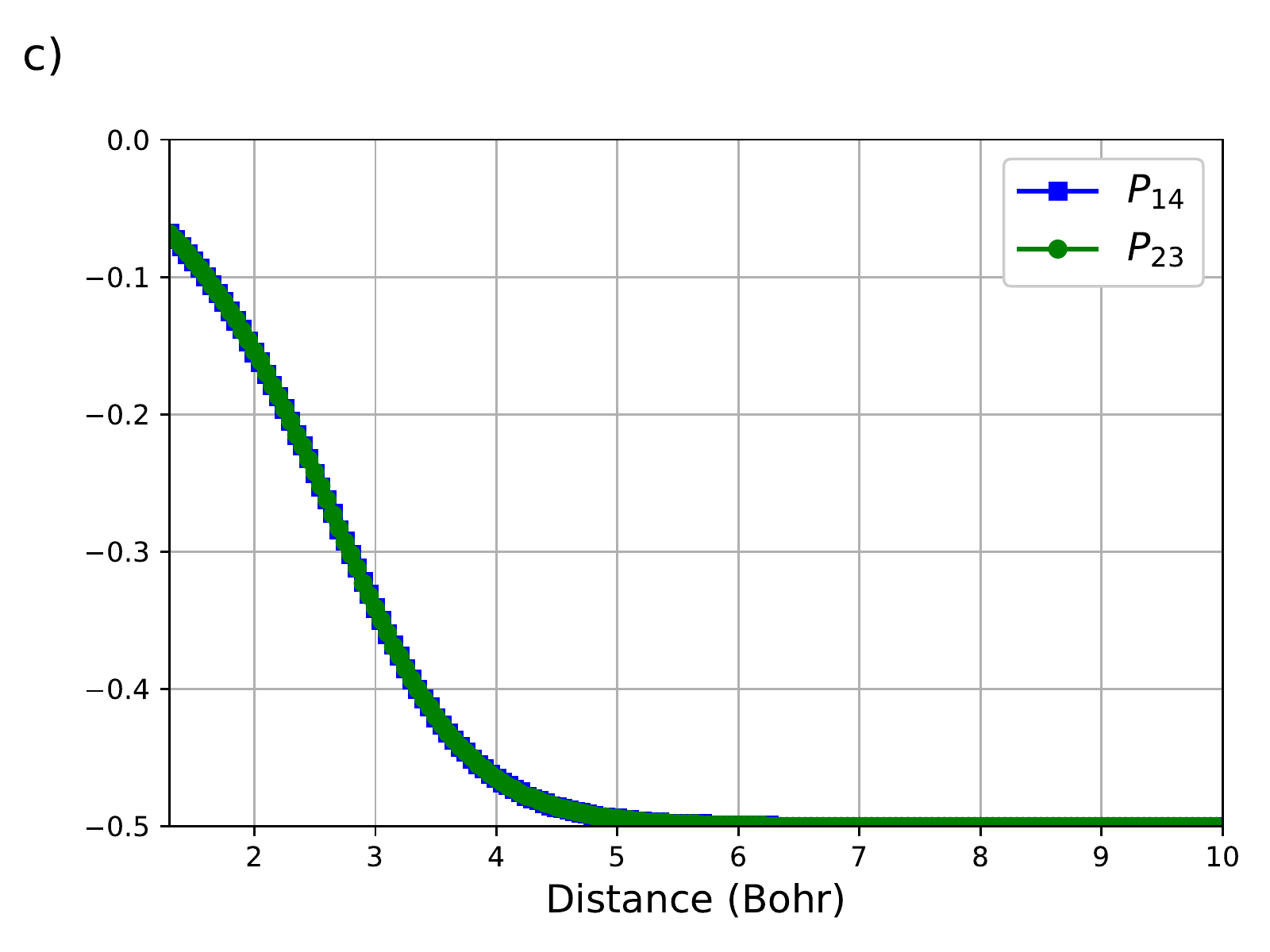}
	\end{subfigure}
		\caption{(a) 1-RDM elements of the 1010 RG state. (b) Non-zero $D_{ij}$ elements of the 1010 RG state. (c) Non-zero $P_{ij}$ elements of the 1010 RG state.}
		\label{fig:H4_rdms}
\end{figure}

It is clear that the 1010 state is qualitatively correct, while the others are not. The 1001 state reproduces the AGP results of ref.\cite{moisset:2022} while the 1100 state corresponds to the H$_8$ result observed in ref.\cite{bytautas:2011} Energy differences with respect to DOCI are presented in figure \ref{fig:H4_gs_all}b. We expected that the optimal choice of state would differ at the minimum and at dissociation, but this does not appear to be the case: the 1010 state is optimal everywhere. Can the choice 1010 be reasoned physically? To do so, we must look at the parameters defining the optimal reduced BCS Hamiltonian for the state 1010, which are shown in figure \ref{fig:H4_BCS_params}.
The single-particle energies $\{\varepsilon\}$ scaled by the absolute value of the pairing interaction are plotted in figure \ref{fig:H4_BCS_params}a. The interaction is always repulsive. The single-particle energies are separated into two groups, with both members of each group being near-degenerate, though distinct. The rapidities corresponding to the 1010 state of this Hamiltonian are shown in figure \ref{fig:H4_BCS_params}b. The rapidities are each trapped between the near-degenerate single-particle energies, i.e. $\varepsilon_1 < u_2 < \varepsilon_4$ and $\varepsilon_2 < u_1 < \varepsilon_3$. The parameters are breaking the system into two sub-systems and the state 1010 puts 1 rapidity in each sub-system. The individual subsystems correspond to H$_2$ dissociations, which the RG state 10 treats exactly.\cite{johnson:2020} The OO-DOCI orbitals, shown in ref. \cite{ward_thesis}, correspond to individual H$_2$ systems. There is a cusp in the optimal parameters near $r=6.0$ bohr, though this is not reflected in the energy of the 1010 state applied to the Coulomb Hamiltonian for H$_4$. In figure \ref{fig:H4_gs_all}a) the axis is cut near $r=6.0$ in order to better show behaviour of the other curves near the minimum. The 1010 result is smooth to $r=10.0$ bohr. The non-zero RDM elements for the 1010 state are shown in figure \ref{fig:H4_rdms}.
They are clearly smooth and approach the correct limits at dissociation: each of the 1-RDM elements is $\frac{1}{2}$, the $D_{ij}$ elements \emph{between} the two sub-systems are $\frac{1}{4}$ and the $P_{ij}$ \emph{within} sub-systems are $-\frac{1}{2}$. The only non-zero two-electron integrals, in physicists' notation, are the direct integrals $V_{ijij}$ between sub-systems and the pair-transfer integrals $V_{iijj}$ within sub-systems (for real orbitals, these are the exchange integrals $V_{ijji}$). Here, the antisymmetric product of strongly-orthogonal geminals (APSG), or equivalently the Piris natural orbital functional PNOF5,\cite{piris:2011,pernal:2013} is exact since for APSG the diagonal-correlation function is the product of the 1-RDM elements
\begin{align}
D^{APSG}_{ij} = \gamma_i \gamma_j
\end{align}
\emph{for $i$ and $j$ belonging to separate sub-systems}, which gives the correct result when $\gamma_i = \frac{1}{2}$. If $i$ and $j$ belong to the sub-system then $D_{ij}=0$, but the pair-correlation function
\begin{align}
P^{APSG}_{ij} = - \sqrt{\gamma_i \gamma_j}
\end{align}
is non-zero. The phase for $P_{ij}$ can be chosen for each pair, but the correct choice here is presented. There are two clear advantages to the present approach over APSG. First, APSG requires manually separating the orbital space into disjoint units, which is difficult. Here, the optimization naturally finds the correct expression. Second, APSG targets the ground state, whereas the reduced BCS Hamiltonian gives a complete set of states which, as we will now show, quite closely match the set of DOCI states.

\begin{figure}[ht!]
	\begin{subfigure}{\textwidth}
		\includegraphics[width=0.3\textwidth]{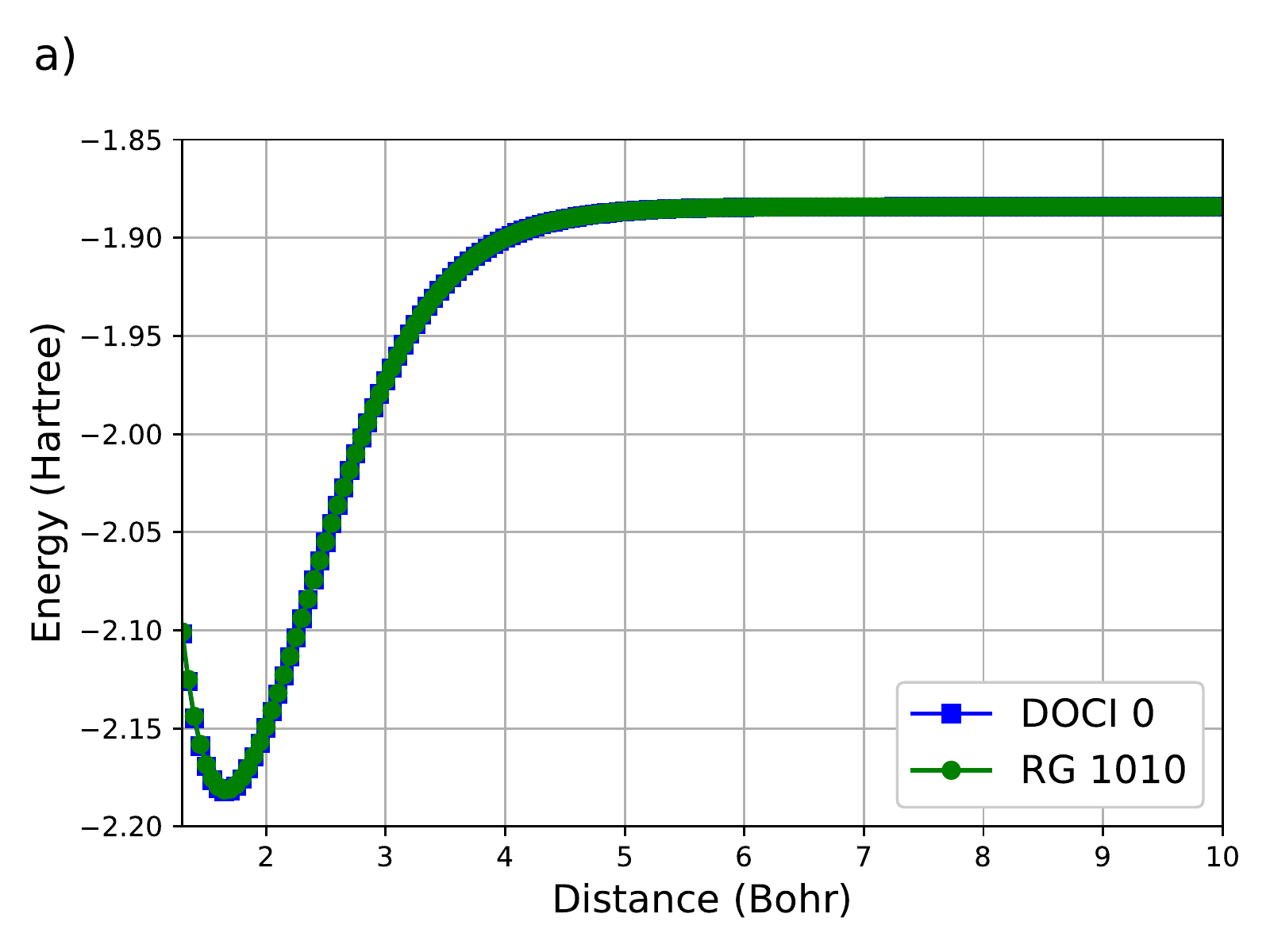} \hfill
		\includegraphics[width=0.3\textwidth]{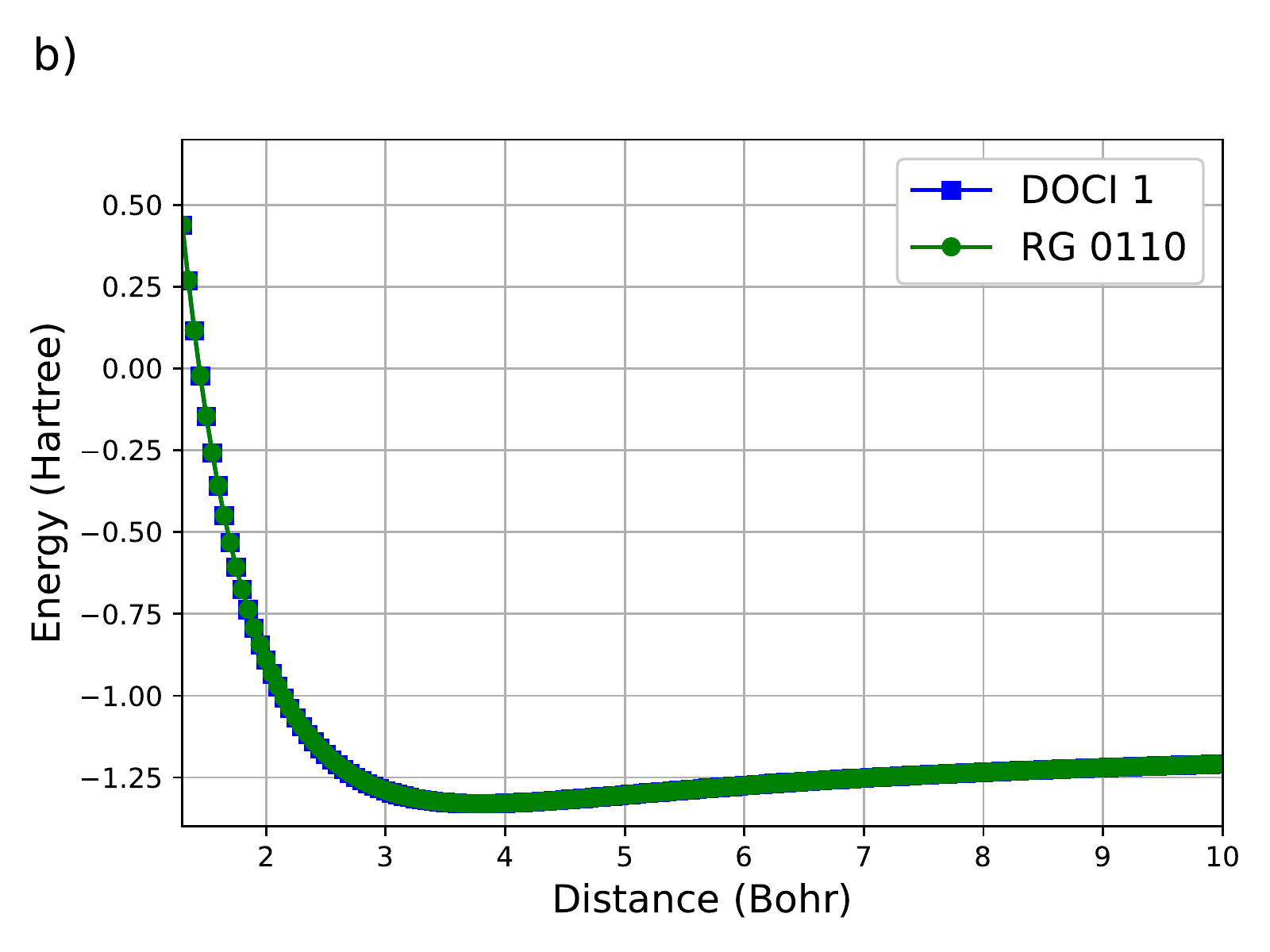} \hfill
		\includegraphics[width=0.3\textwidth]{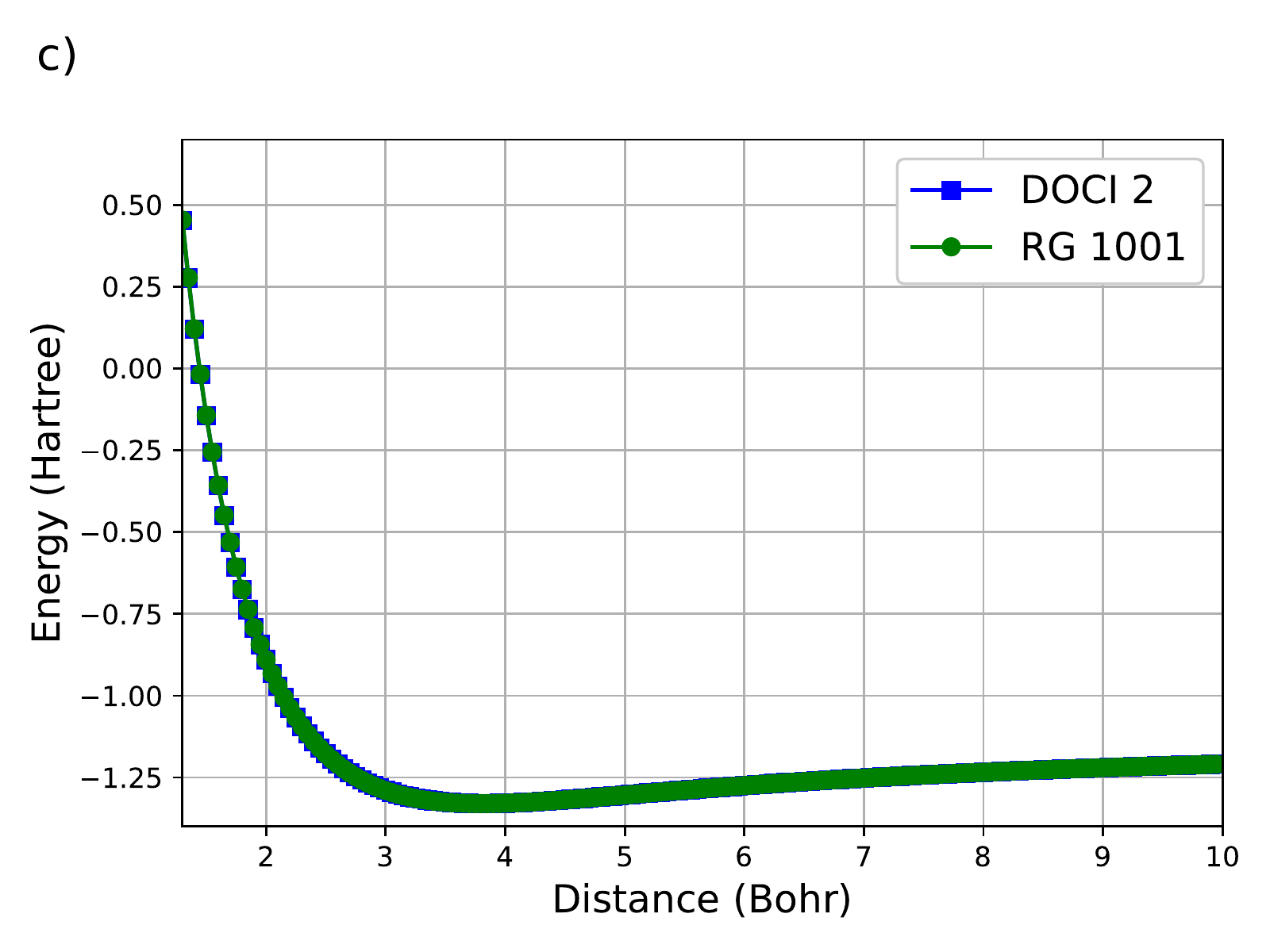}
	\end{subfigure}
	\begin{subfigure}{\textwidth}
		\includegraphics[width=0.3\textwidth]{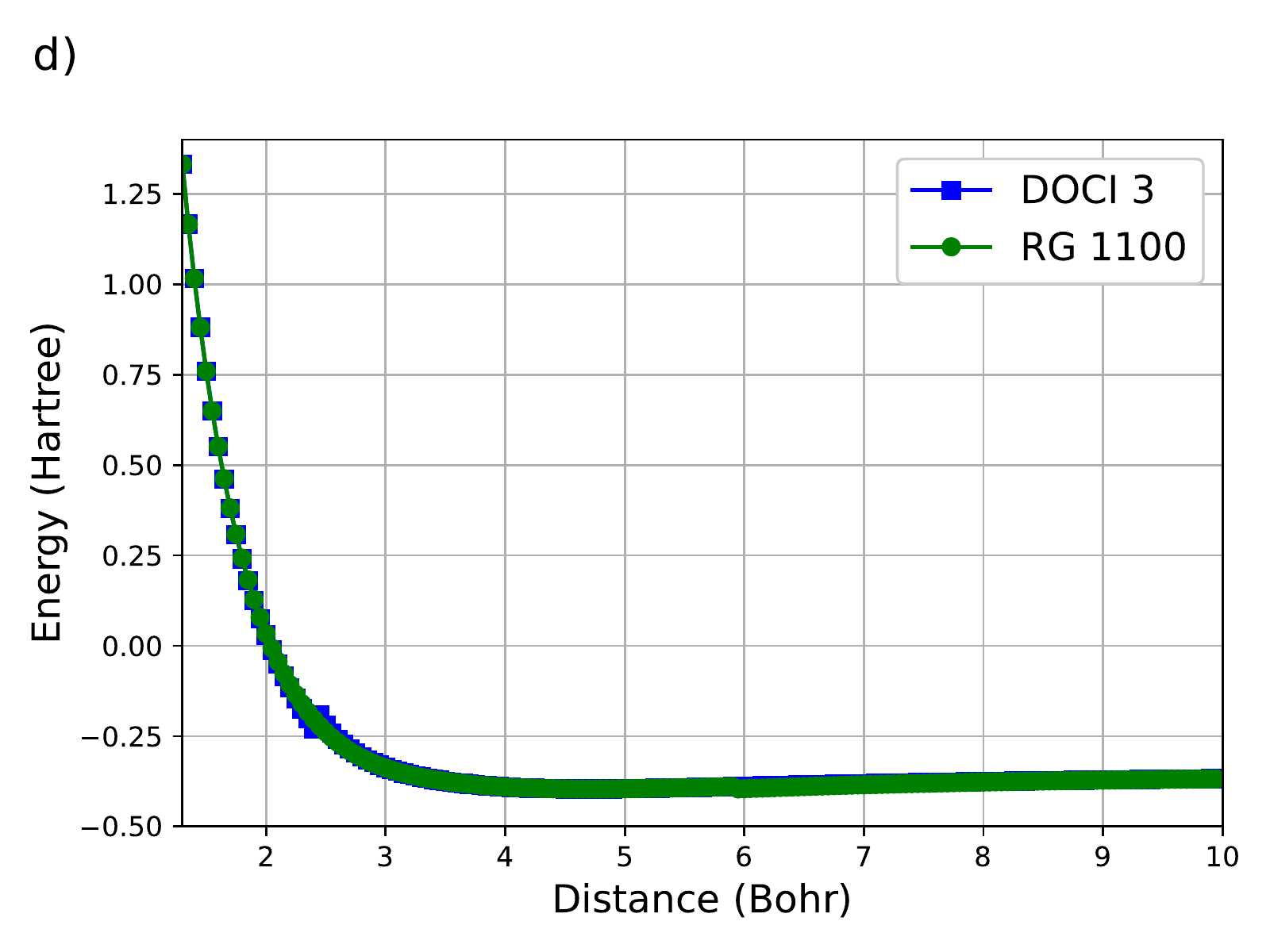} \hfill
		\includegraphics[width=0.3\textwidth]{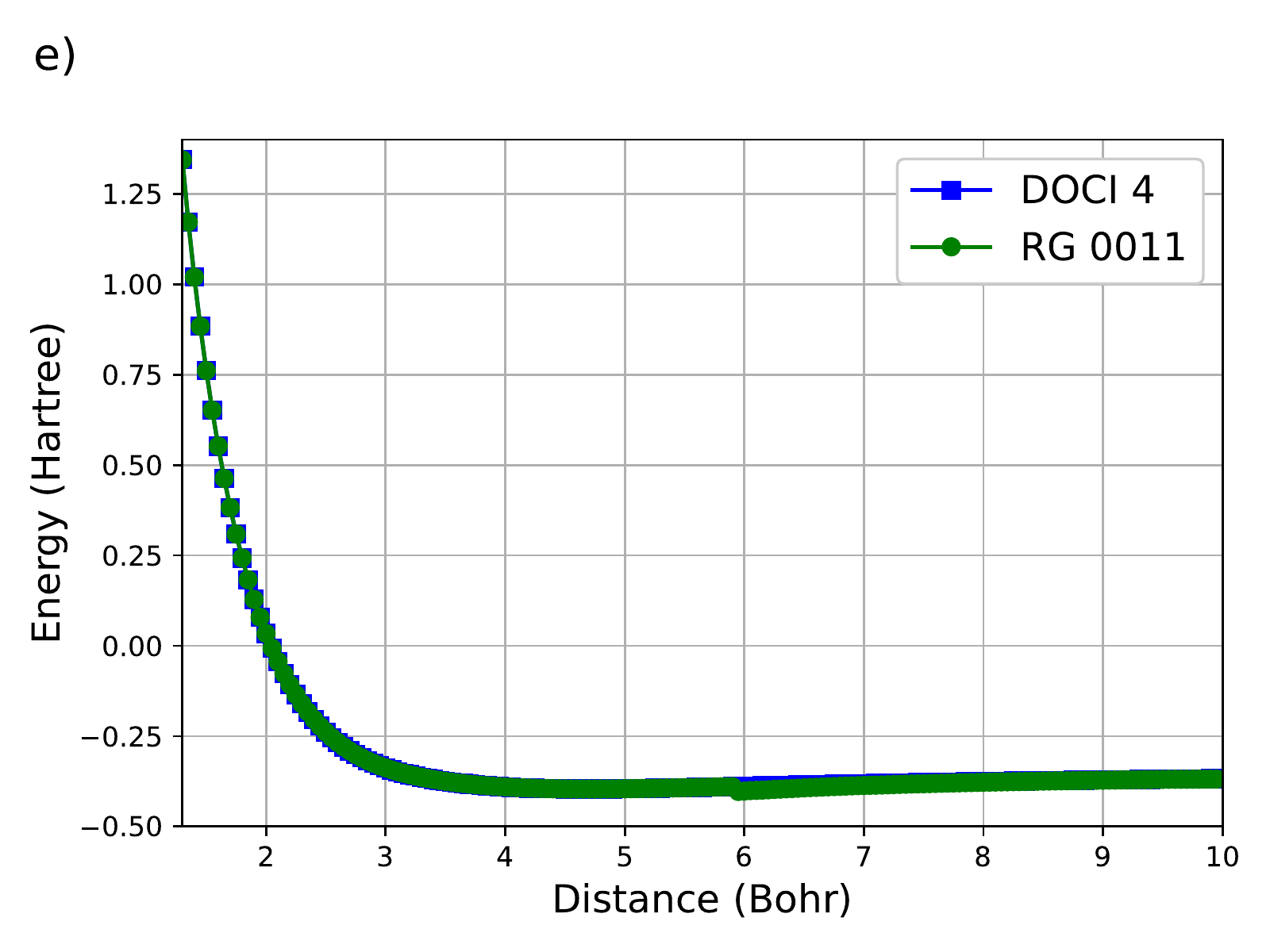} \hfill
		\includegraphics[width=0.3\textwidth]{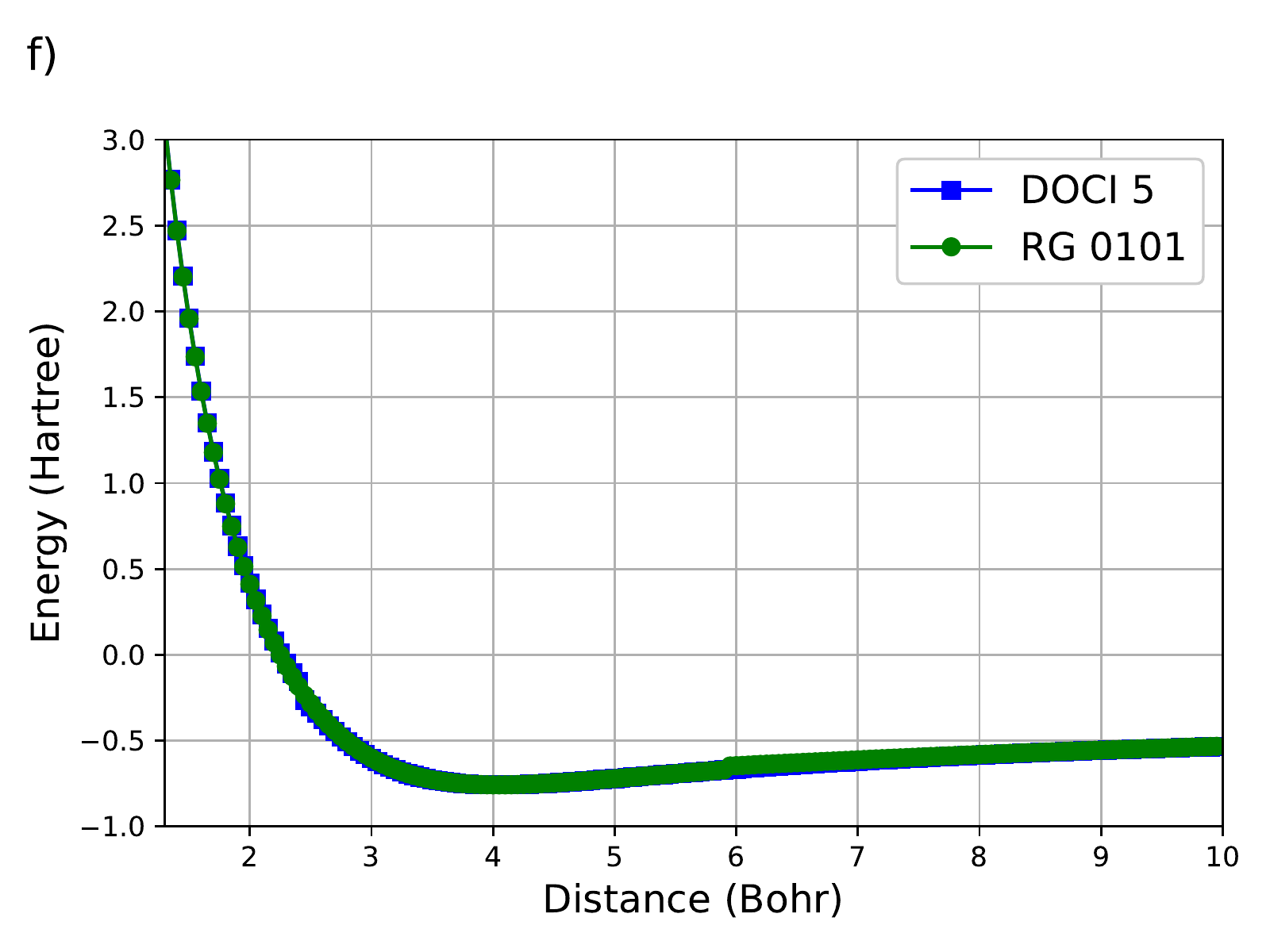}
	\end{subfigure}
		\caption{RG states computed with reduced BCS parameters defining the Hamiltonian corresponding to the optimal 1010 solution and DOCI states. All results computed in the STO-6G basis set in the basis of OO-DOCI orbitals.}
		\label{fig:H4_DOCI_curves}
\end{figure}

\begin{figure}[ht!]
	\begin{subfigure}{\textwidth}
		\includegraphics[width=0.3\textwidth]{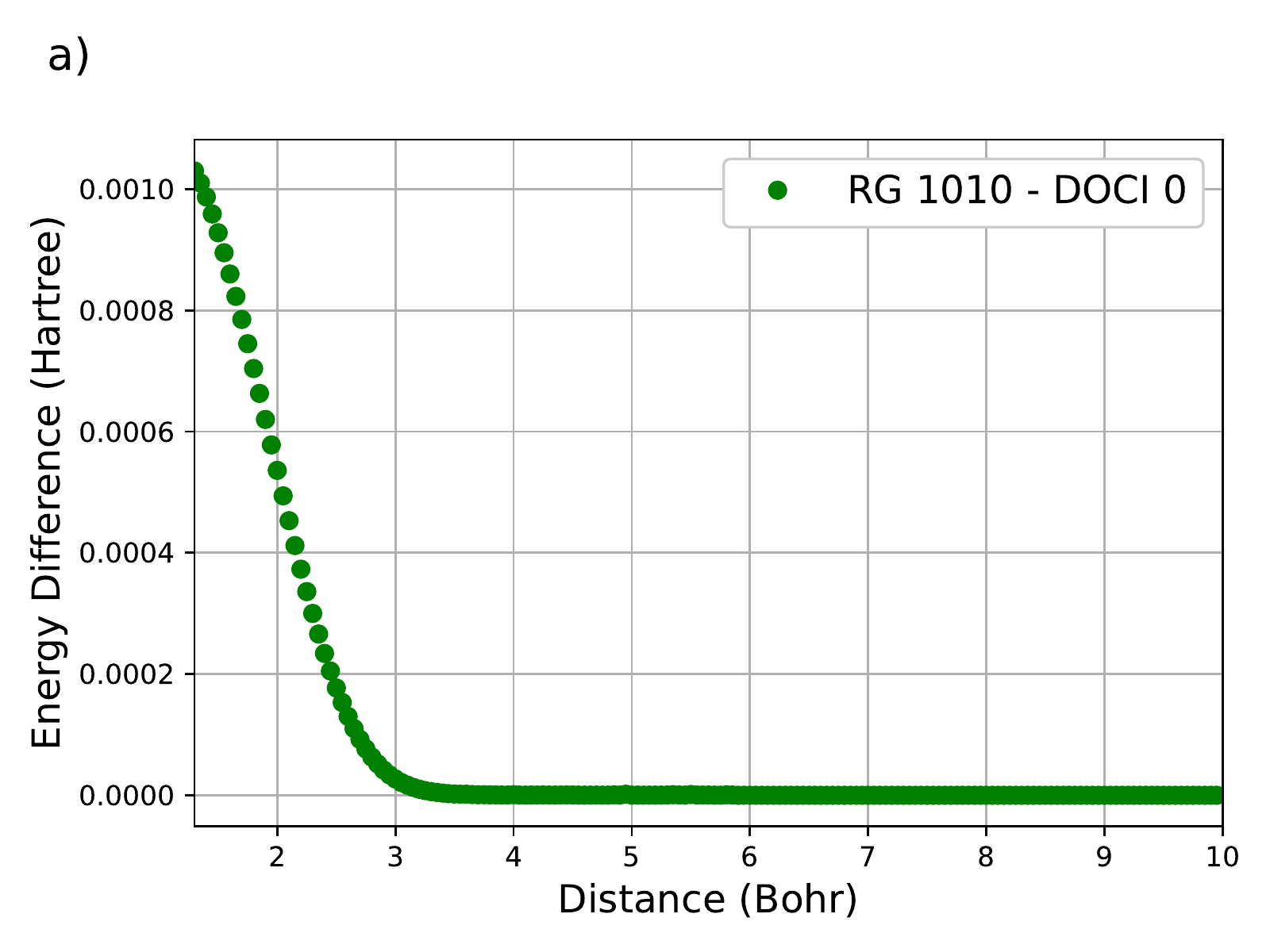} \hfill
		\includegraphics[width=0.3\textwidth]{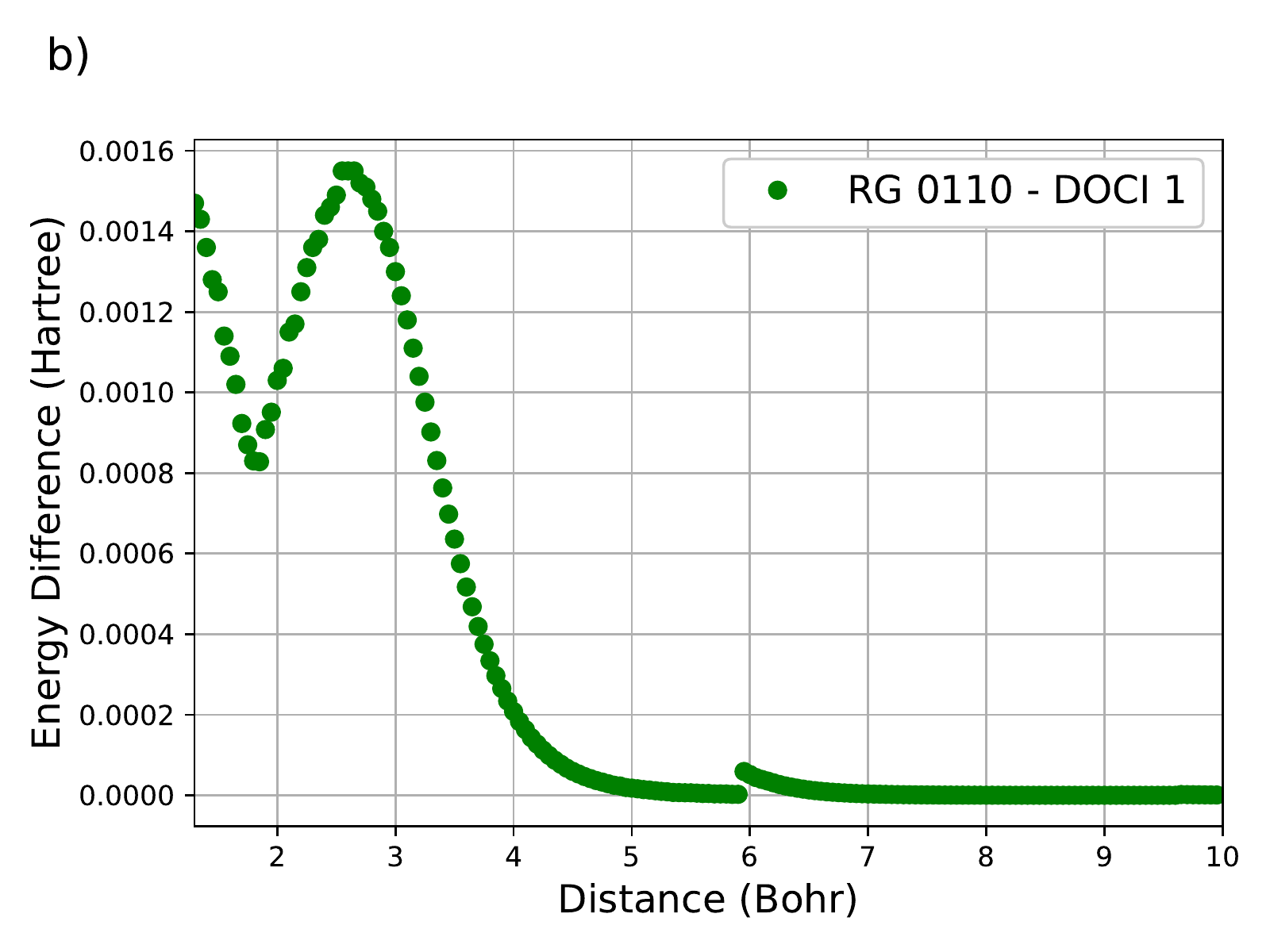} \hfill
		\includegraphics[width=0.3\textwidth]{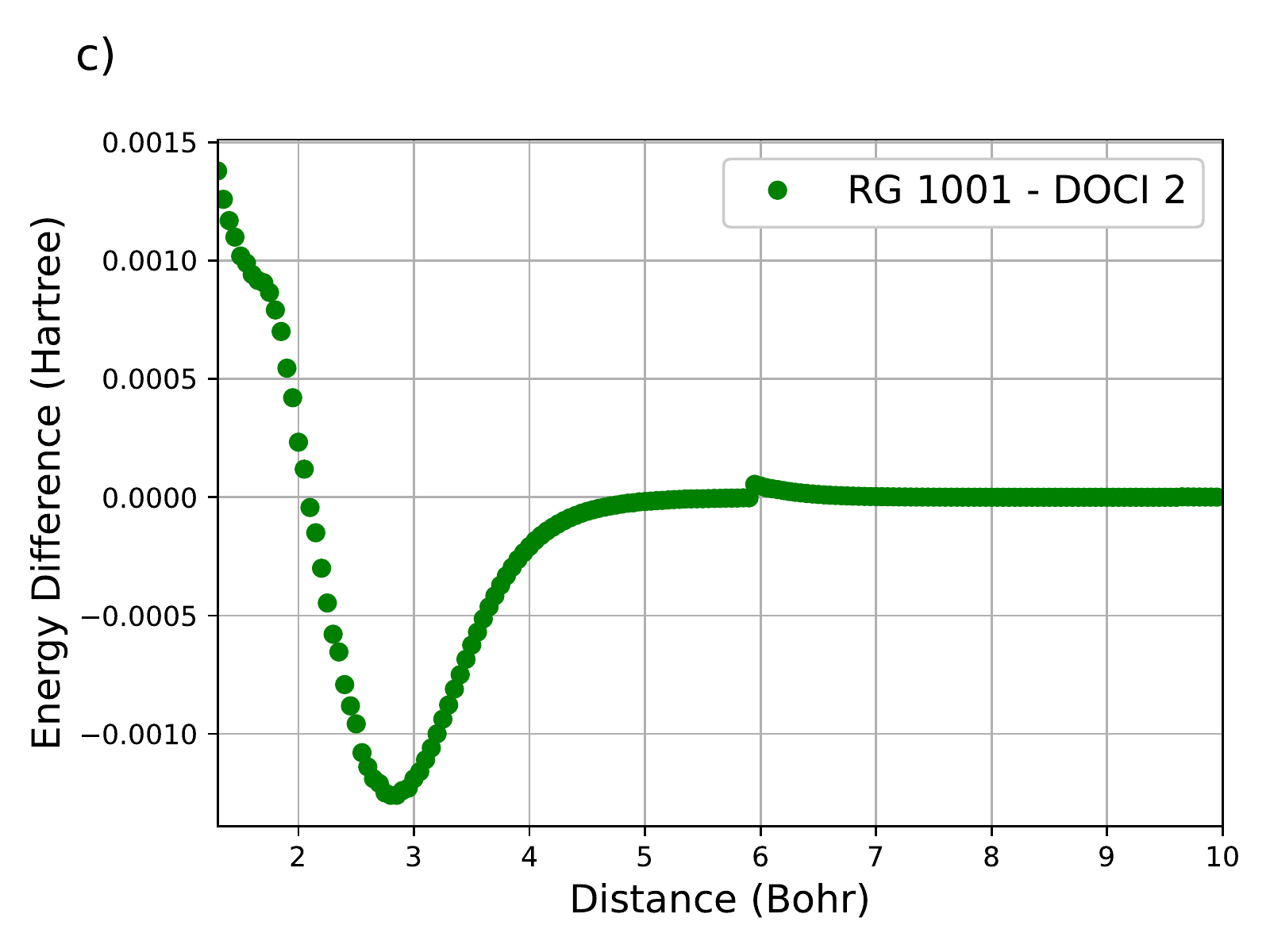}
	\end{subfigure}
	\begin{subfigure}{\textwidth}
		\includegraphics[width=0.3\textwidth]{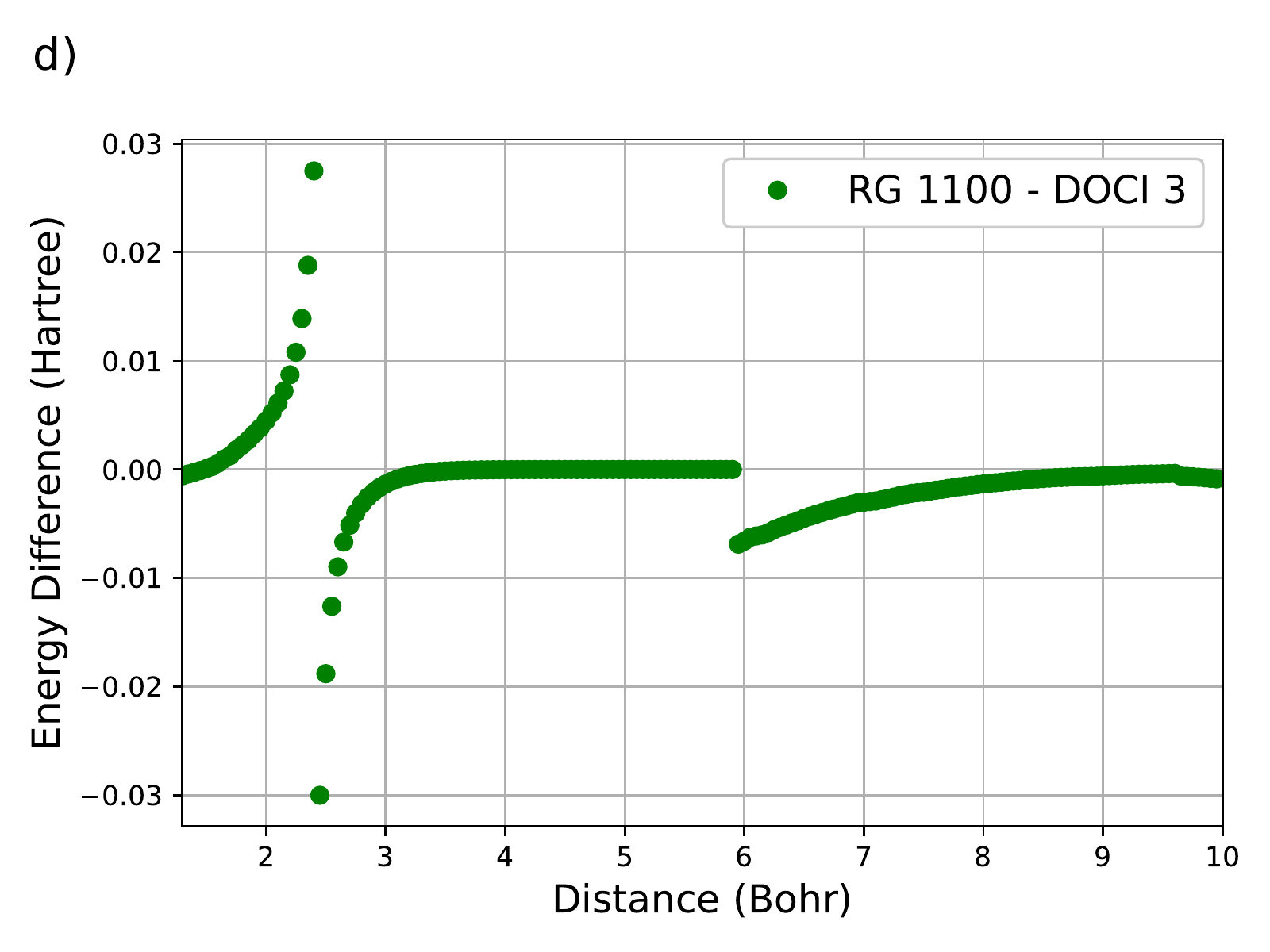} \hfill
		\includegraphics[width=0.3\textwidth]{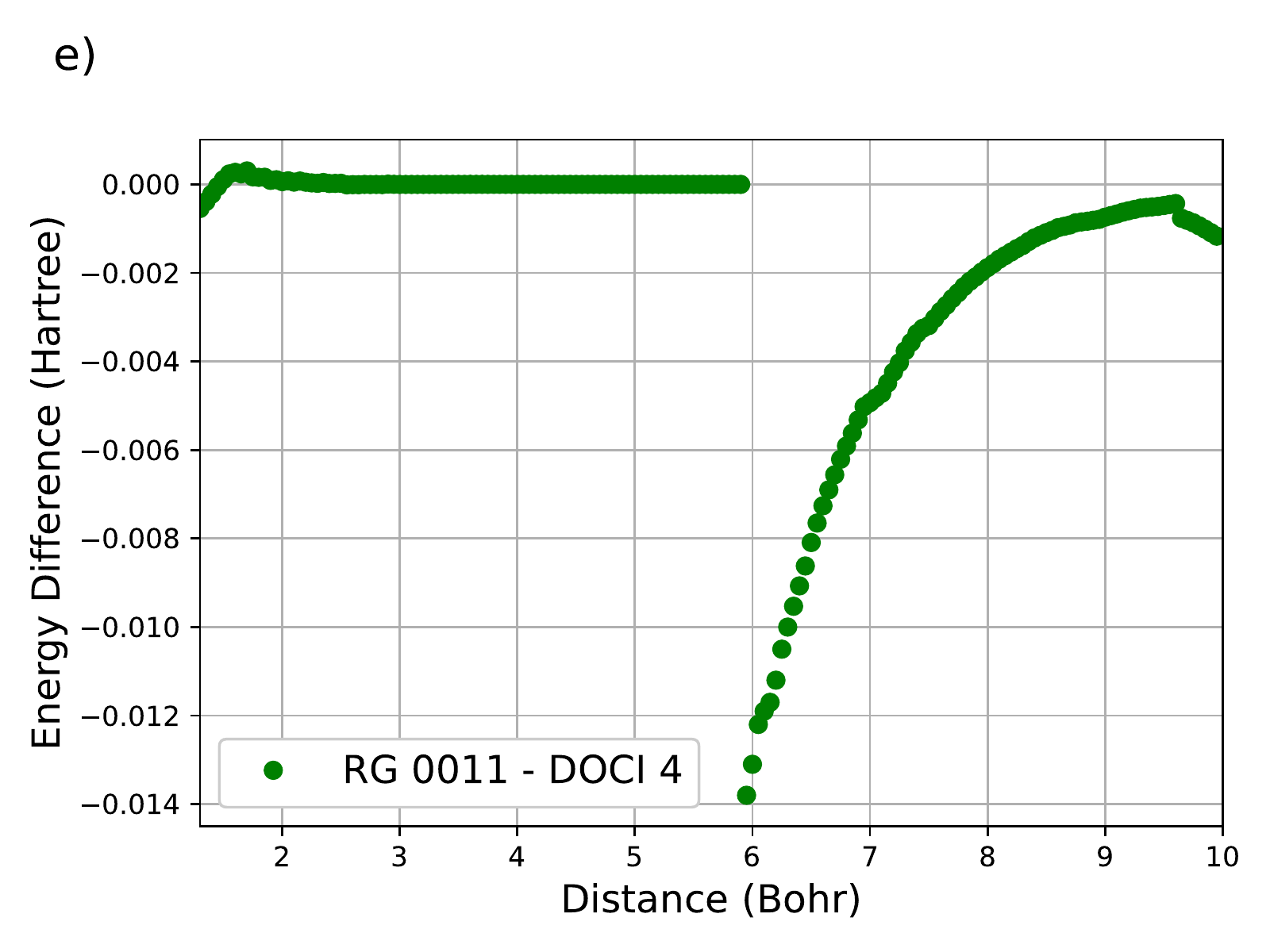} \hfill
		\includegraphics[width=0.3\textwidth]{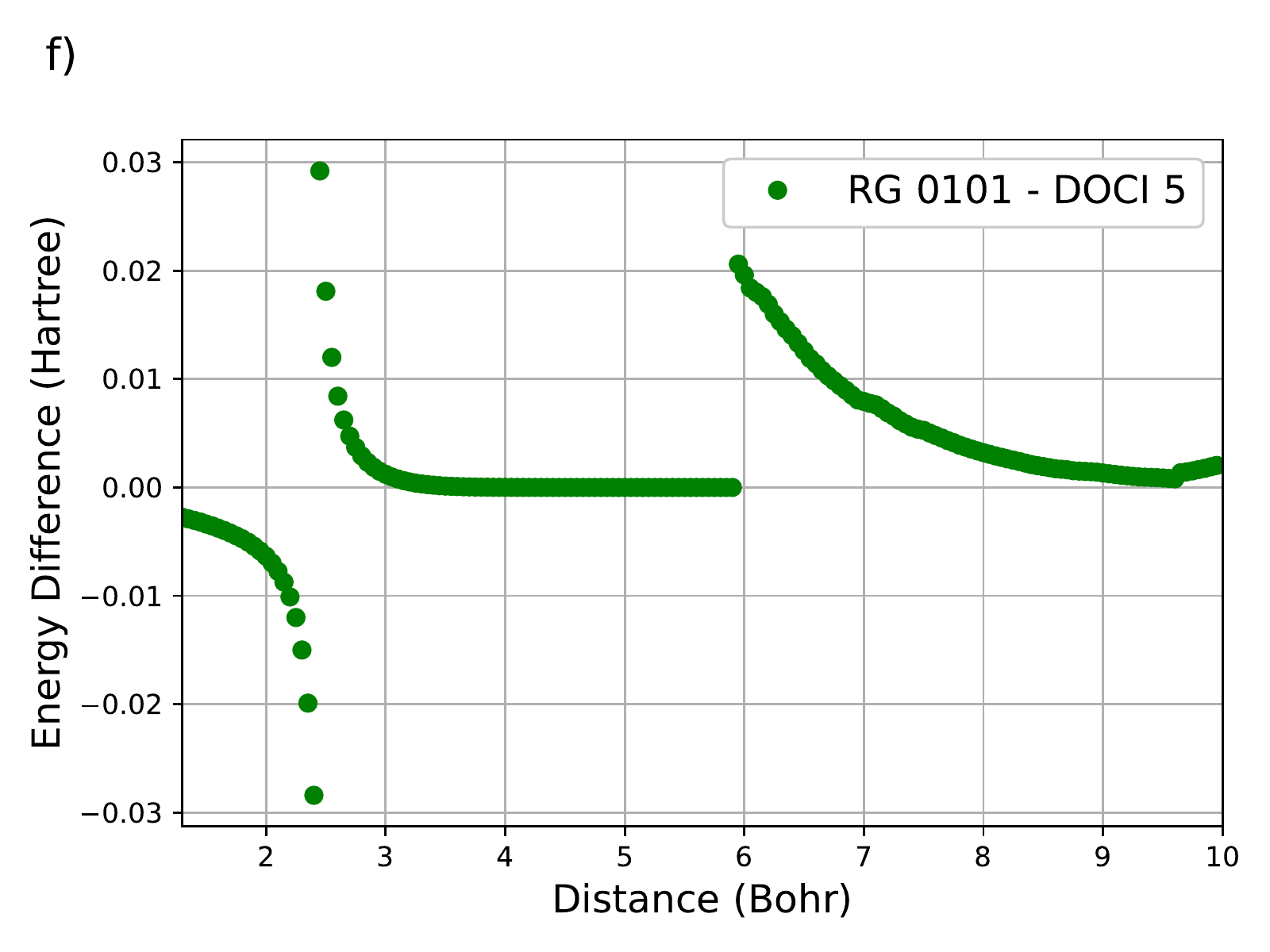}
	\end{subfigure}
	\caption{Energy differences between RG states computed with reduced BCS parameters defining the Hamiltonian corresponding to the optimal 1010 solution and DOCI states. All results computed in the STO-6G basis set in the basis of OO-DOCI orbitals.}
	\label{fig:H4_DOCI_diffs}
\end{figure}

We have chosen the 1010 state, and optimized the parameters $\{\varepsilon\}$ and $g$ to minimize the energy of the Coulomb Hamiltonian. The optimized parameters define a reduced BCS Hamiltonian, for which all the eigenvectors are known as the distinct solutions of Richardson's equations. Each solution was computed and their corresponding Coulomb Hamiltonian energies are plotted in figure \ref{fig:H4_DOCI_curves}. It should be highlighted that the DOCI states are in the orbitals optimized for the DOCI ground state, and are labelled by their energetic ordering near the minimum.
The 3rd and 5th DOCI excited states have an avoided crossing near $r=2.45$ bohr, and to match the RG curves we have manually switched the curves near these points. As can be seen, the RG states correspond \emph{directly} to the DOCI states. The difference in energy between the RG states and the DOCI states are plotted in figure \ref{fig:H4_DOCI_diffs}.
The 1010 state is essentially the same as the DOCI ground state everywhere except the minimum. There, the system is weakly-correlated, the RG states are more or less Slater determinants, and a short perturbative expansion would yield the correct energy. It is remarkable how well the RG states match the DOCI states. They do not capture the avoided crossing near $r =2.45$, but this would be asking too much. The RG states are computed from the parameters defining the optimal ground state mean-field. Avoided crossings are ``correlation'' effects of the mean-field: to capture them, we would need to diagonalize the Hamiltonian matrix in the basis of RG states. There is also deviation from the DOCI states around $r=6.0$ bohr, where the cusp in the optimal reduced BCS Hamiltonian occurs. Our hypothesis was that RG states would be a better basis for DOCI than Slater determinants, but these results tend to suggest that the RG states \emph{are} the DOCI states, except near particular points. We would have been satisfied with the conclusion that the 1010 state is essentially the DOCI ground state, but the matching of all states is strongly suggestive that the problem has been correctly understood. 1010 always represents the ground state, but the energetic ordering of the excited RG states changes over the course of the dissociation. Near the minimum, excitations are ordered by their rank in the sense that the ``double'' excitation 0101 is the highest in energy. At dissociation, the ordering is different, though we will postpone the discussion to the results for H$_6$ as it will be more clear.

\begin{figure}[ht!]
	\begin{subfigure}{\textwidth}
		\includegraphics[width=0.475\textwidth]{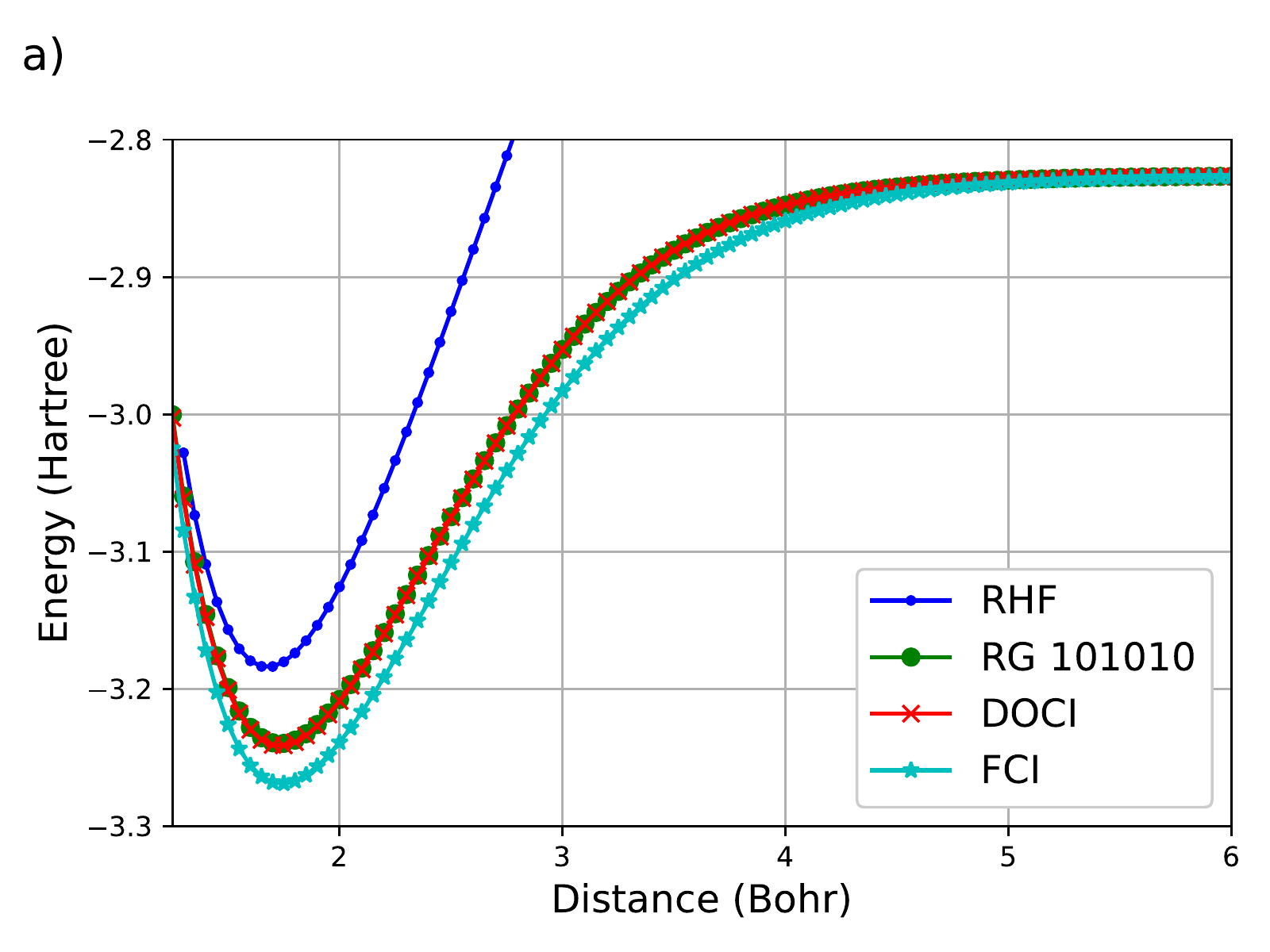} \hfill
		\includegraphics[width=0.475\textwidth]{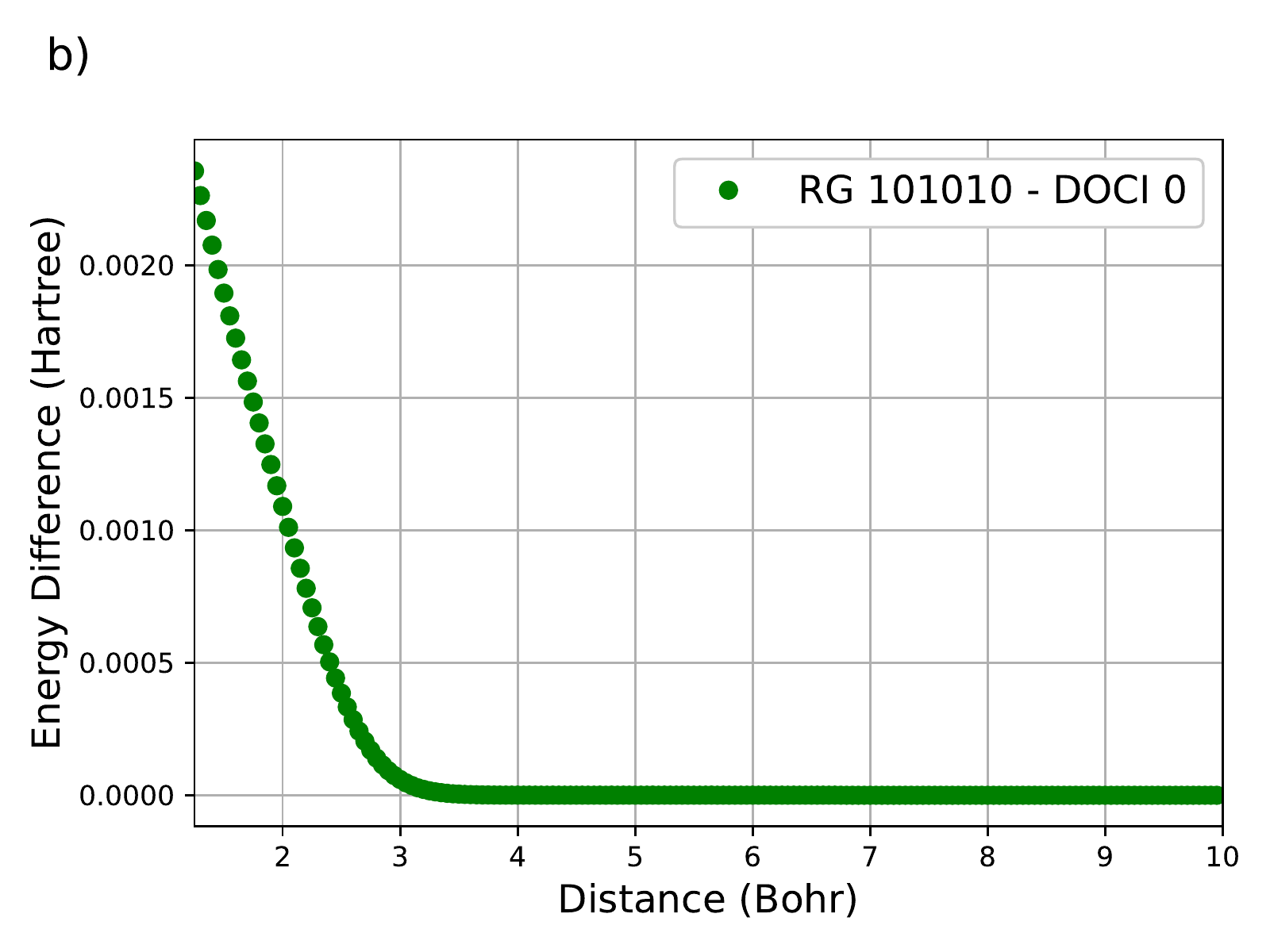}
	\end{subfigure}
	\caption{(a) Bond dissociation curves for H$_6$. (b) Energy difference between the 101010 RG state and the DOCI ground state. All results computed with the STO-6G basis set in the basis of OO-DOCI orbitals.}
		\label{fig:H6_gs_curves}
\end{figure}

For H$_6$ we proceeded directly to using the 101010 state, presented in figure \ref{fig:H6_gs_curves}.
It is clear that the 101010 RG state is very close to the DOCI ground state everywhere except the minimum where a short expansion would fix the problem completely. With the parameters defining the optimal Hamiltonian for the 101010 state, we computed all the RG states and compared them with the DOCI states. The DOCI spectrum is a little more complicated, to say the least, though after resolving many crossings and avoided crossings, the RG states again seem to match the DOCI states very well. The results are plotted in appendix \ref{ap:H6_ex}. As for H$_4$, there is a cusp in the optimal reduced BCS Hamiltonian parameters though it occurs near $r=6.80$ bohr. This does not affect the smoothness of the 101010 approximation to the DOCI ground state, but is visible in other states.

The energetic ordering of the RG states is different near the minimum and at dissociation, though both follow clear patterns. Near the minimum, the RG states closely resemble Slater determinants, and thus the usual rules apply with the clear distinction being that the 101010 state is the minimum. Generally, the next lowest in energy are the single pair excitations, corresponding to the exchange of a 1 and a 0 in the state. Double-pair excitations are generally next, followed by the triple-excitation 010101 at the top of the spectrum. At dissociation the rules are different. The state 101010 at dissociation is essentially 3 distinct subsystems, each of which has a state 10. It is useful to separate excitations into two types. First, there are \emph{swaps} for which a subsystem switches 10 to 01. Second, there are \emph{transfers} which correspond more or less to a transfer of a pair from one subsystem to another: one 10 becomes 00 while another becomes 11. With these ideas, the energetic ordering of the states is clearer. The lowest excited states are 011010, 100110, and 101001. They are degenerate and are ``single-swaps'' of the ground state 101010. The next states are the double-swaps 011001, 010110, and 100101 which are nearly degenerate (they differ by less than 2 milli-Hartree). Next are the 6 single-transfer states 100011, 101100, 001011, 111000, 001110, and 110010 which occur in 3 near-degenerate pairs. The triple-swap 010101 is next, followed finally by the single-swap-single-transfer states 010011, 011100, 110100, 000111, 001101, and 110001 which also occur in three near-degenerate pairs. 

\begin{figure}[ht!]
	\begin{subfigure}{\textwidth}
		\includegraphics[width=0.475\textwidth]{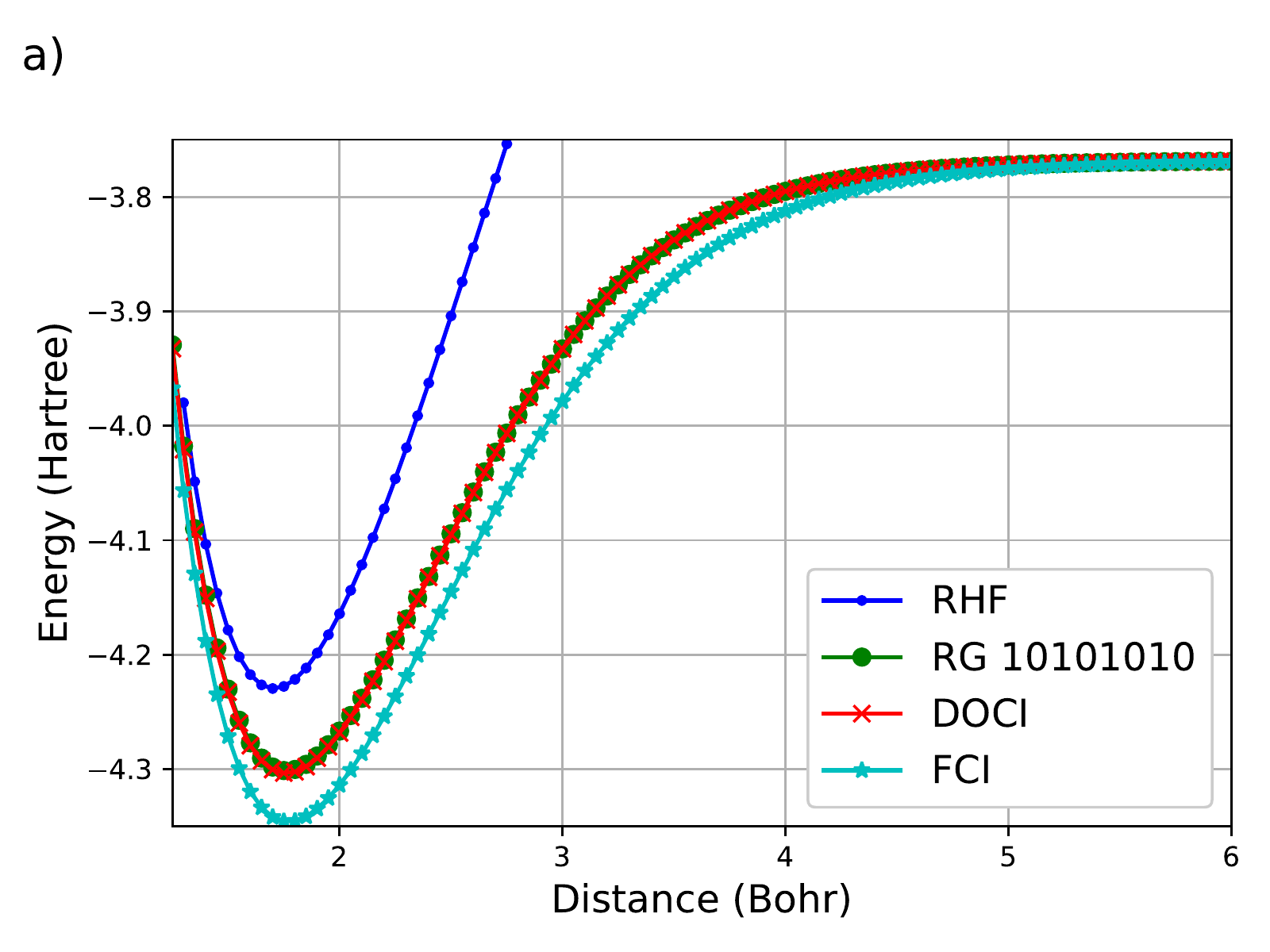} \hfill
		\includegraphics[width=0.475\textwidth]{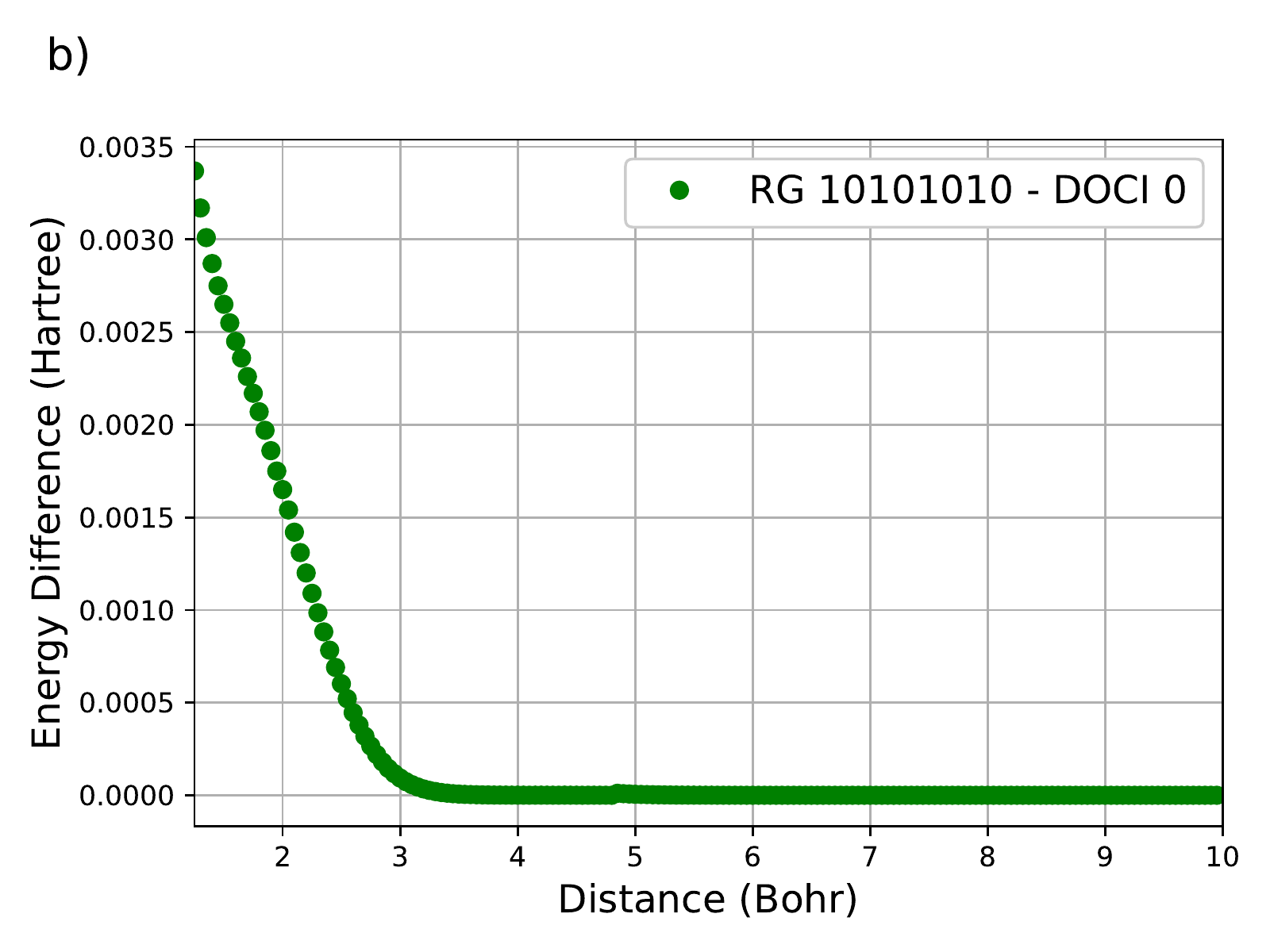}
	\end{subfigure}
	\caption{(a) Bond dissociation curves for H$_8$. (b) Energy difference between the 10101010 RG state and the DOCI ground state. All results computed with the STO-6G basis set in the basis of OO-DOCI orbitals.}
		\label{fig:H8_gs_curves}
\end{figure}

For H$_8$ the results are more or less the same. The ground state is very-well described by the RG state 10101010 as shown in figure \ref{fig:H8_gs_curves}.
We did not plot the $\binom{8}{4}=70$ states to compare with the DOCI states, but at dissociation the energetic ordering follows the same pattern as for H$_6$. The lowest excited states are the degenerate single-swaps, followed by double-swaps, the single-transfers, the triple-swaps, and the single-swap-single-transfer states. Next are the quadruple swap 01010101, the double-swap-single-transfer states followed finally by the double-transfer states. The two double-transfer states 00111100 and 11000011 occur in the middle of the double-swap-single-transfer states. These are the only exceptions to the ordering pattern.

\begin{figure}[ht!]
	\begin{subfigure}{\textwidth}
		\includegraphics[width=0.475\textwidth]{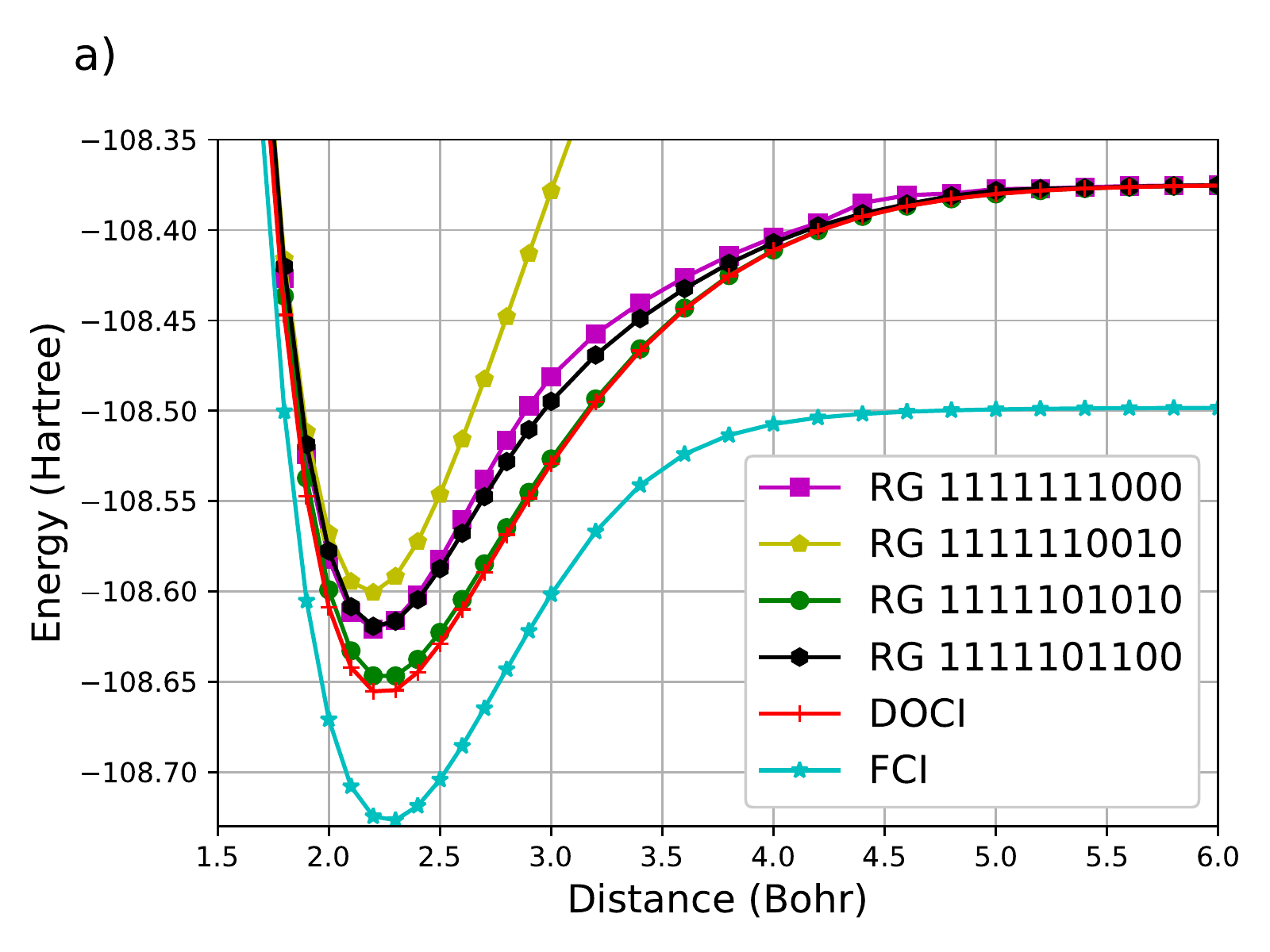} \hfill
		\includegraphics[width=0.475\textwidth]{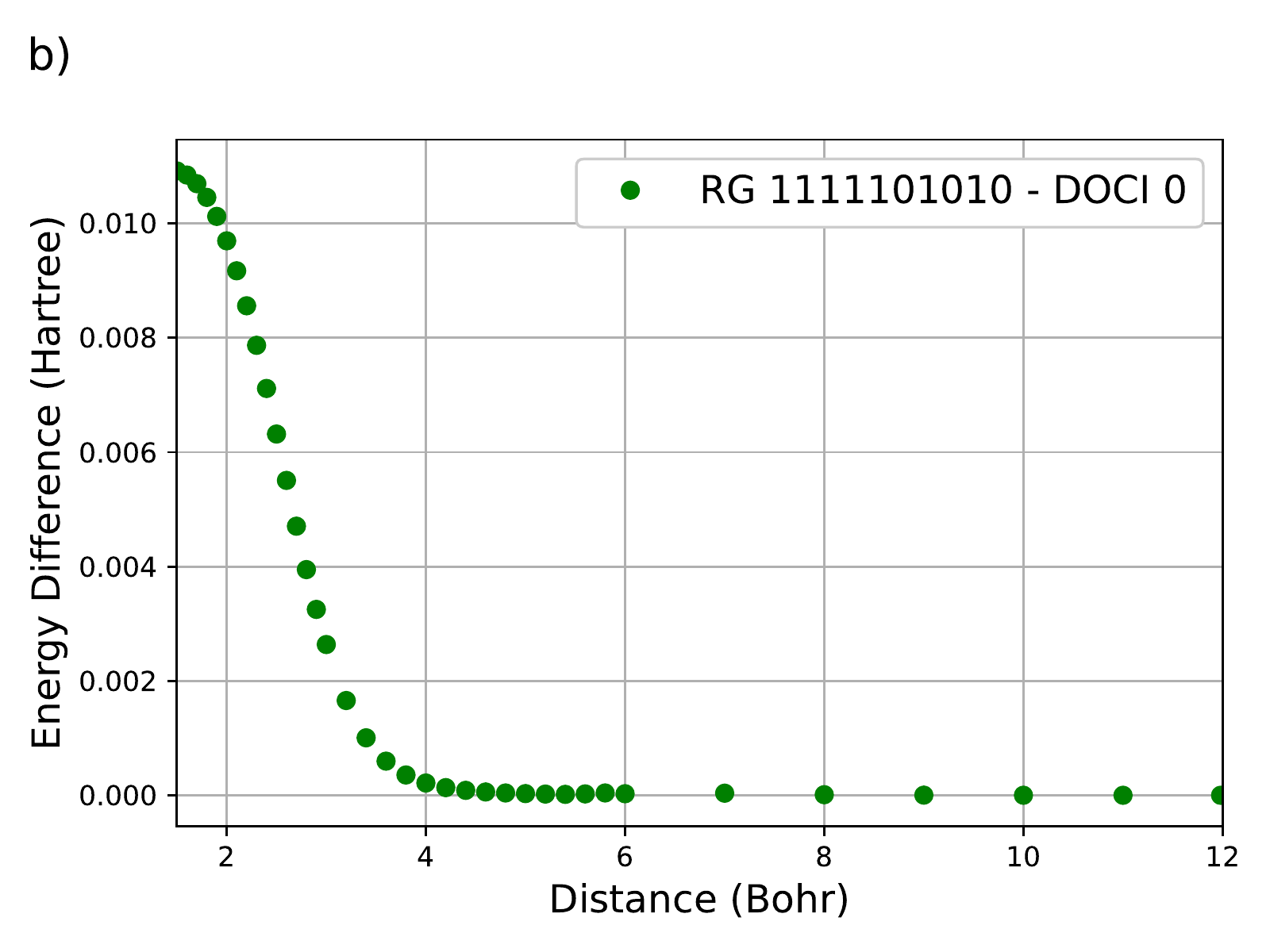}
	\end{subfigure}
	\caption{(a) Bond dissociation curves for N$_2$. (b) Energy difference between the 1111101010 RG state and the DOCI ground state. All results computed with the STO-6G basis set in the basis of OO-DOCI orbitals.}
		\label{fig:N2_gs_curves}
\end{figure}

Finally, we computed results for the dissociation of N$_2$. It was not obvious which distribution would lead to the best RG treatment of the DOCI ground state, so several were computed and are plotted in figure \ref{fig:N2_gs_curves}.
It is clear that the distribution 1111101010 is close to the DOCI ground state, with a deviation that follows the same pattern seen for the H-chains: they agree quite well everywhere except near the minimum. This distribution is easily reasoned physically: the core is always filled, hence the string 1111, while the valence orbitals are half-filled at dissociation, hence 101010. It is interesting that most of the distributions we tried are correct at dissociation. The ordering of the excited states is less clear.

The numerical results herein presented clearly demonstrate that an RG mean-field is very close to the DOCI ground state, and the corresponding RG excited states closely match the DOCI excited states, provided that the correct RG mean-field is chosen. On its face, choosing the right state is an NP-complete problem, though our results should suggest that physical reasoning should lead directly to the correct choice, or at least narrow the choice down enough that a greedy algorithm or dynamic programming would find the correct state. It is clear enough that for any system at half-filling, the correct choice is the N\'{e}el state 1010...10. Based on the results for N$_2$, we suggest a hypothesis based on splitting the spatial orbital basis into core, valence and Rydberg orbitals. The core is always full, and the Rydbergs are always empty, so will correspond to 11..11 and 00..00 respectively. The valence orbitals will be partially filled, and each pair of partial fillings correspond to 10. For N$_2$ in a minimal basis, the valence is half-filled, so the choice 101010 is obvious and unambiguous. For C$_2$, the valence orbitals are less than half-filled so it is less clear which distribution is optimal. Further calculations are obviously necessary to confirm or refine this hypothesis, which we have begun and will report in a subsequent contribution. Choosing the correct distribution is not the same as choosing an active space. The distribution is \emph{not} tied to specific orbitals since the single-particle energies must be sorted for the Richardson's equations solver. Thus, to pick the correct distribution, we do not need to know which specific orbitals are partially occupied, only where they would occur in an energetic ordering. 

\section{Conclusion}
Variational calculations for bond-dissociation processes were performed using RG states as a mean-field. It is evident that a particular choice of RG state is very close to the DOCI ground state. Remarkably, the other RG states match the DOCI excited states quite well. Our hypothesis was that RG states are a better basis for DOCI than Slater determinants, though it seems that the states are near-identical. For hydrogen chains, the N\'{e}el state, corresponding to the distribution 1010..10 is always the optimal state. For N$_2$, the state 1111101010 was found to be optimal. Both choices can be reasoned physically. Our hypothesis for choosing the correct state is based on separating the orbital spectrum into core, valence and Rydberg spaces. This does \emph{not} require identifying which specific orbitals are in which space, only the number that are in each space. Obviously, further results are necessary to confirm or refute this hypothesis, and we will report our further results as soon as they are available.

The optimal variational RG curves are nearly identical to DOCI everywhere except near the minimum, where the correct result is weakly-correlated electrons. There, the RG states have a weak pairing strength $g$, and are thus quite close to Slater determinants. A short perturbative expansion, or CI diagonalization, will account for the missing energy.

With this we consider the principle proven. A single RG state effectively reproduces DOCI for Coulomb systems. This approach is a physical solution to a physical problem. While AP1roG / pCCD is currently the method to beat, the advantages of RG states are clear: AP1roG / pCCD is state-specific and must be solved by projection whereas the RG mean-field for the physical ground state gives a complete set of states which approximate the seniority-zero excited states quite well. Both methods have $\mathcal{O}(N^4)$ scaling in a given basis, or $\mathcal{O}(N^5)$ with orbital-optimization. Obviously non-zero seniority sectors will need to be treated separately. In the near term this may be accomplished in any number of ways by introducing non-zero seniorities post-hoc. In the longer term, this should be done correctly at the mean-field level which is substantially more difficult.

It is now the time to adapt this approach to larger systems for which the next steps are clear. An effective numerical procedure requires three components: an initial guess for the parameters close enough to the minimum, an effective to compute the objective function, and an algorithm to find the minimum. It is hard to imagine that the energy expression could be made more efficient in terms of rapidities, though it would be preferable to not use rapidities at all. Unfortunately, for the moment the RDM elements require them. If the RDM elements could be computed directly in terms of the EBV then substantial effort and potential numerical headaches would be avoided. The best way to avoid problems with floating point operations is to avoid performing floating point operations. The algorithm would be easier in terms of EBV as well, since their dependence on the variables $\{\varepsilon\}$ is much less non-linear for the rapidities. Finally, the orbital-optimization is of central importance. Orbital-optimized AP1roG is able to find the optimal orbitals reliably, so we are not so concerned. These issues are short-term stepping stones to a general purpose RG mean-field approach for weakly-correlated pairs of electrons. All are details we will address in a series of contributions. The most important point is that it is worth pursuing this line of research. RG states \emph{are} effective.

\section{Acknowledgements}
P.A.J. was supported by NSERC and Compute Canada. C.-\'{E}.F. is grateful for funding from the Vanier Canada Graduate Scholarships. We thank Paul W. Ayers, Dimitri Van Neck and Stijn De Baerdemacker for many helpful discussions.

\appendix
\section{H$_6$ RG Excited States} \label{ap:H6_ex}

\begin{figure}[ht!]
	\begin{subfigure}{\textwidth}
		\includegraphics[width=0.24\textwidth]{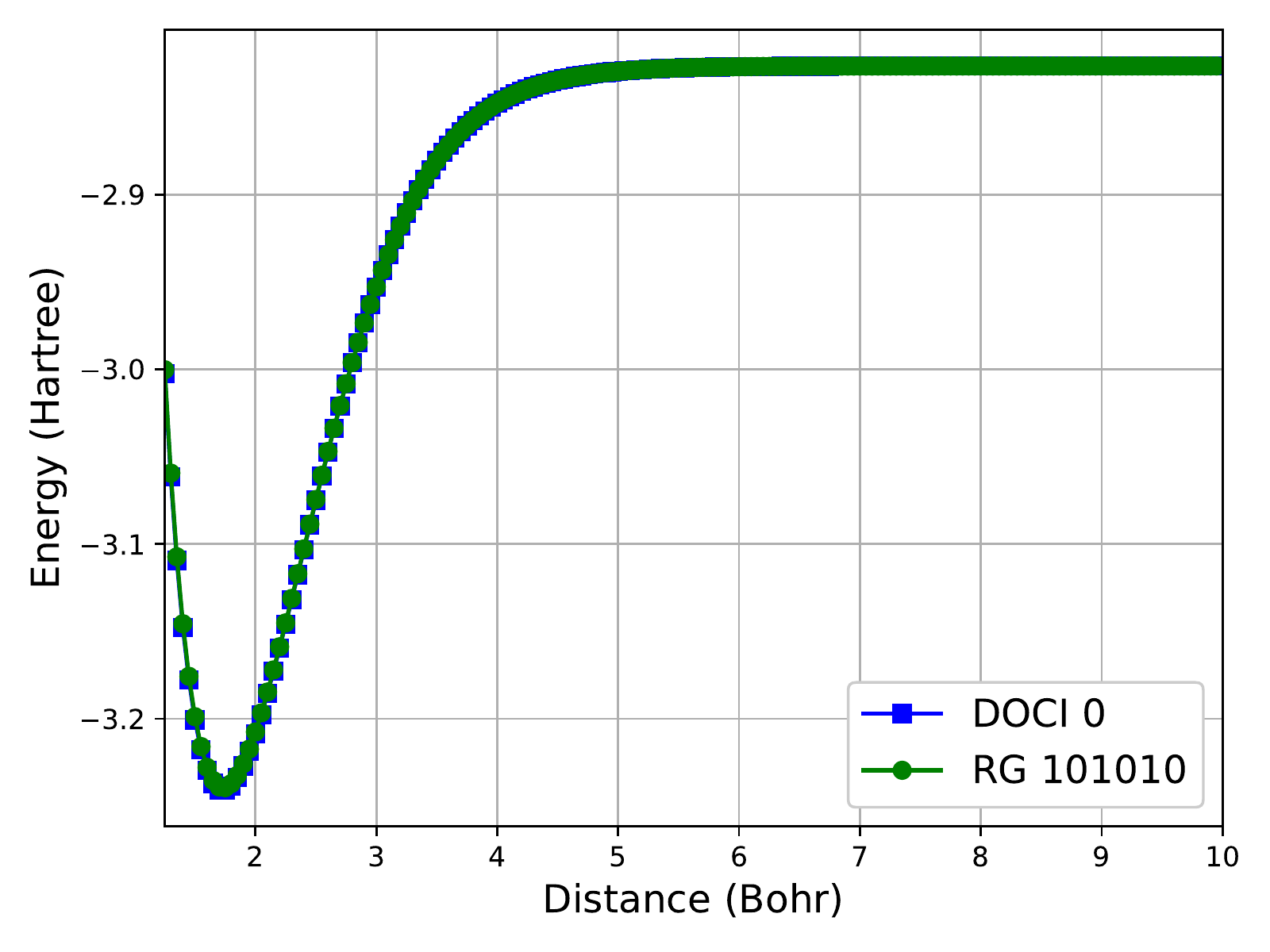} \hfill
		\includegraphics[width=0.24\textwidth]{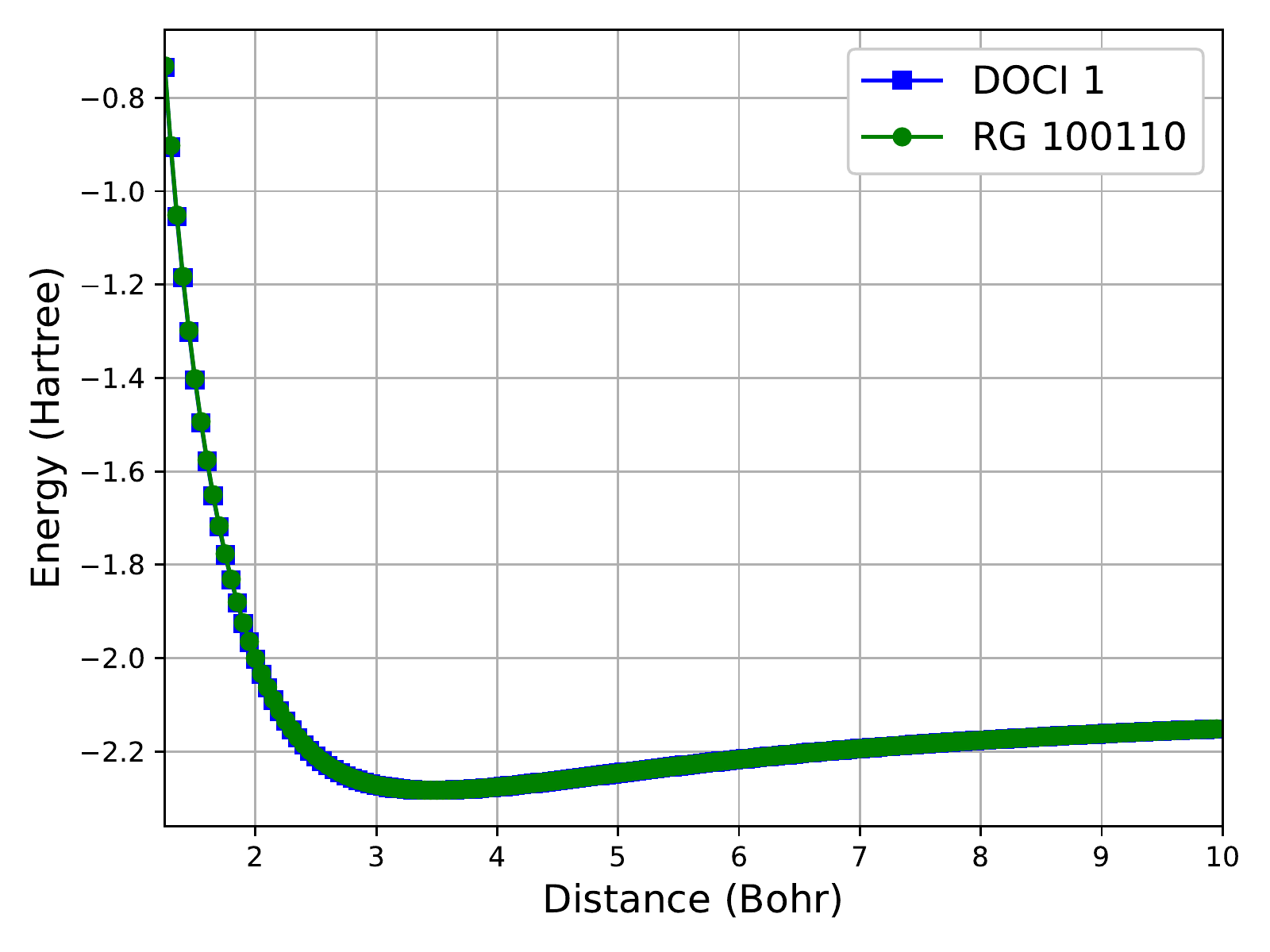} \hfill
		\includegraphics[width=0.24\textwidth]{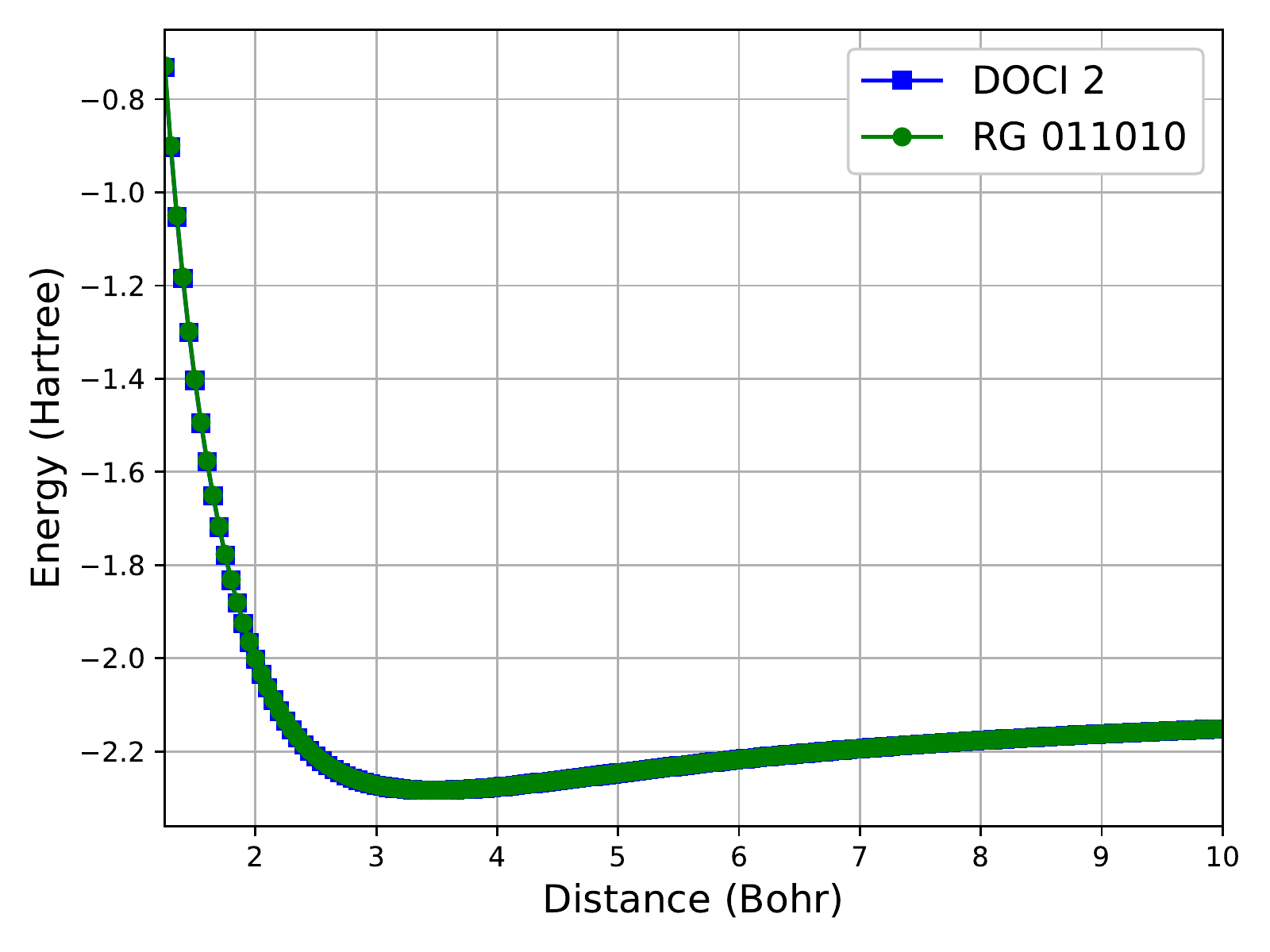} \hfill
		\includegraphics[width=0.24\textwidth]{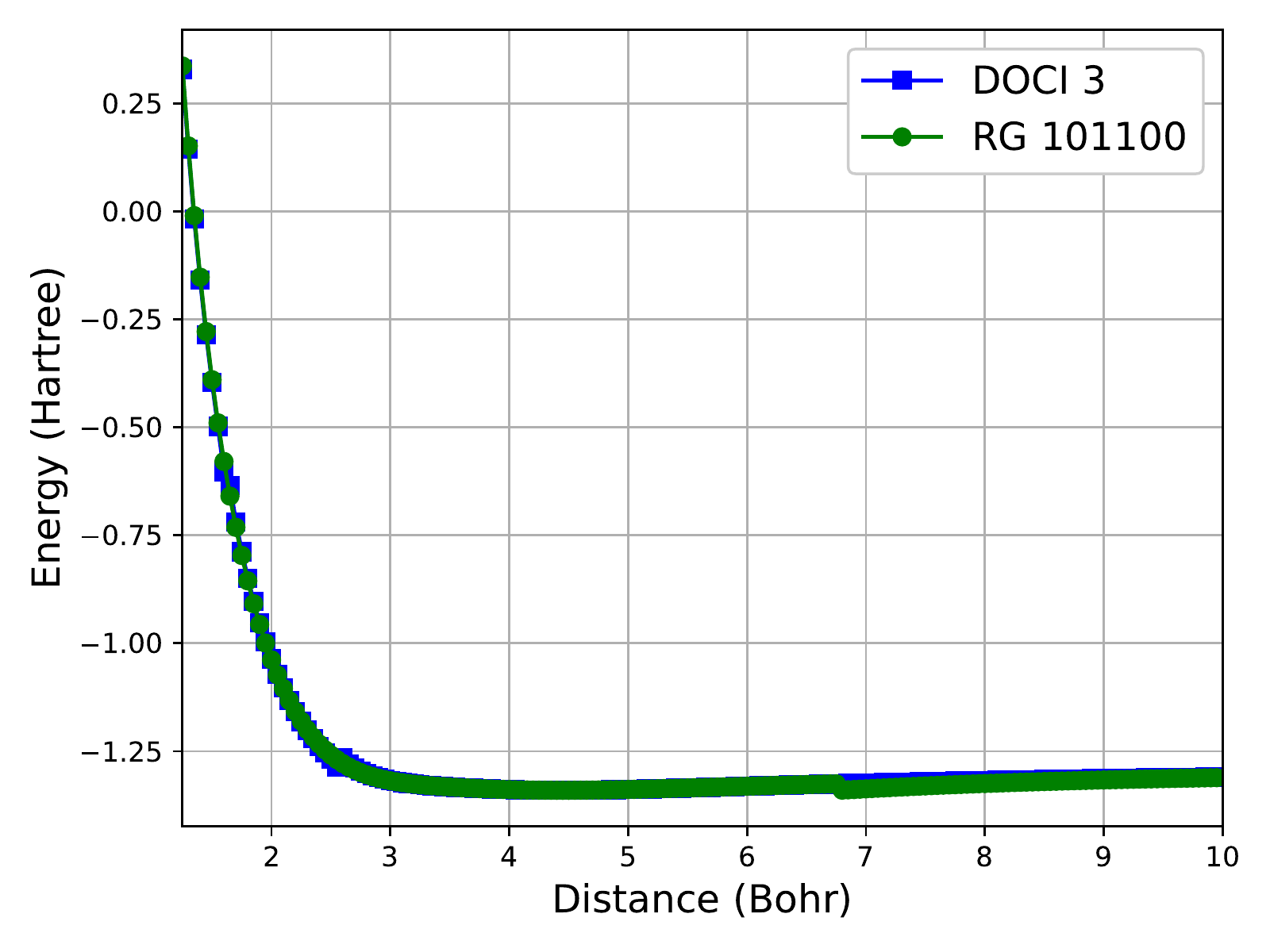}
	\end{subfigure}
	\begin{subfigure}{\textwidth}
		\includegraphics[width=0.24\textwidth]{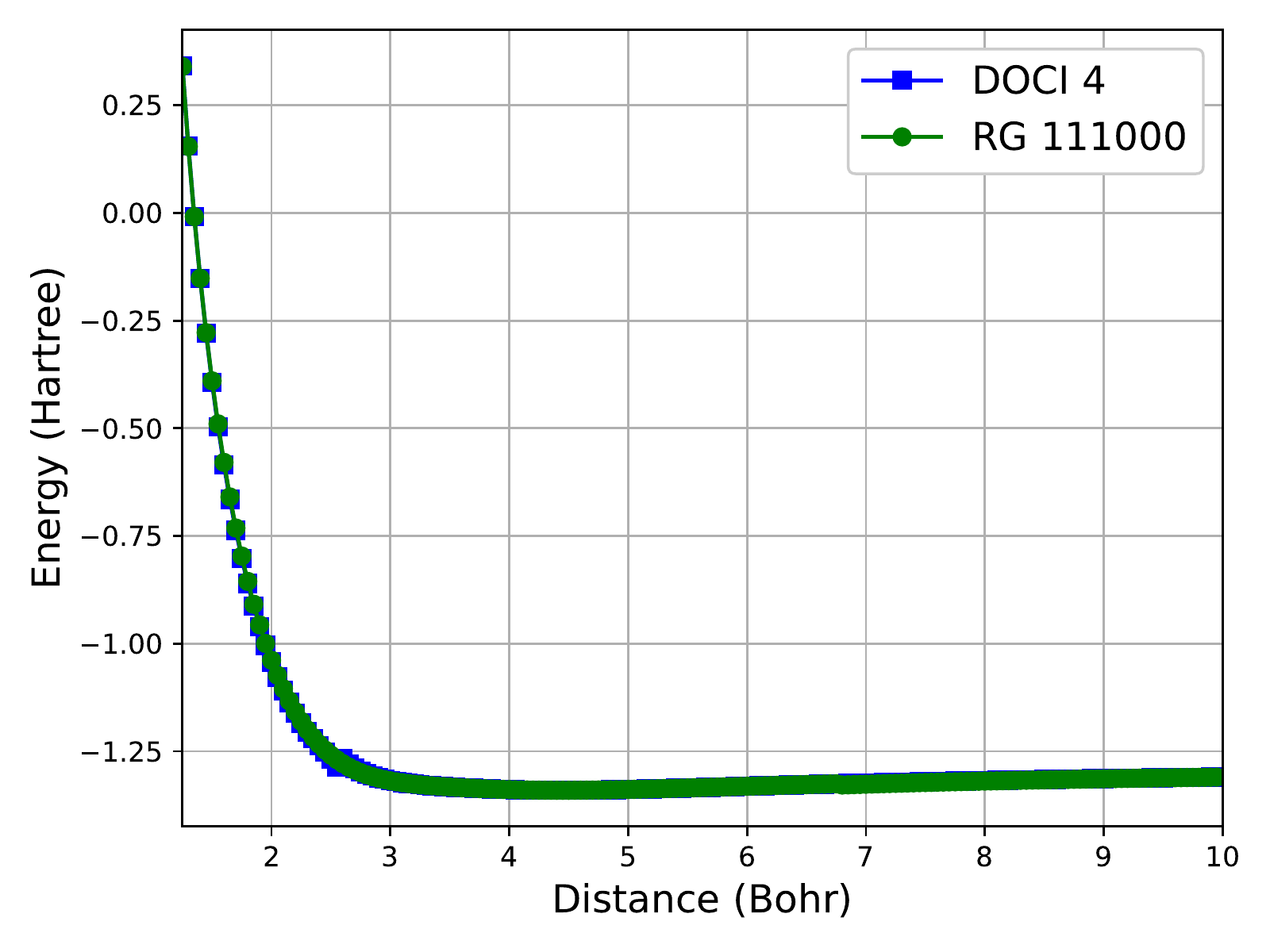} \hfill
		\includegraphics[width=0.24\textwidth]{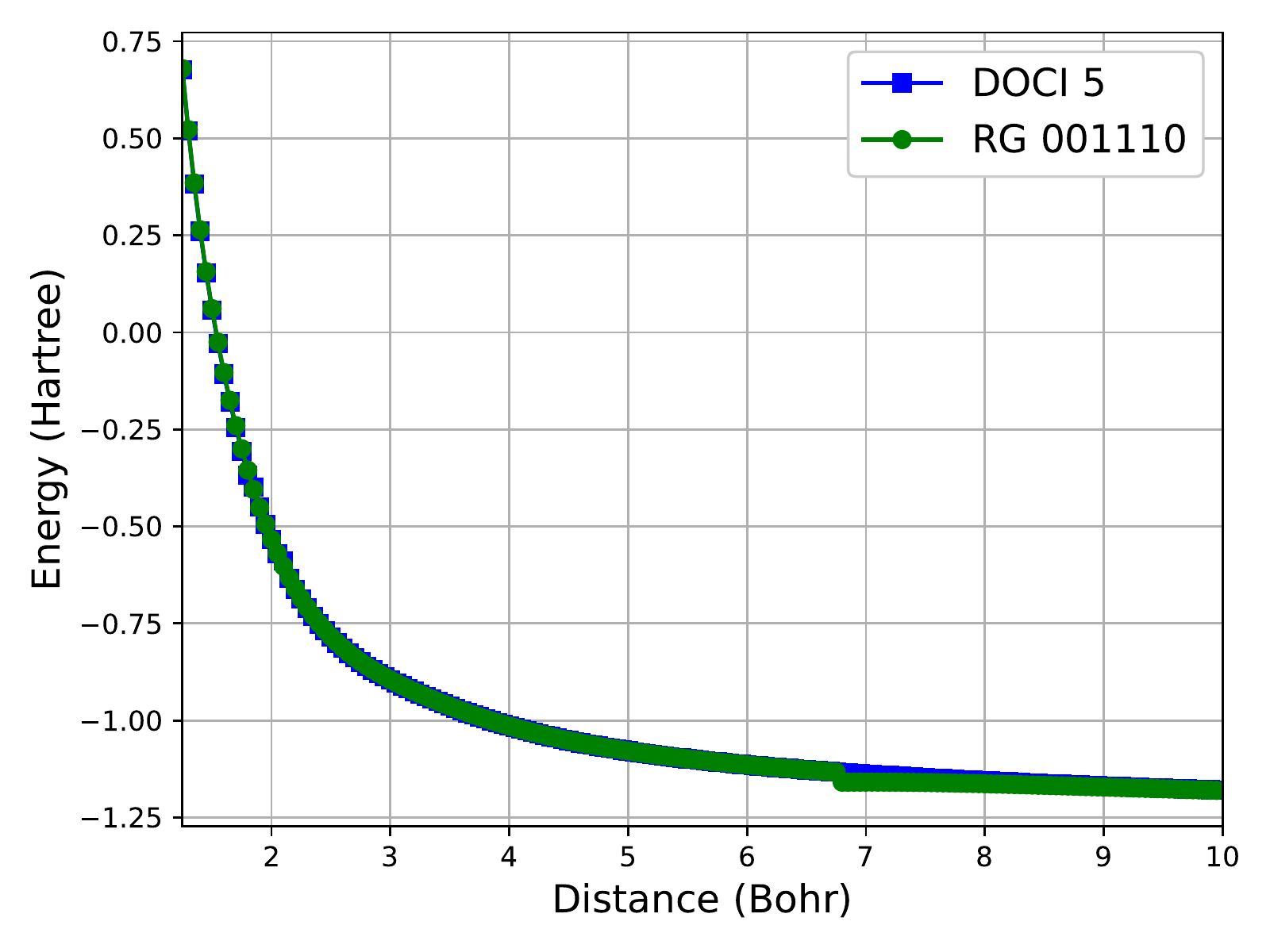} \hfill
		\includegraphics[width=0.24\textwidth]{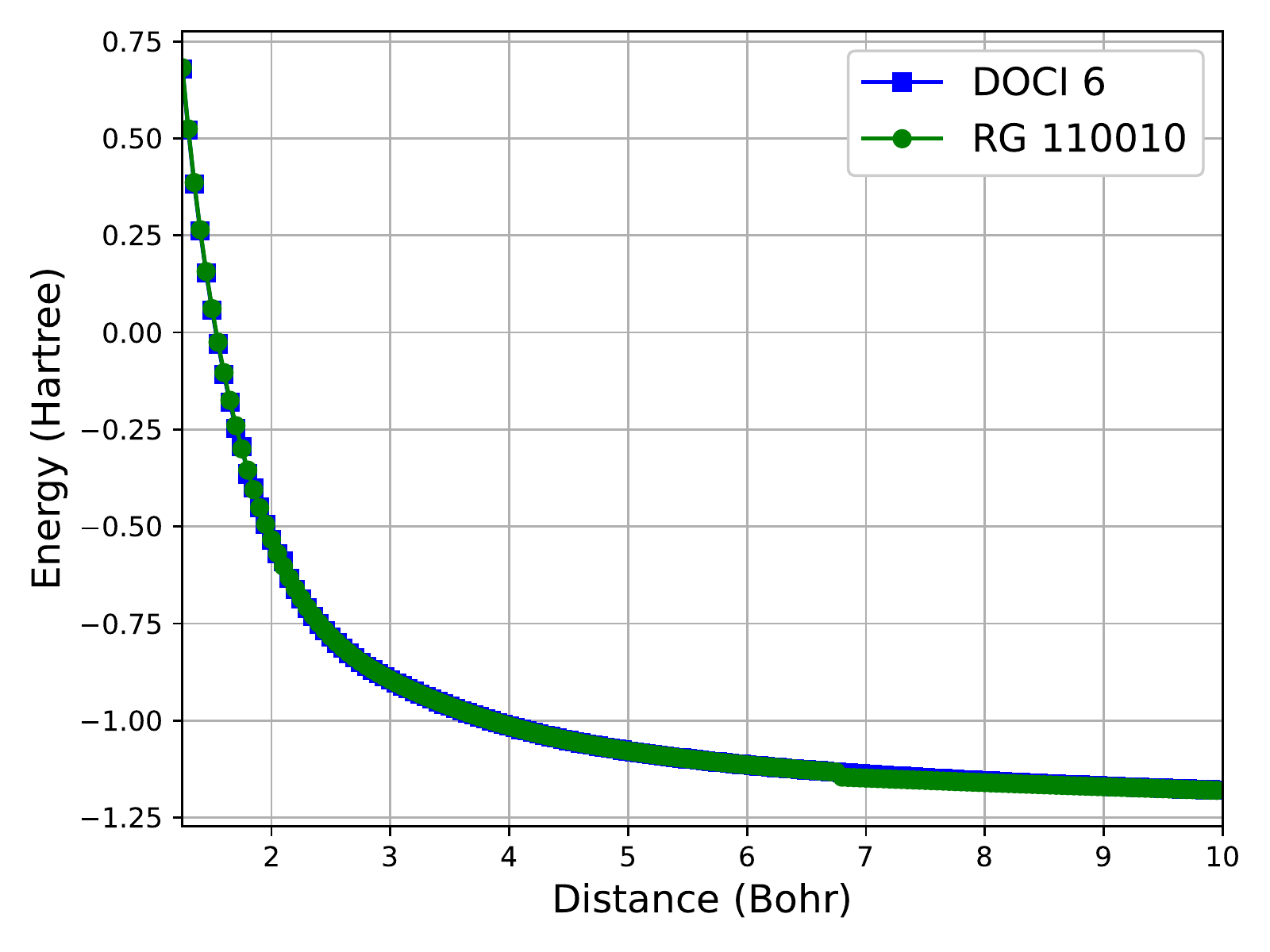} \hfill
		\includegraphics[width=0.24\textwidth]{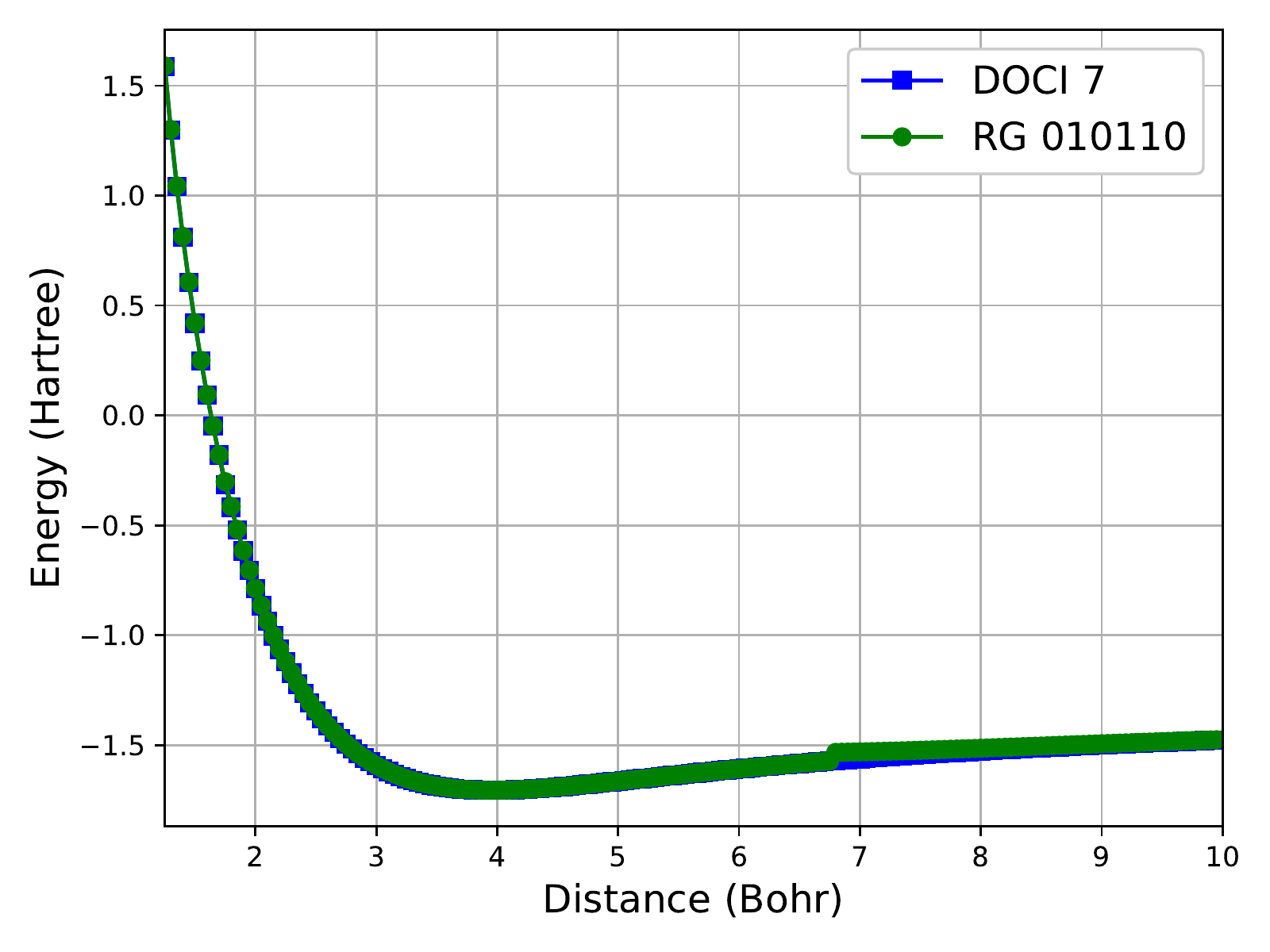}
	\end{subfigure}
	\begin{subfigure}{\textwidth}
		\includegraphics[width=0.24\textwidth]{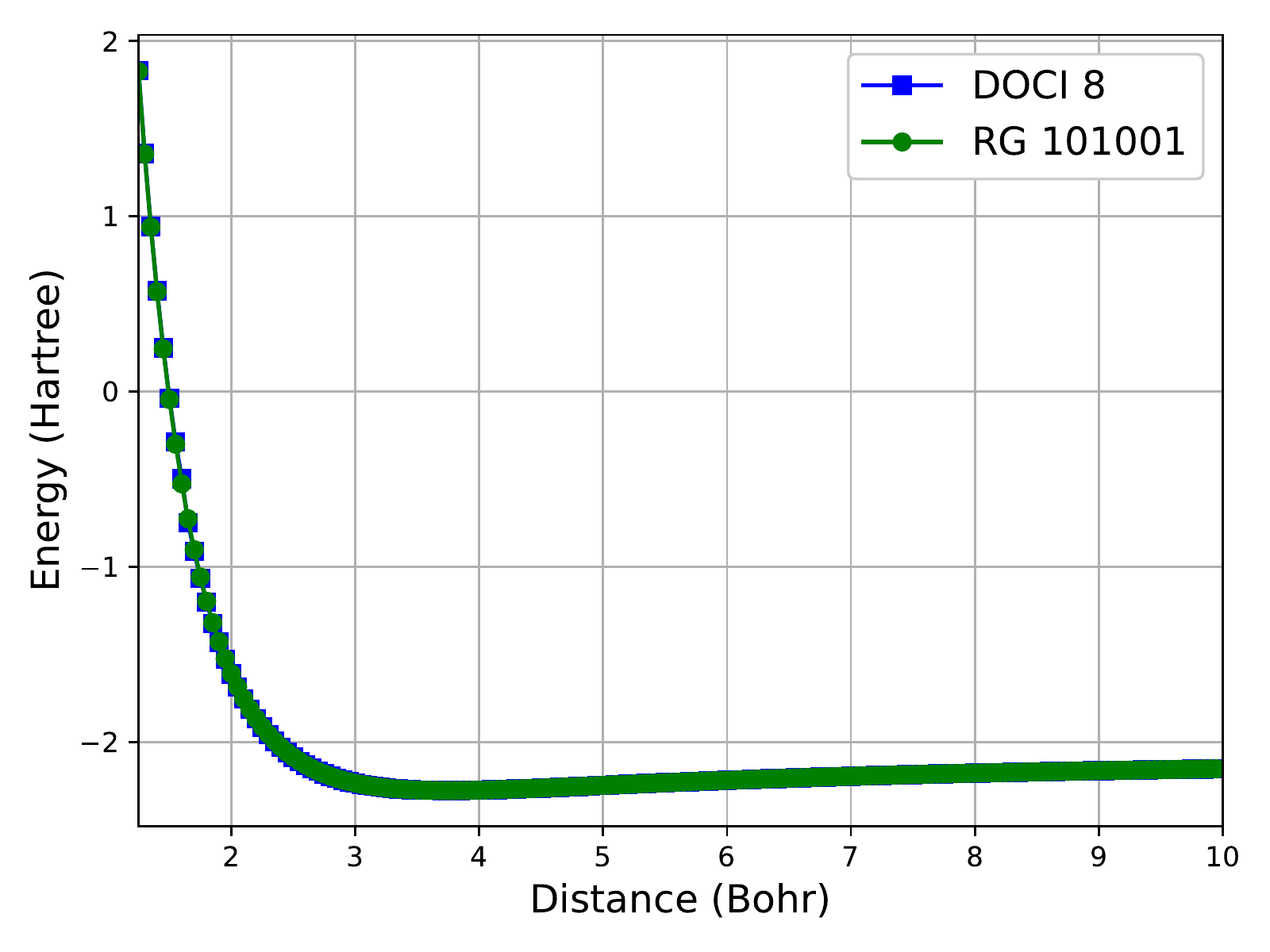} \hfill
		\includegraphics[width=0.24\textwidth]{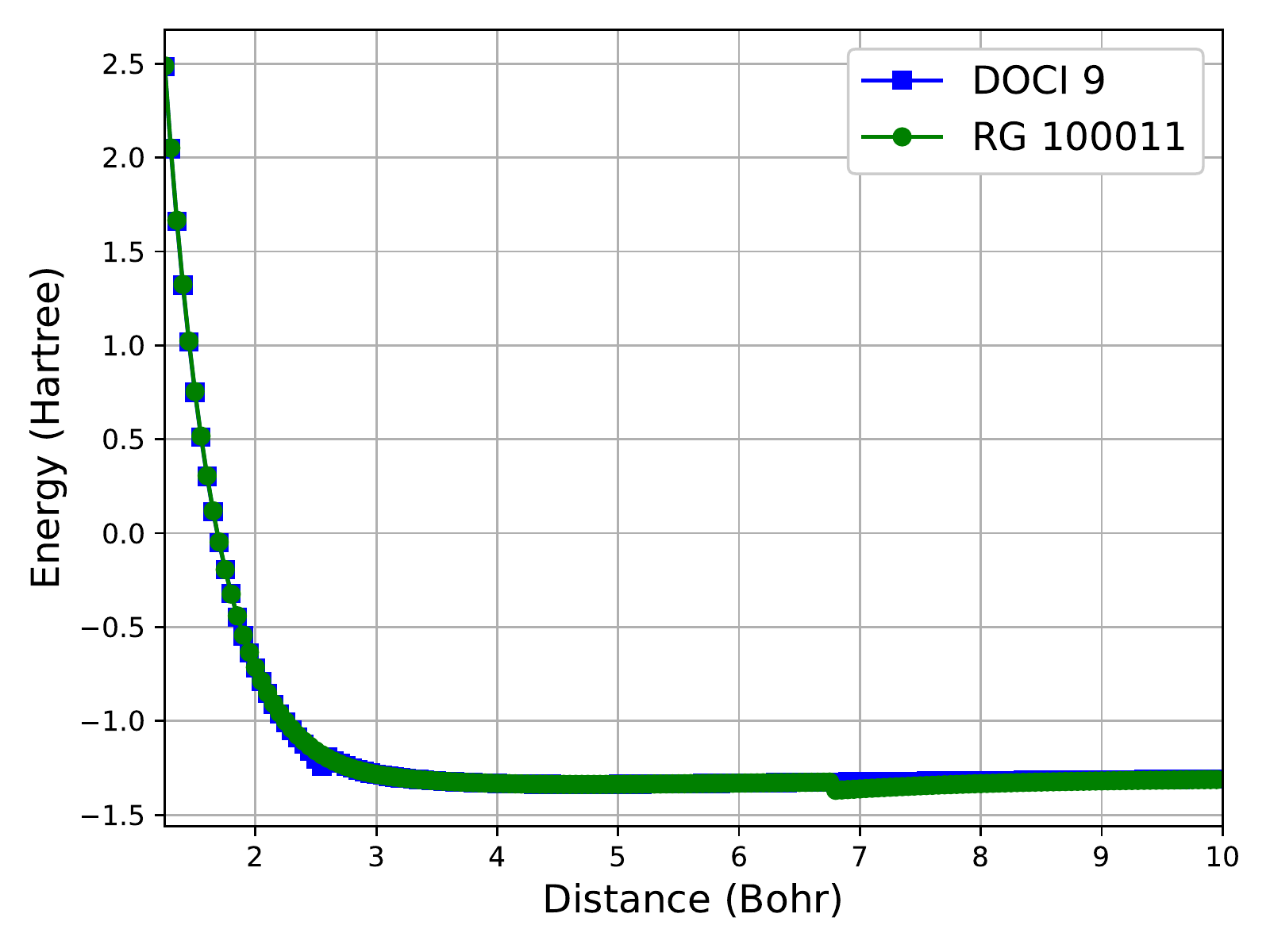} \hfill
		\includegraphics[width=0.24\textwidth]{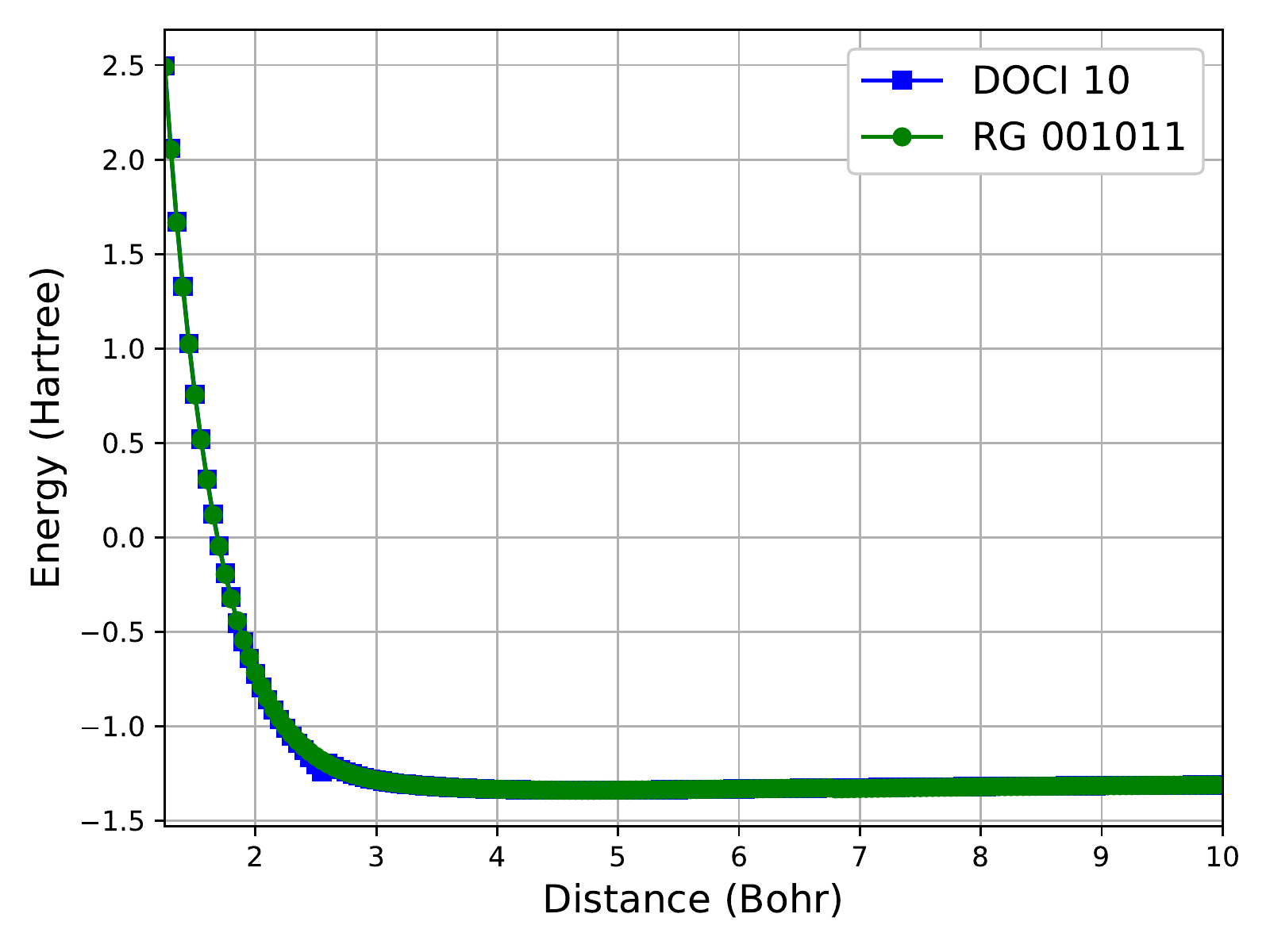} \hfill
		\includegraphics[width=0.24\textwidth]{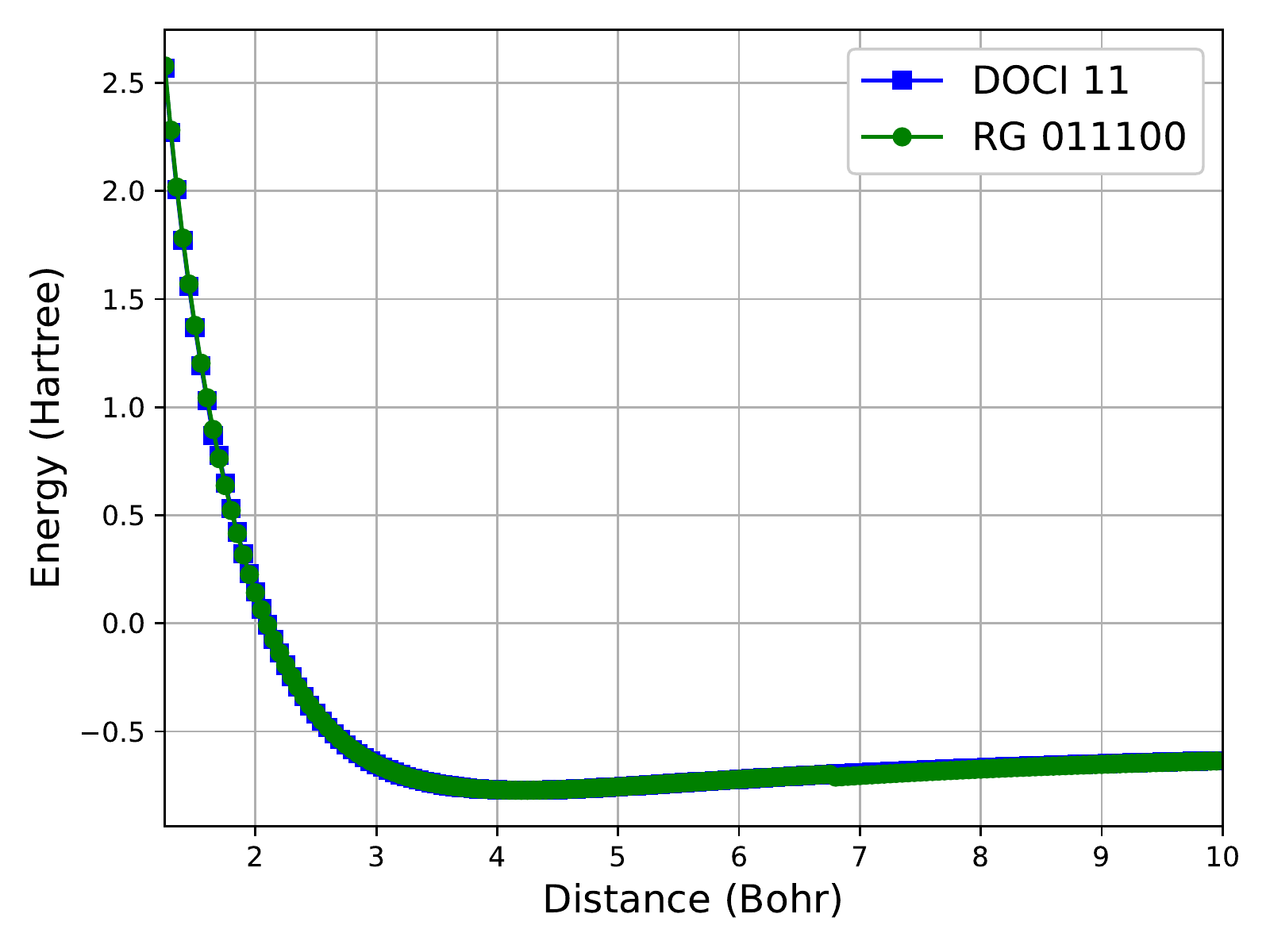}
	\end{subfigure}
	\begin{subfigure}{\textwidth}
		\includegraphics[width=0.24\textwidth]{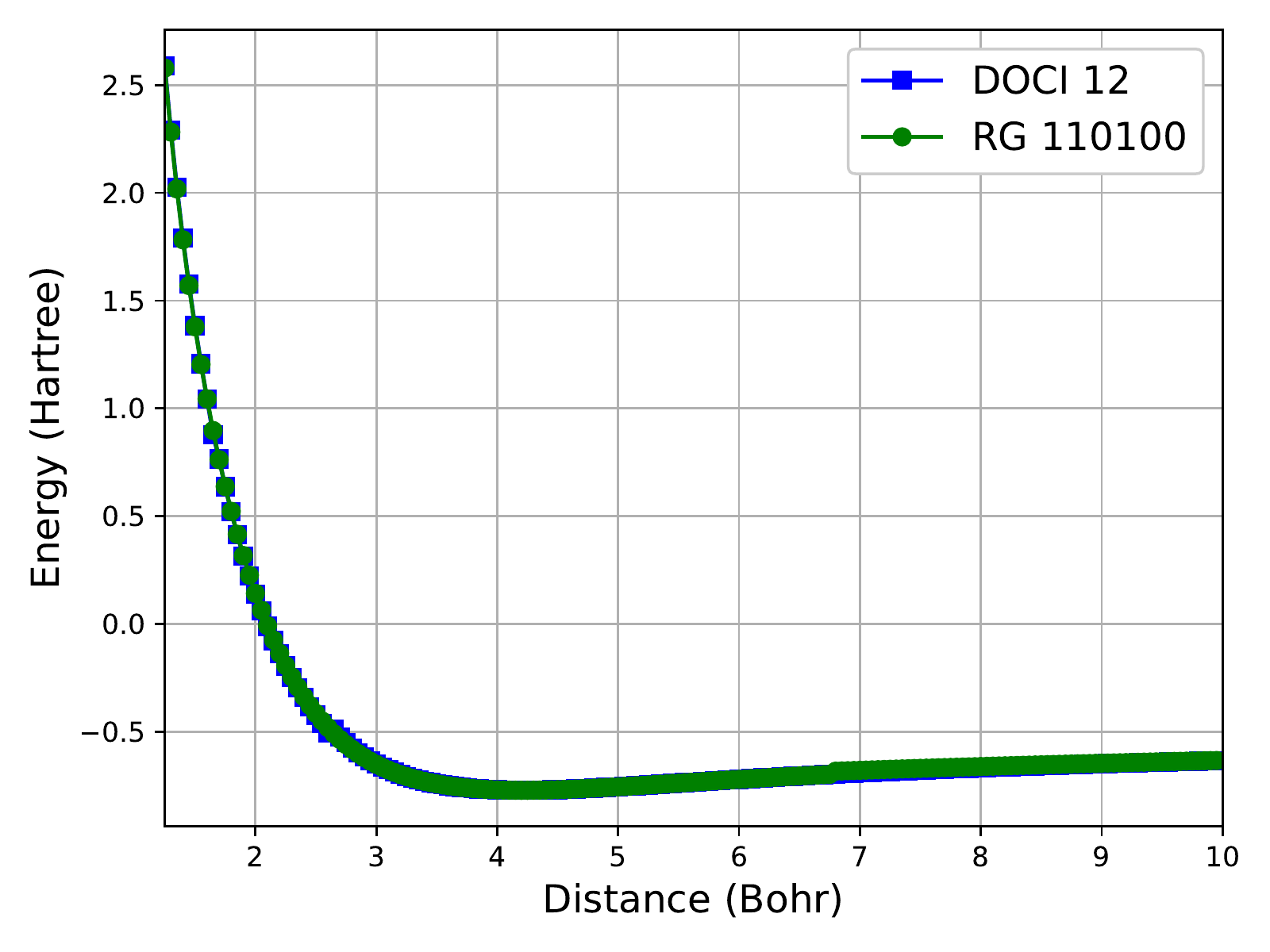} \hfill
		\includegraphics[width=0.24\textwidth]{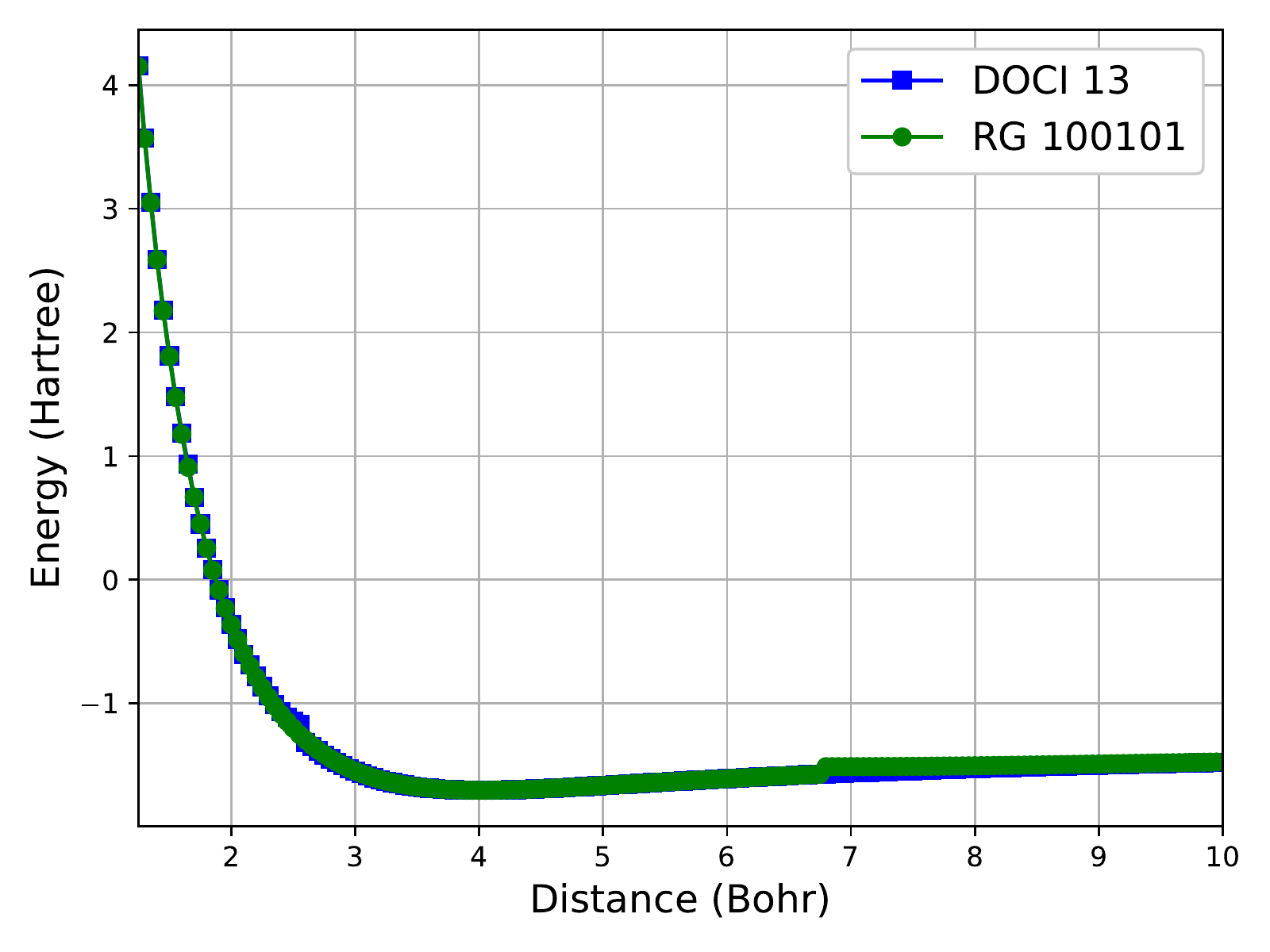} \hfill
		\includegraphics[width=0.24\textwidth]{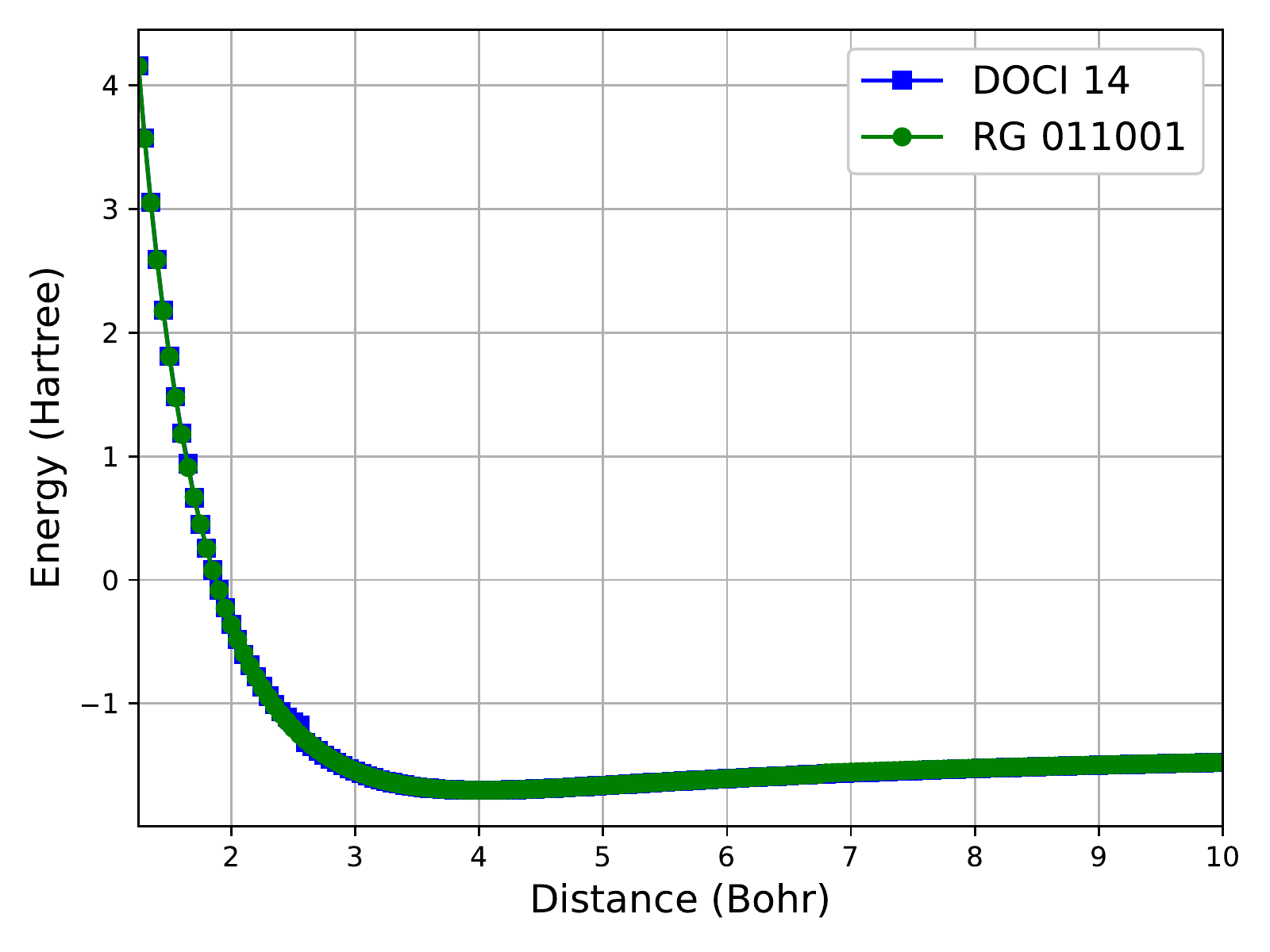} \hfill
		\includegraphics[width=0.24\textwidth]{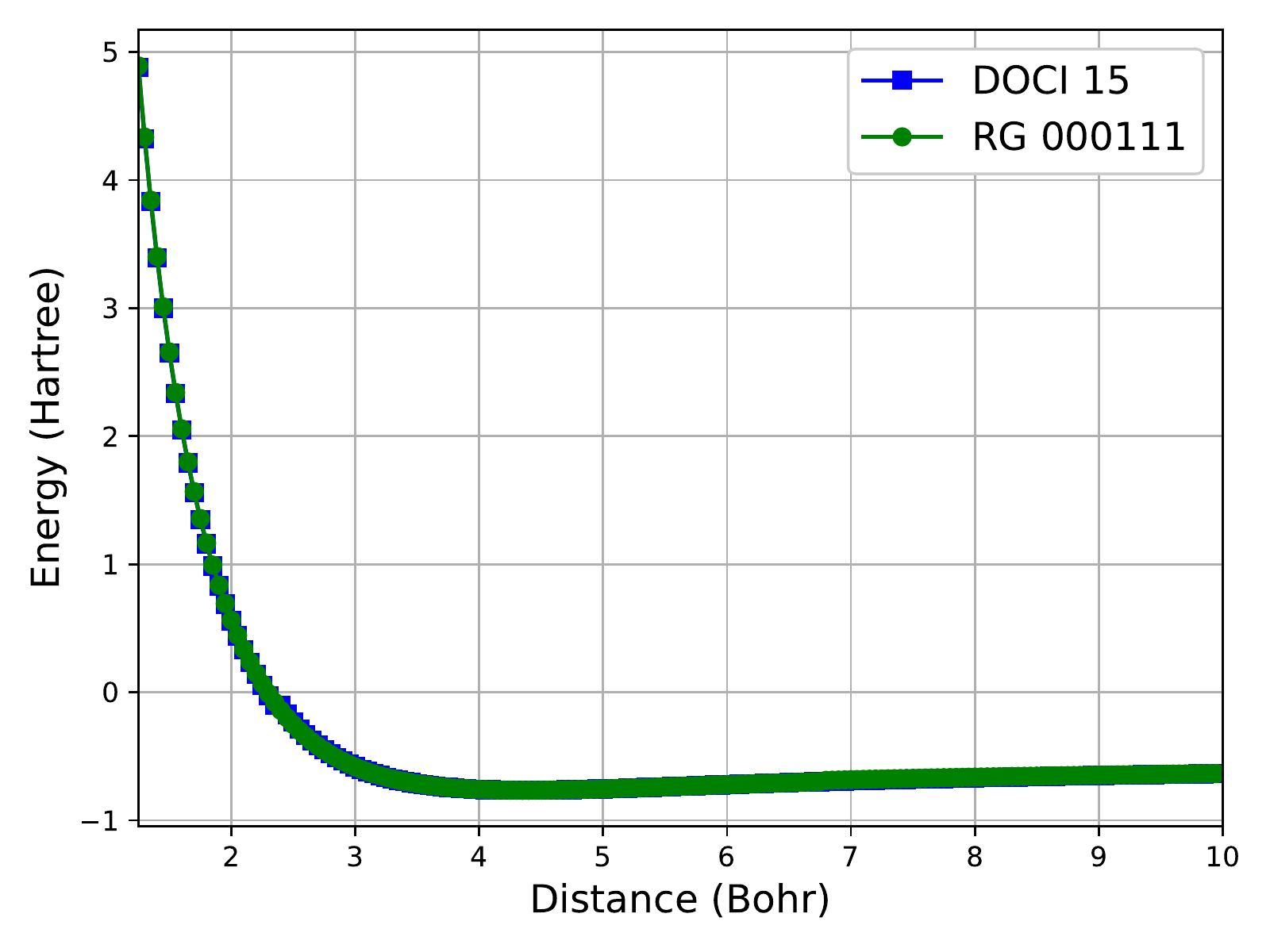}
	\end{subfigure}
	\begin{subfigure}{\textwidth}
		\includegraphics[width=0.24\textwidth]{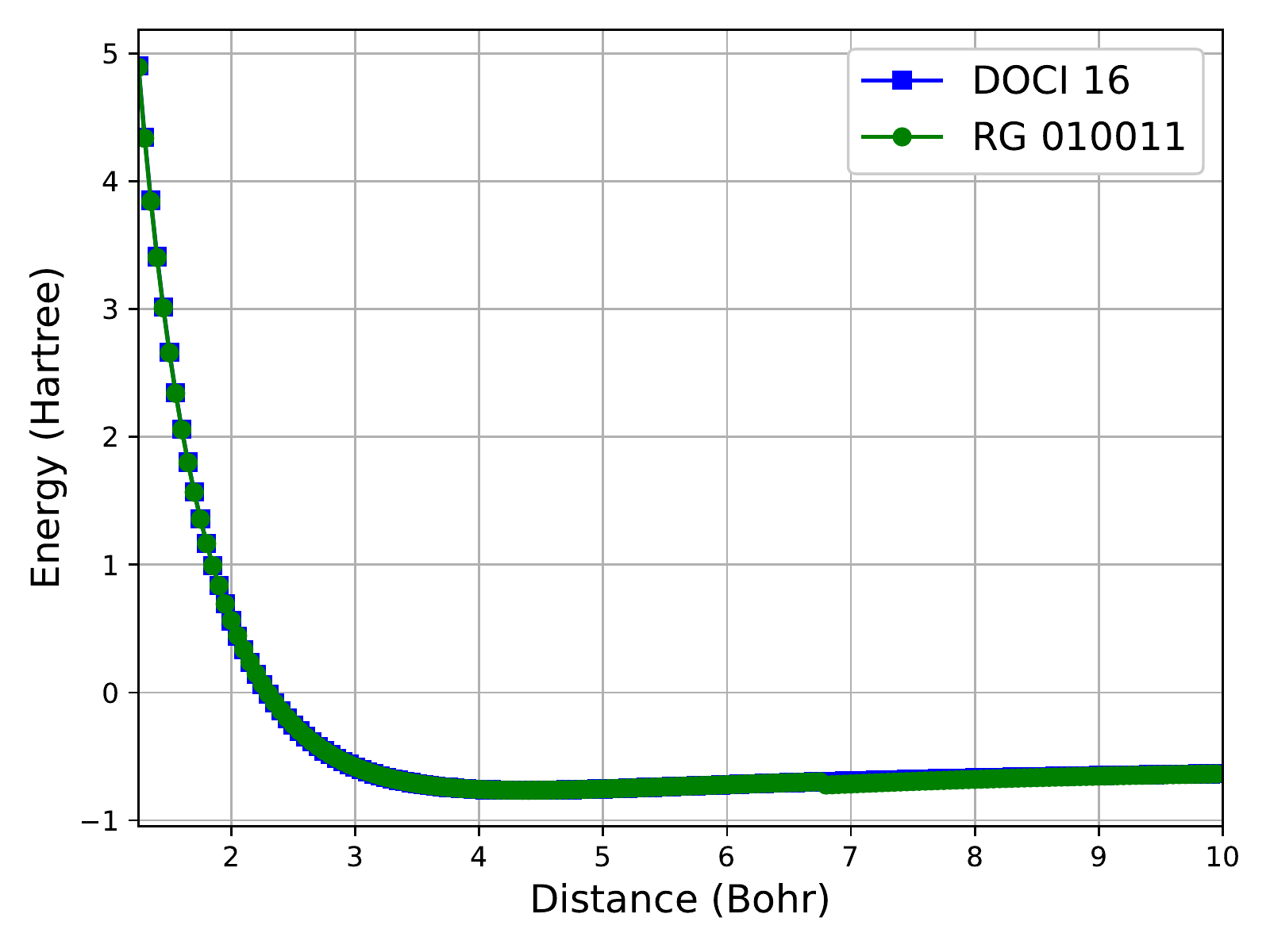} \hfill
		\includegraphics[width=0.24\textwidth]{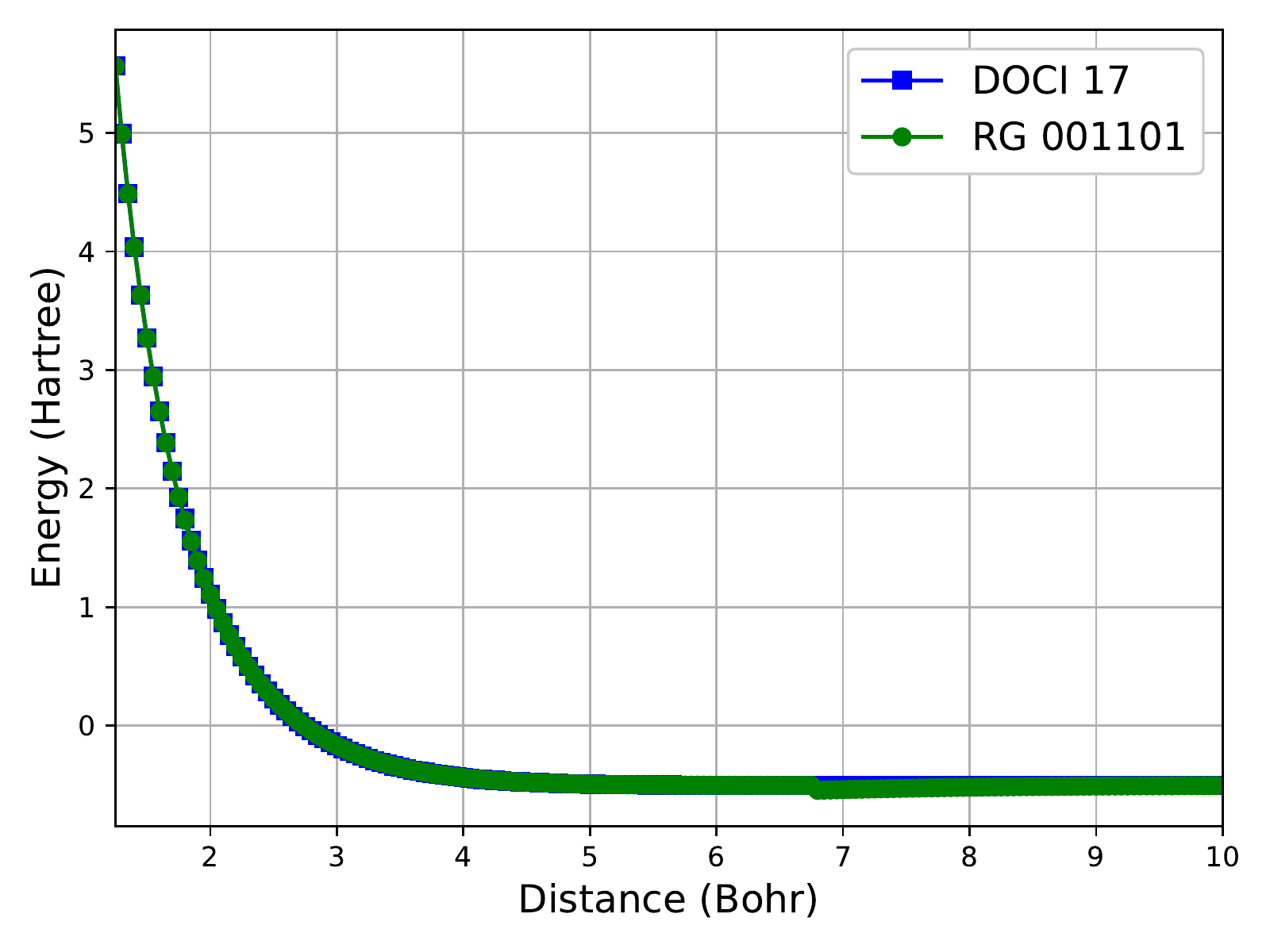} \hfill
		\includegraphics[width=0.24\textwidth]{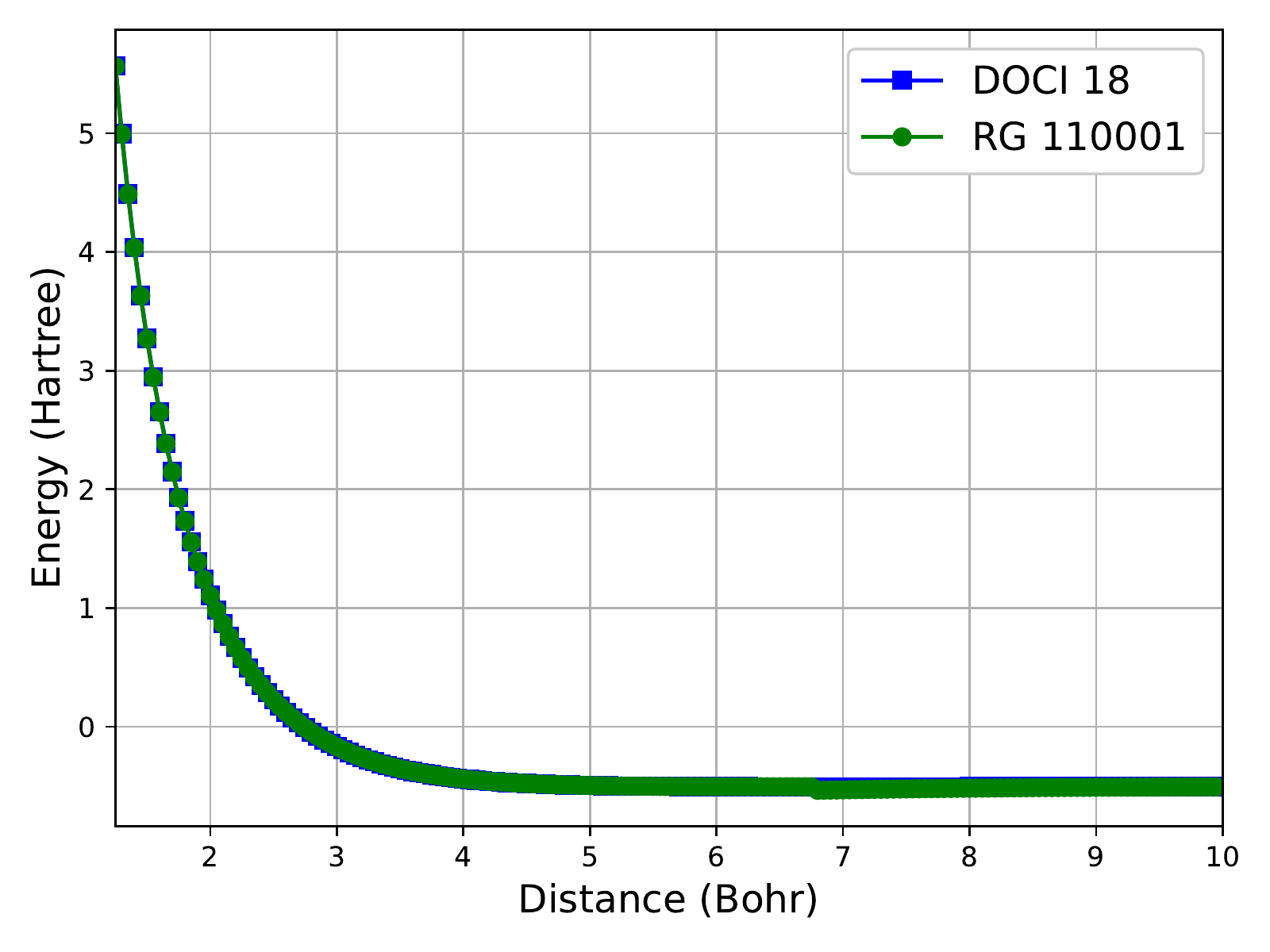} \hfill
		\includegraphics[width=0.24\textwidth]{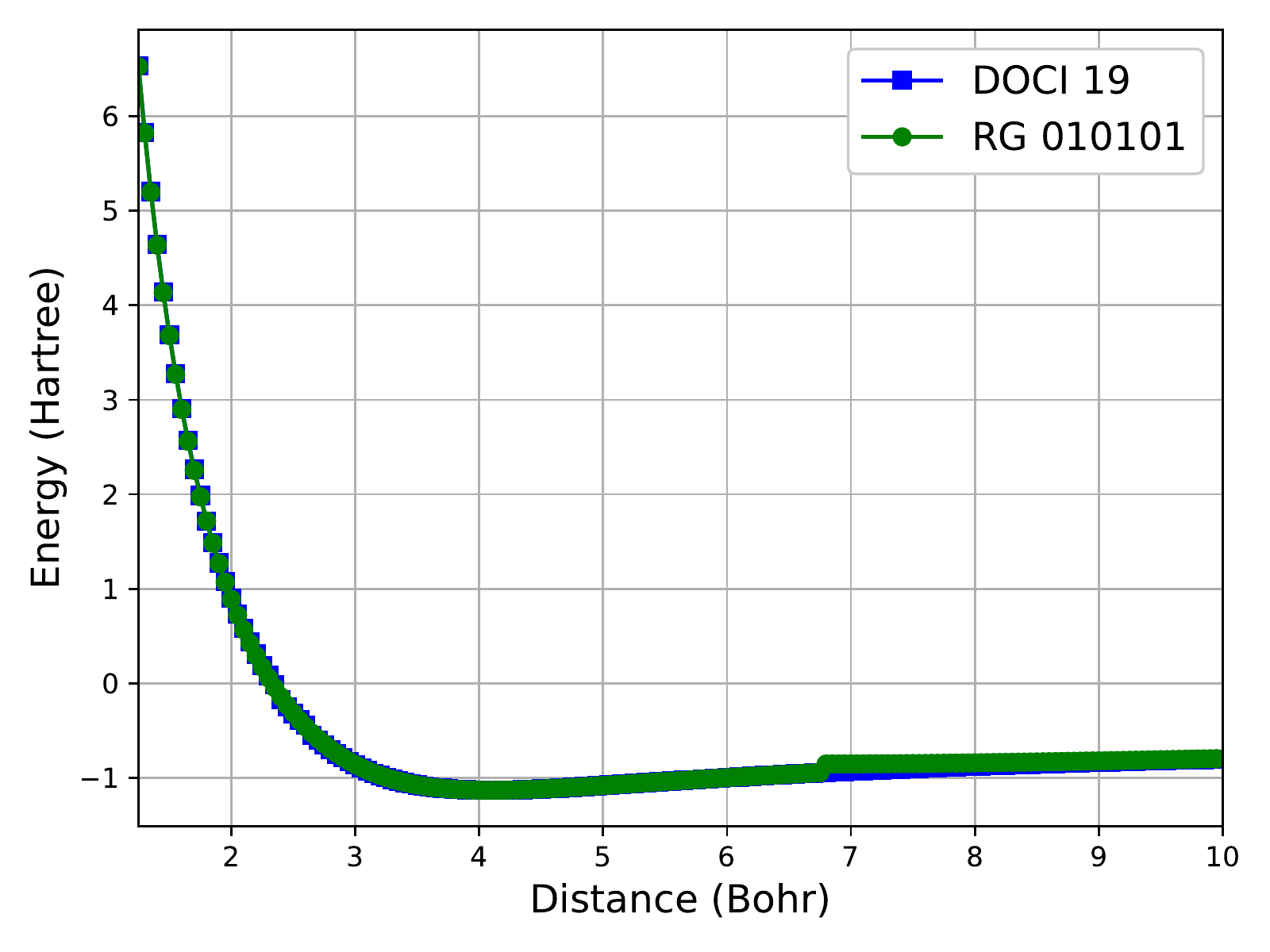}
	\end{subfigure}
	\caption{RG states computed with reduced BCS parameters defining the Hamiltonian corresponding to the optimal 101010 solution and DOCI states. All results computed in the STO-6G basis set in the basis of OO-DOCI orbitals.}
\end{figure}

\begin{figure}[ht!]
	\begin{subfigure}{\textwidth}
		\includegraphics[width=0.24\textwidth]{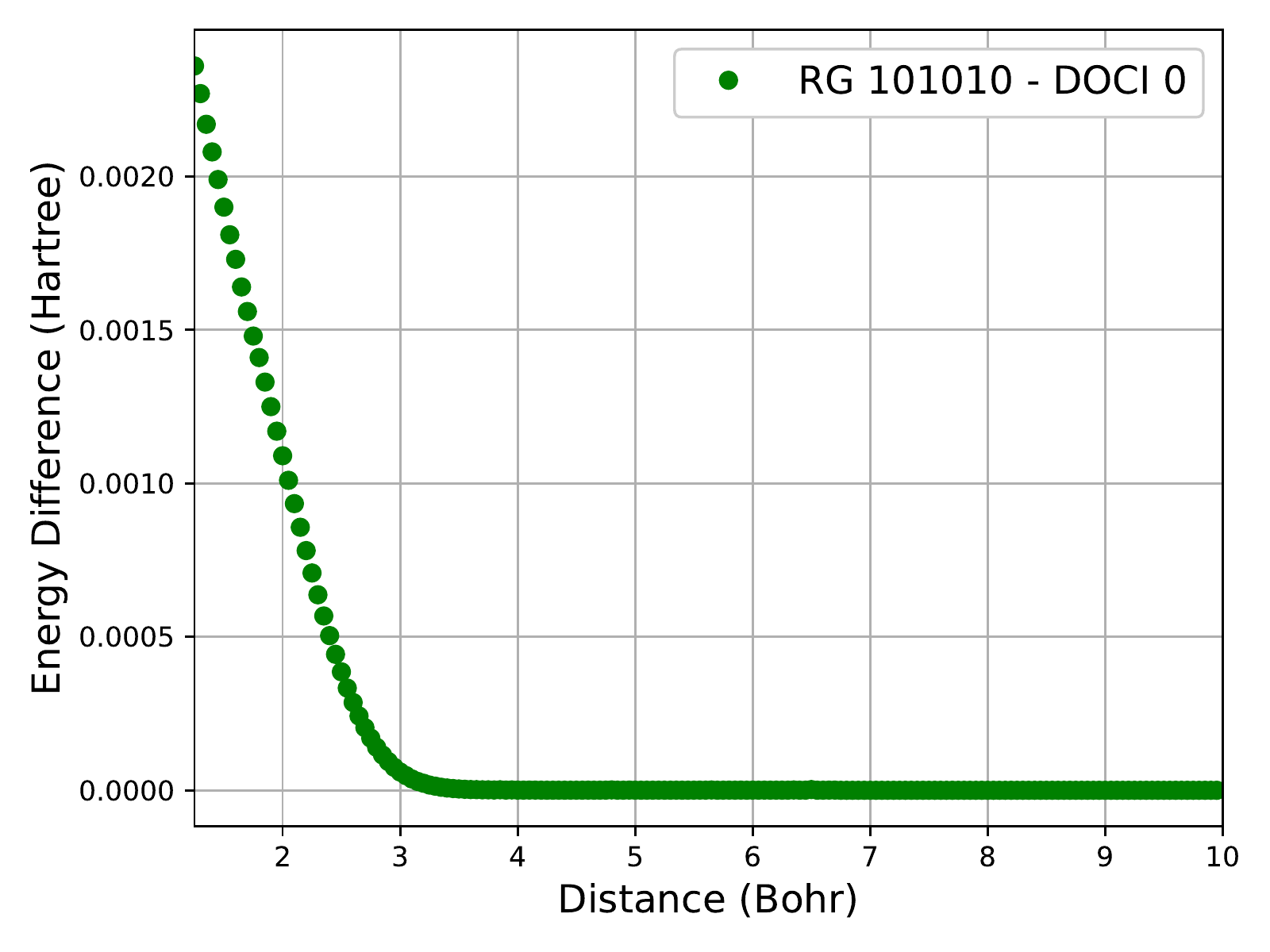} \hfill
		\includegraphics[width=0.24\textwidth]{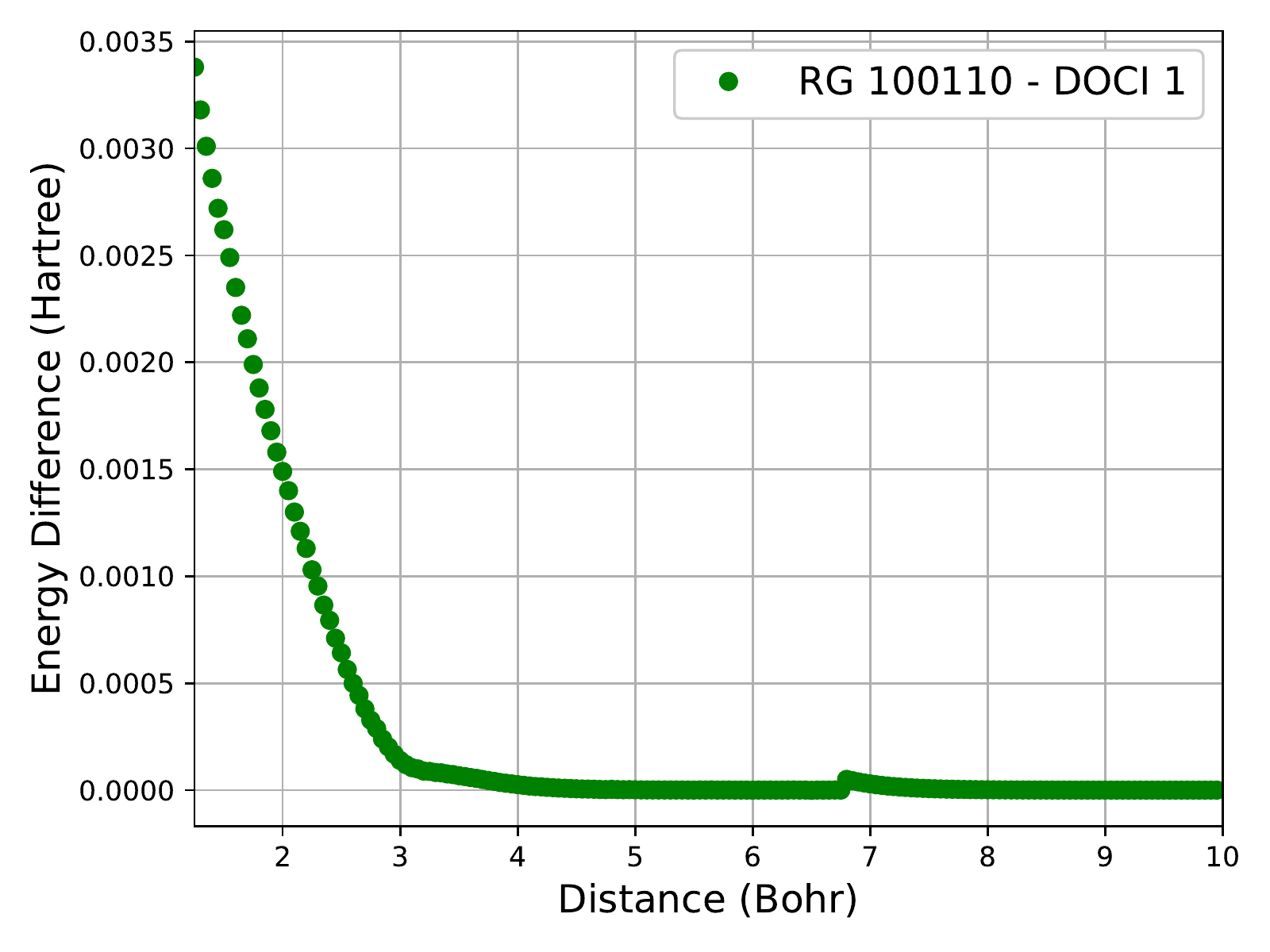} \hfill
		\includegraphics[width=0.24\textwidth]{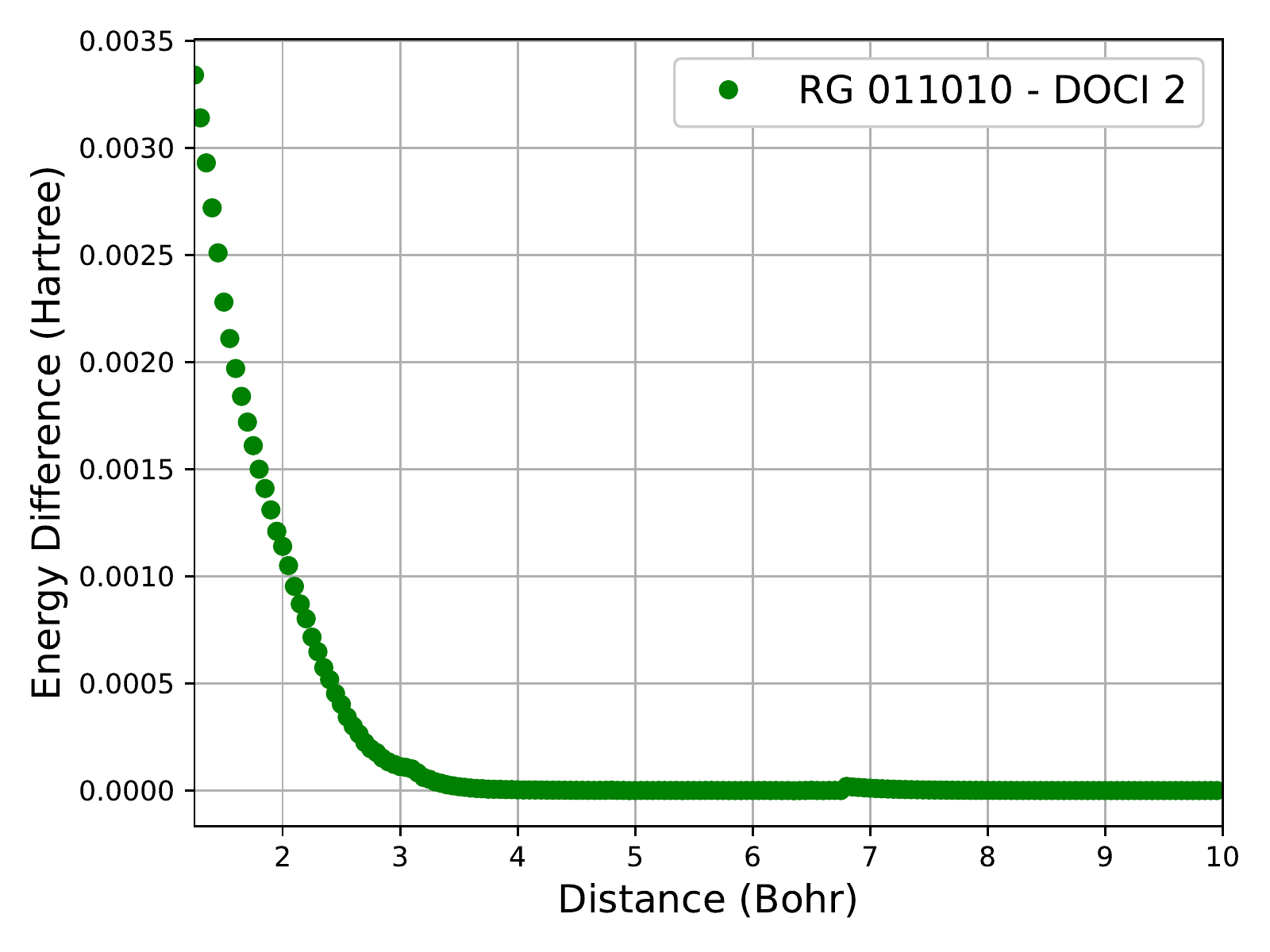} \hfill
		\includegraphics[width=0.24\textwidth]{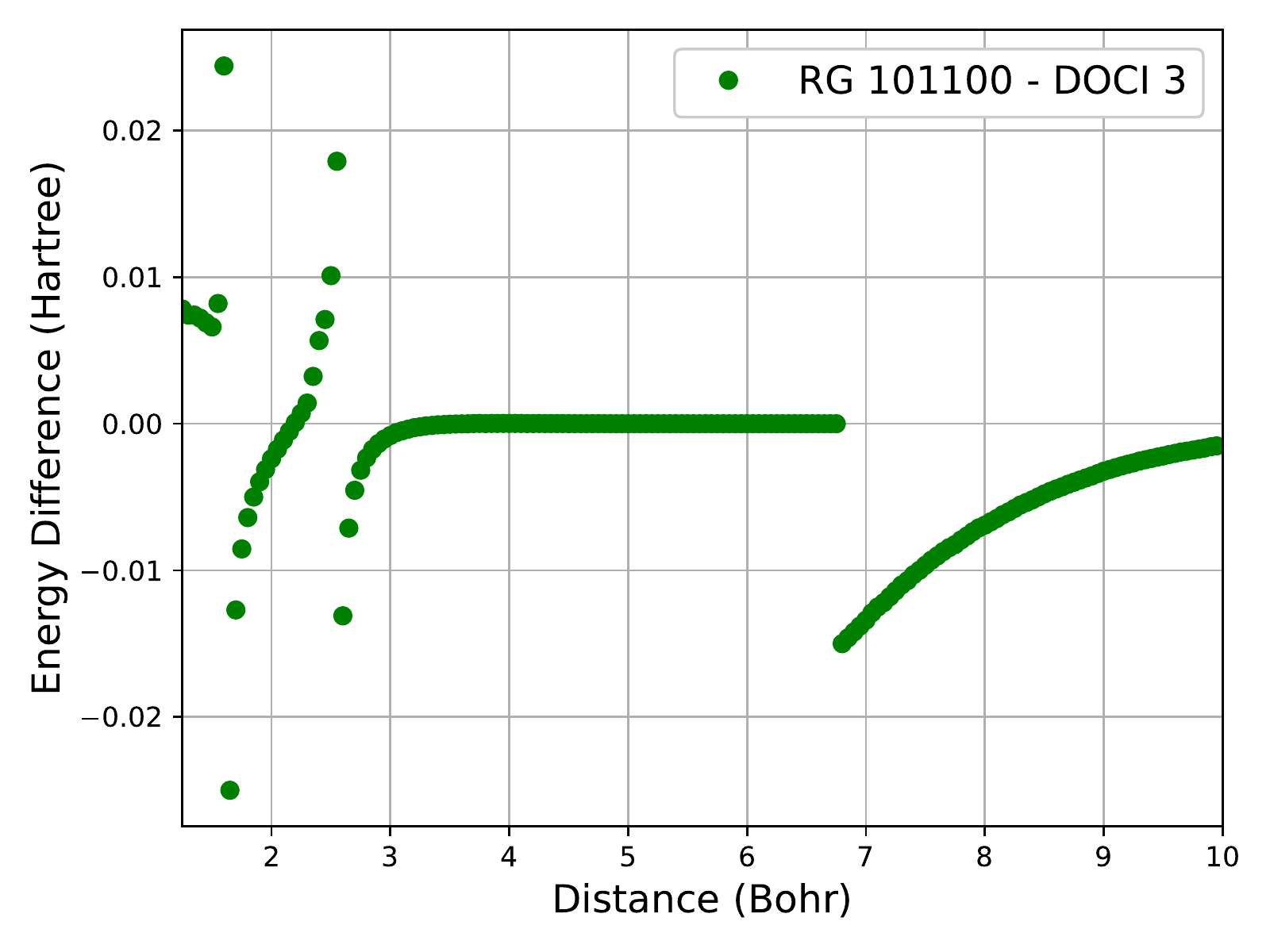}
	\end{subfigure}
	\begin{subfigure}{\textwidth}
		\includegraphics[width=0.24\textwidth]{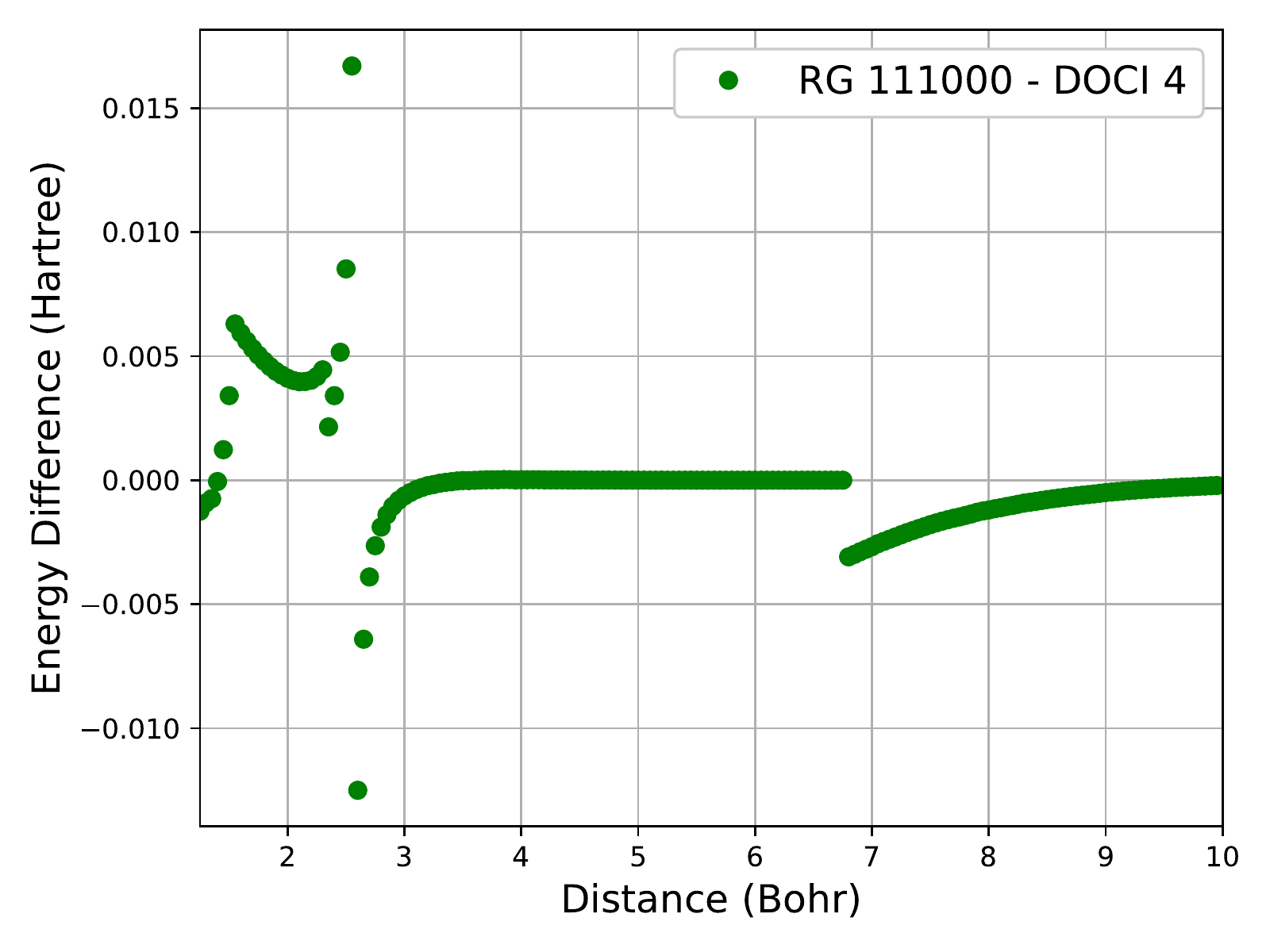} \hfill
		\includegraphics[width=0.24\textwidth]{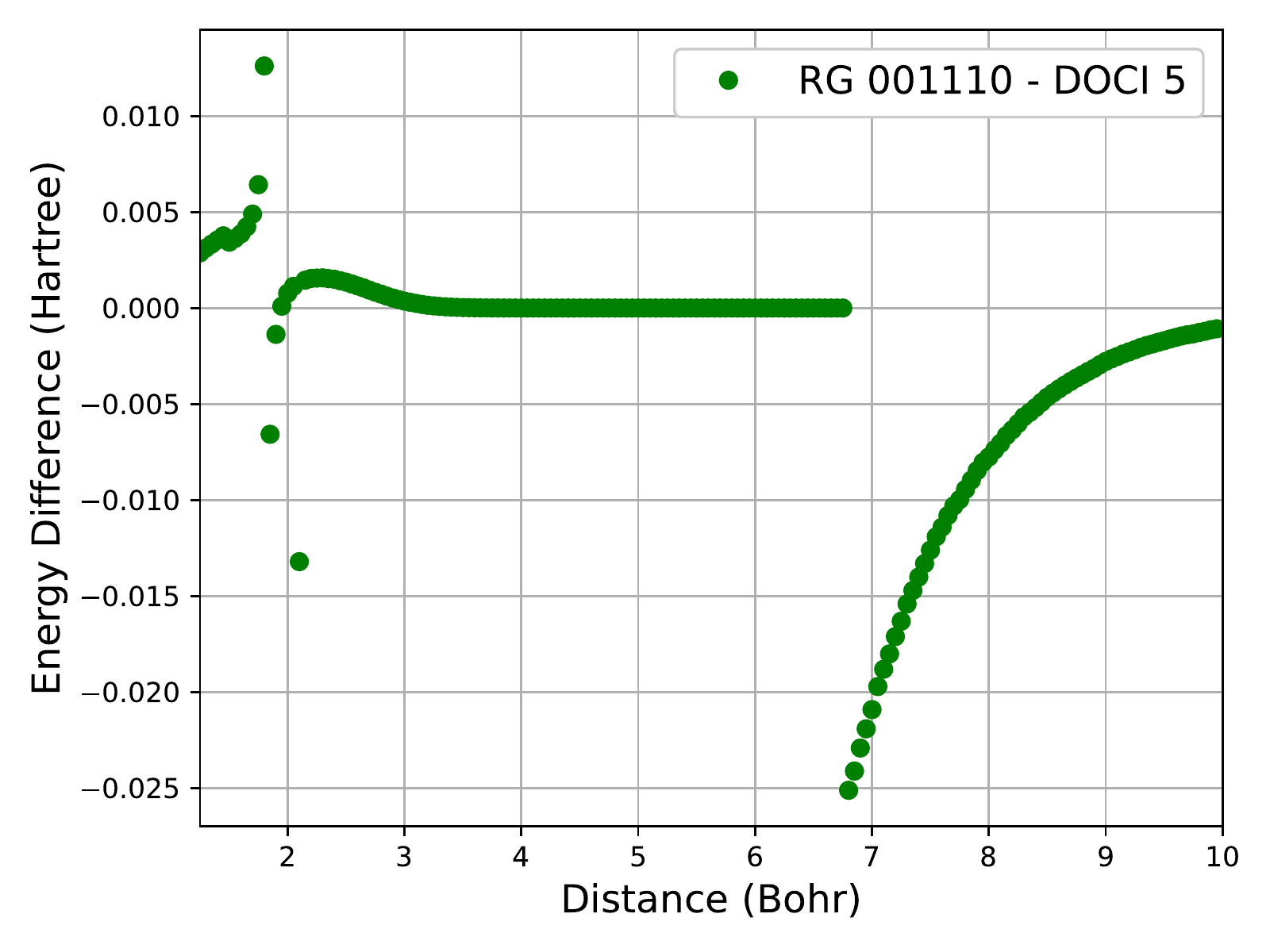} \hfill
		\includegraphics[width=0.24\textwidth]{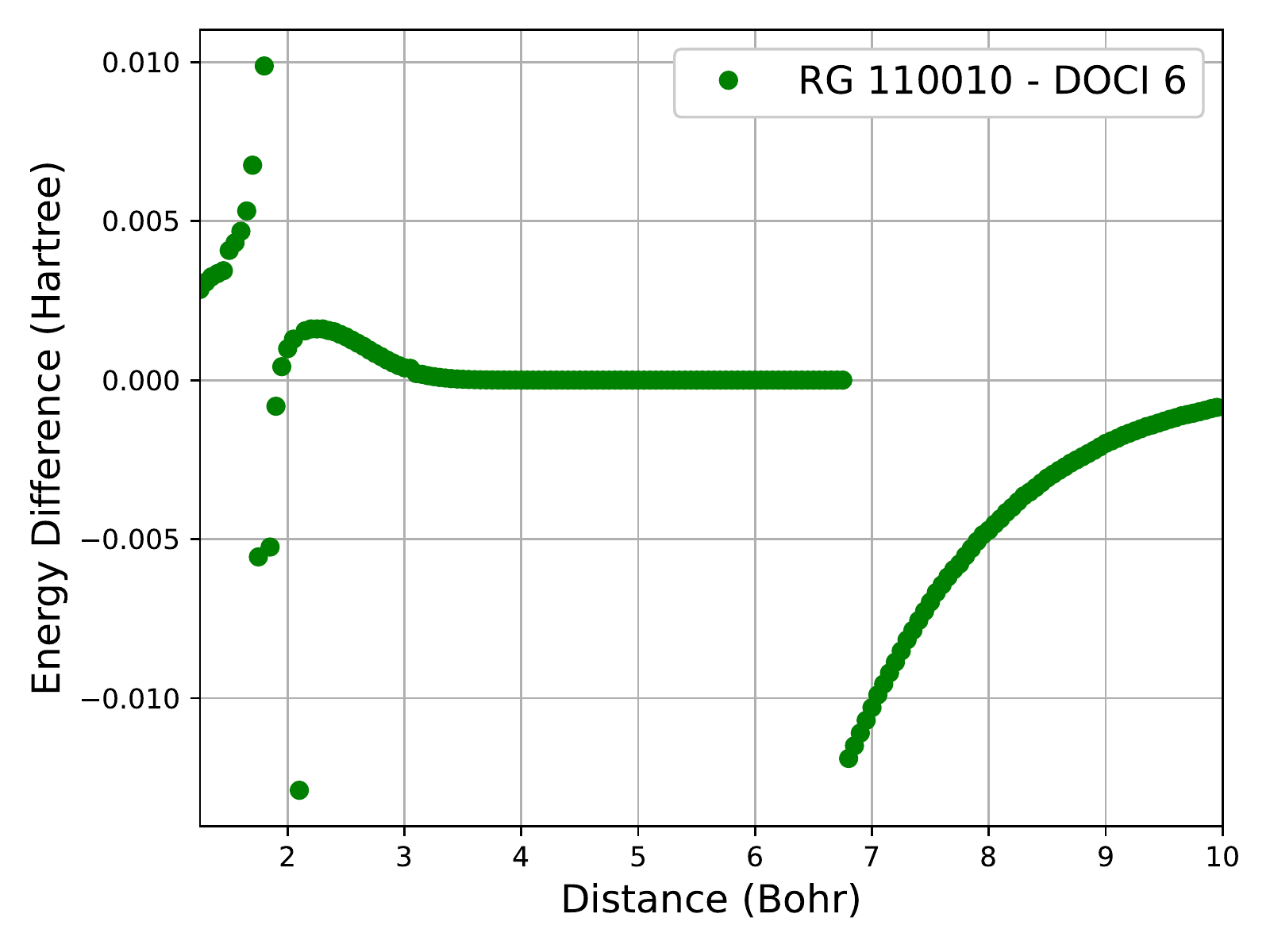} \hfill
		\includegraphics[width=0.24\textwidth]{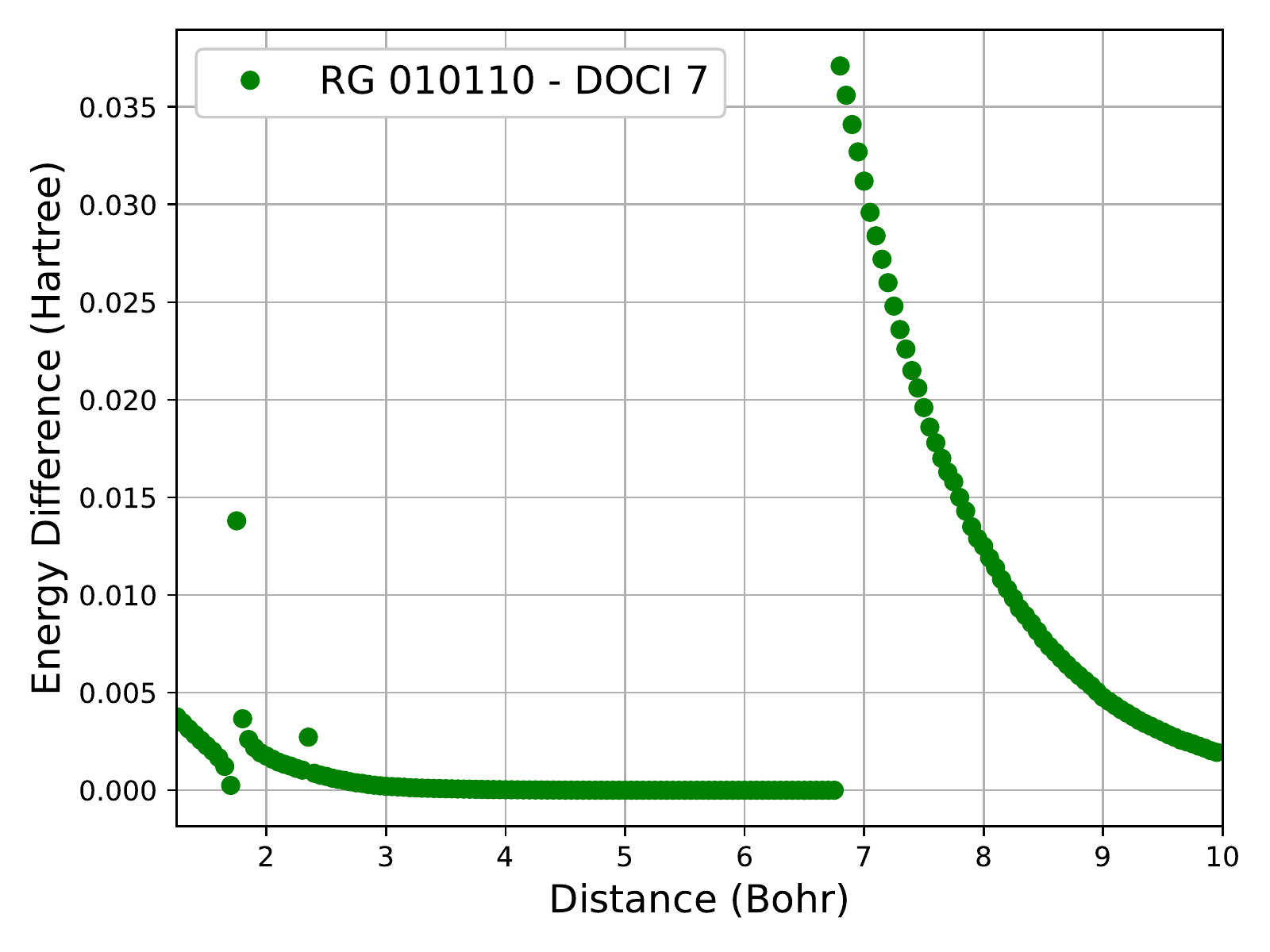}
	\end{subfigure}
	\begin{subfigure}{\textwidth}
		\includegraphics[width=0.24\textwidth]{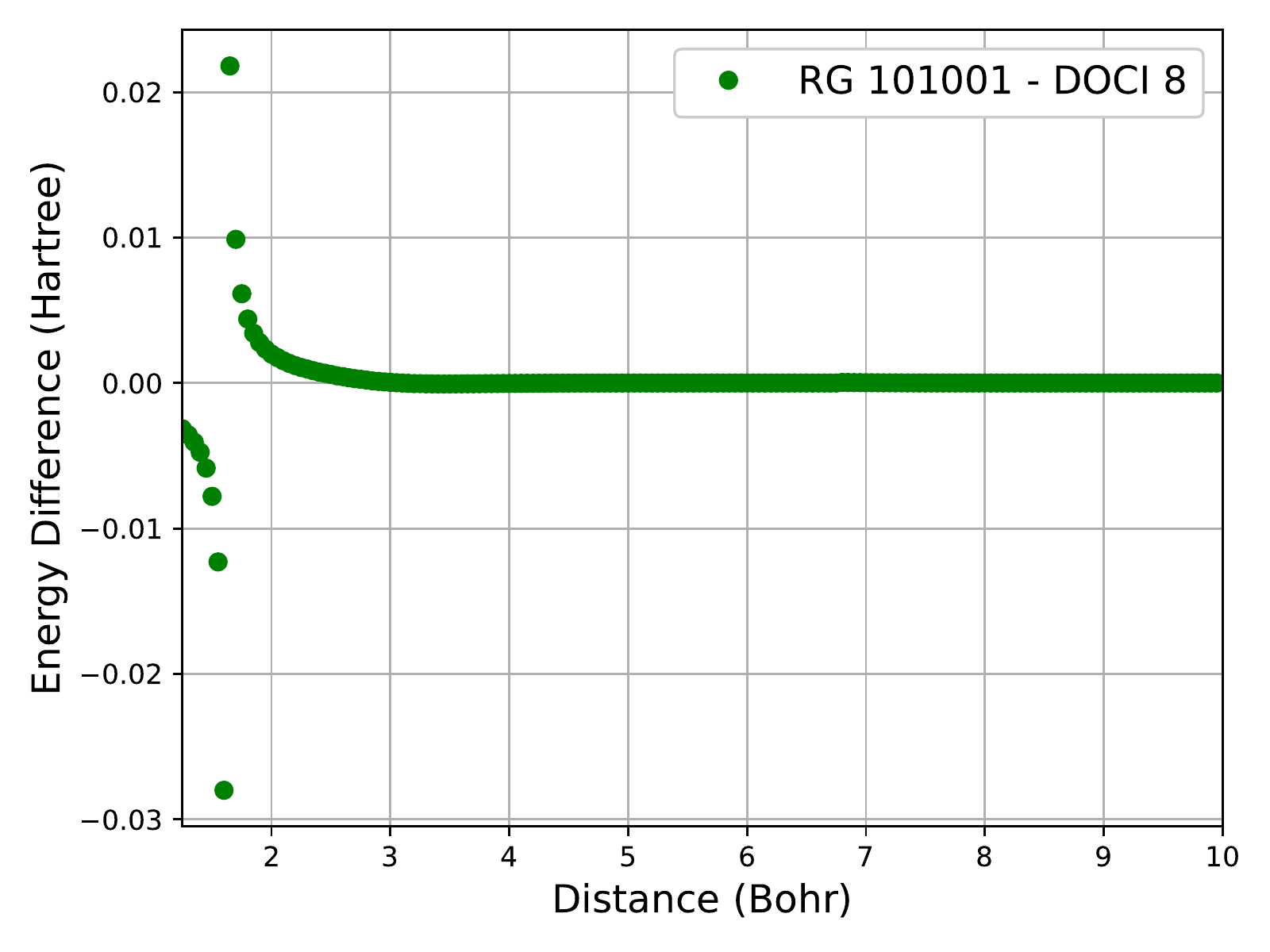} \hfill
		\includegraphics[width=0.24\textwidth]{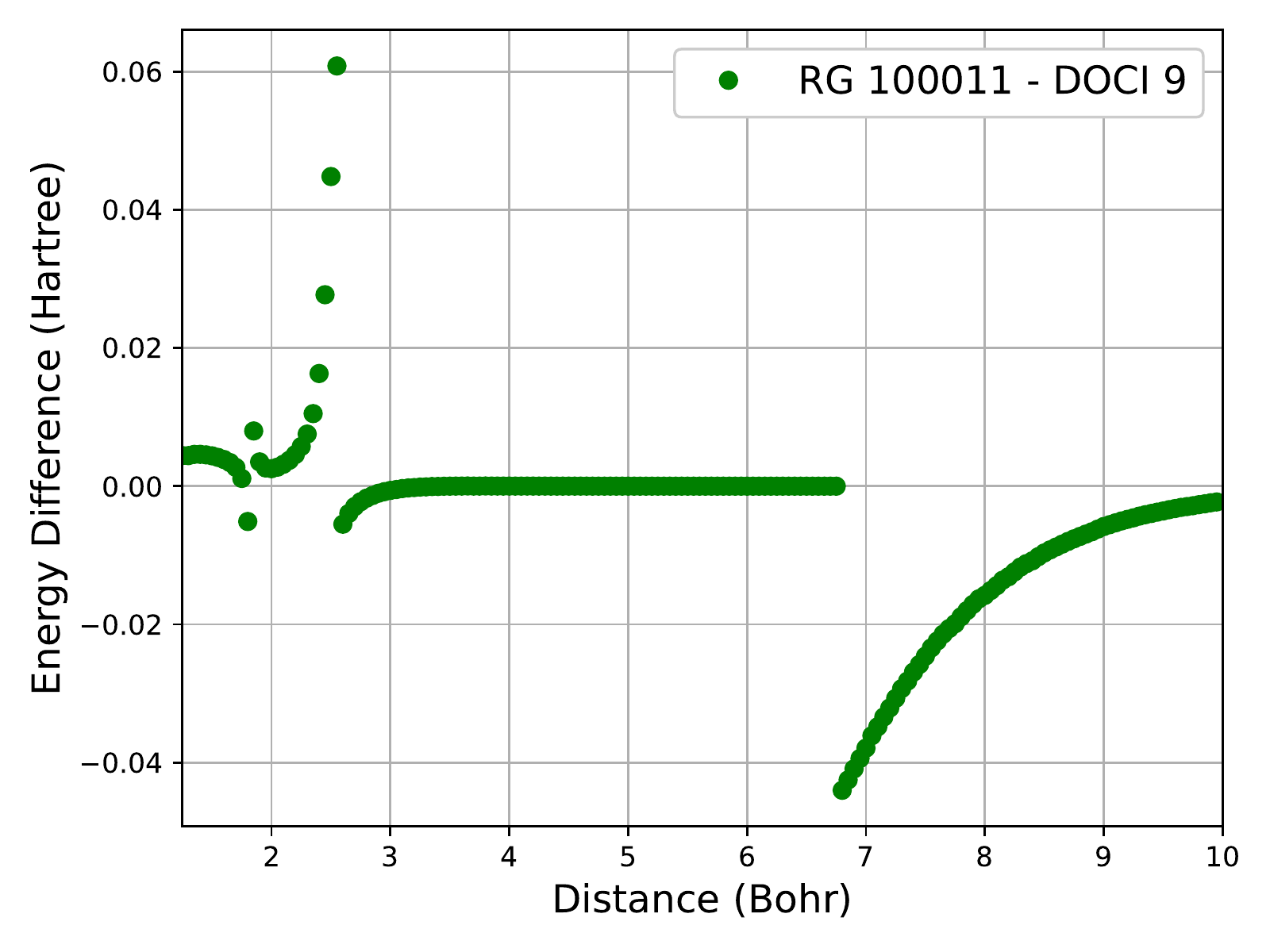} \hfill
		\includegraphics[width=0.24\textwidth]{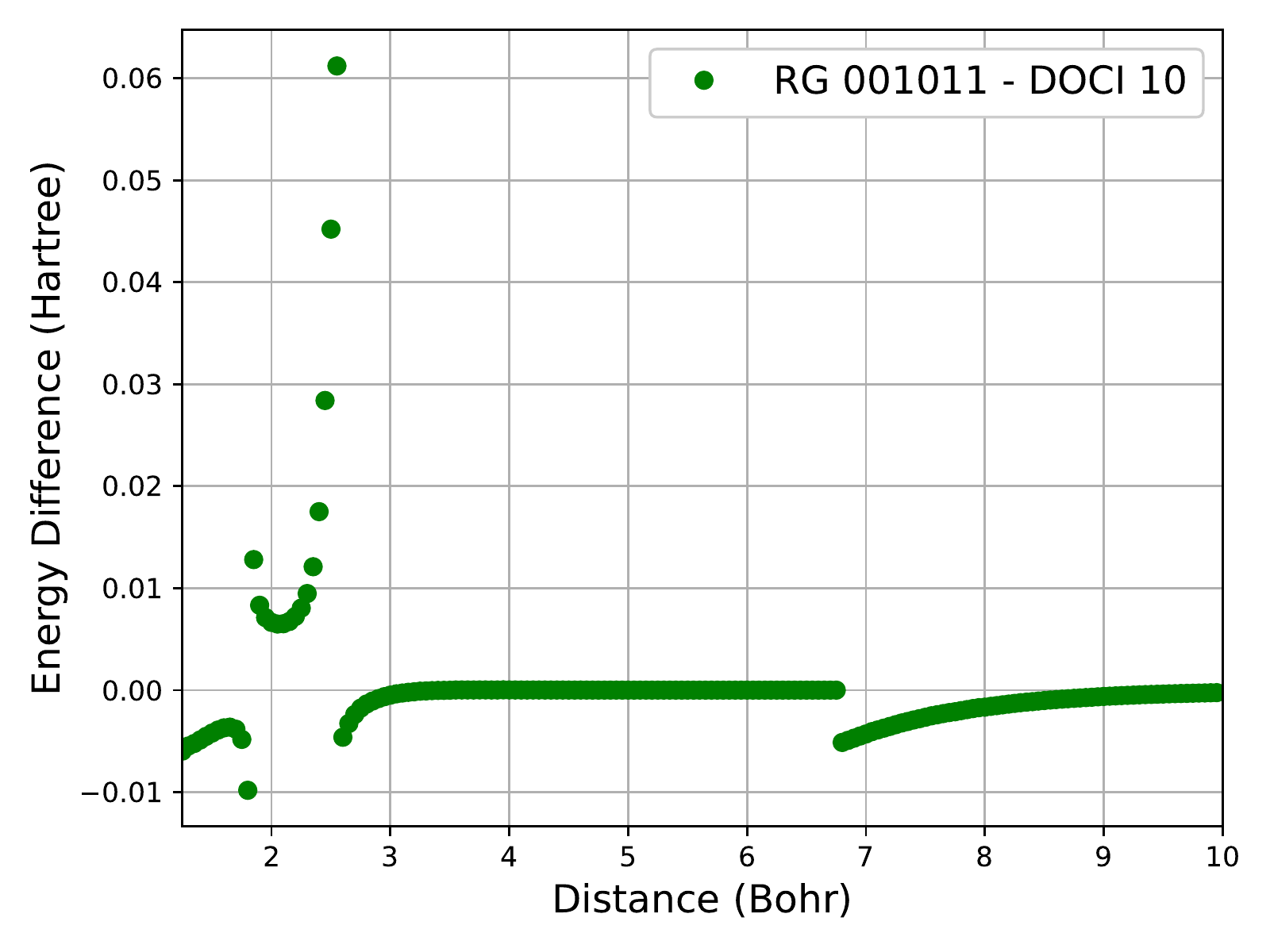} \hfill
		\includegraphics[width=0.24\textwidth]{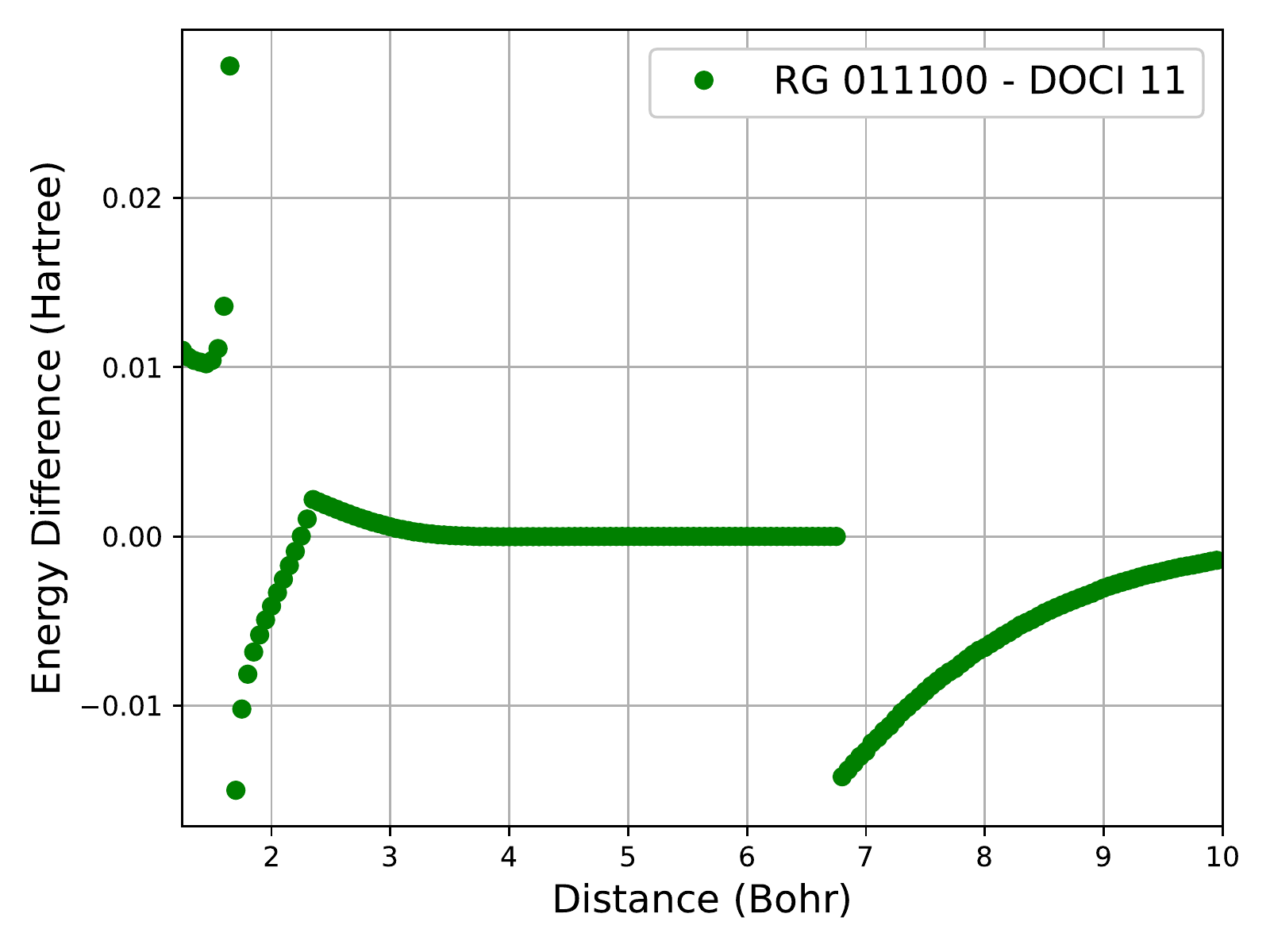}
	\end{subfigure}
	\begin{subfigure}{\textwidth}
		\includegraphics[width=0.24\textwidth]{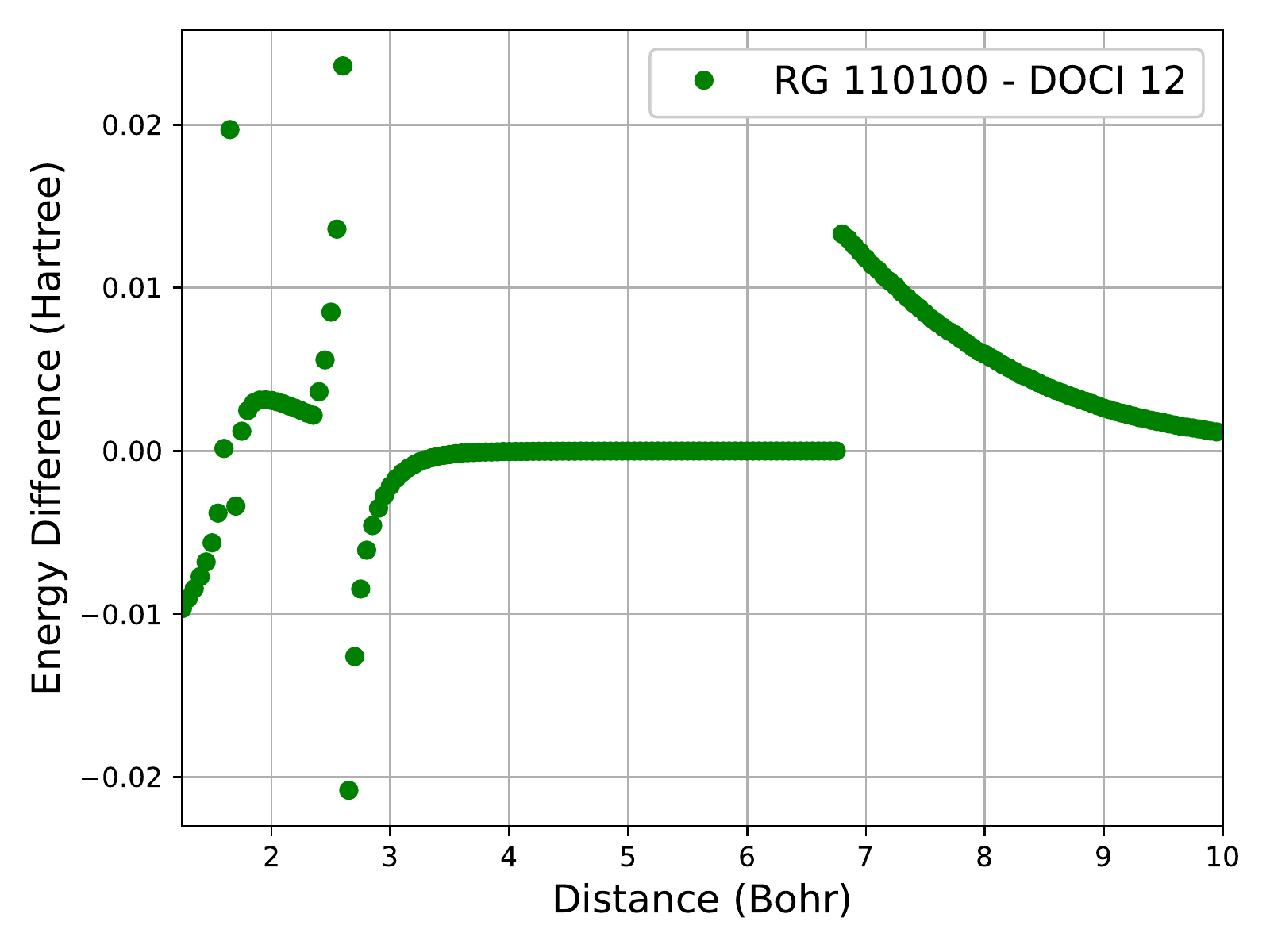} \hfill
		\includegraphics[width=0.24\textwidth]{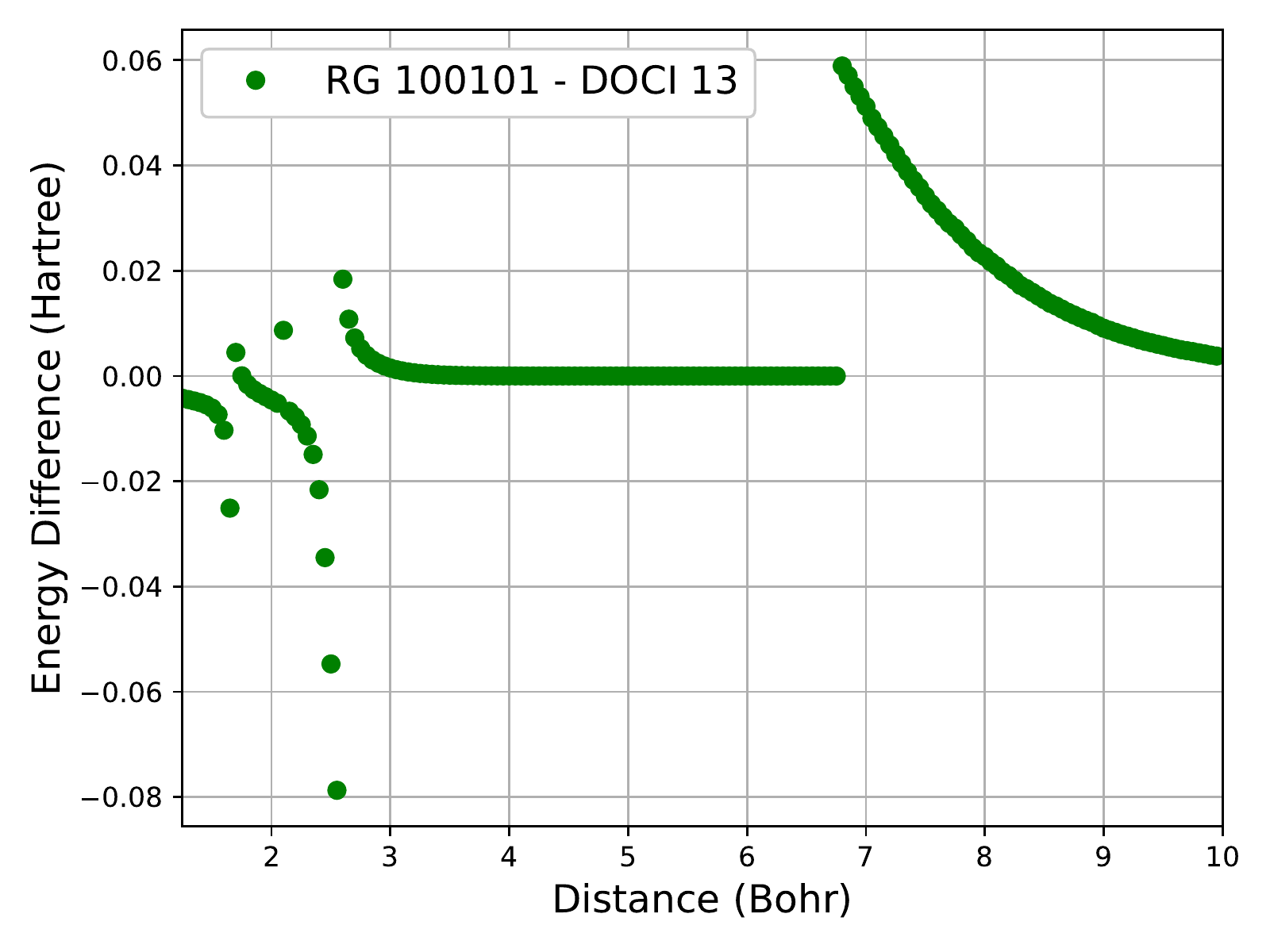} \hfill
		\includegraphics[width=0.24\textwidth]{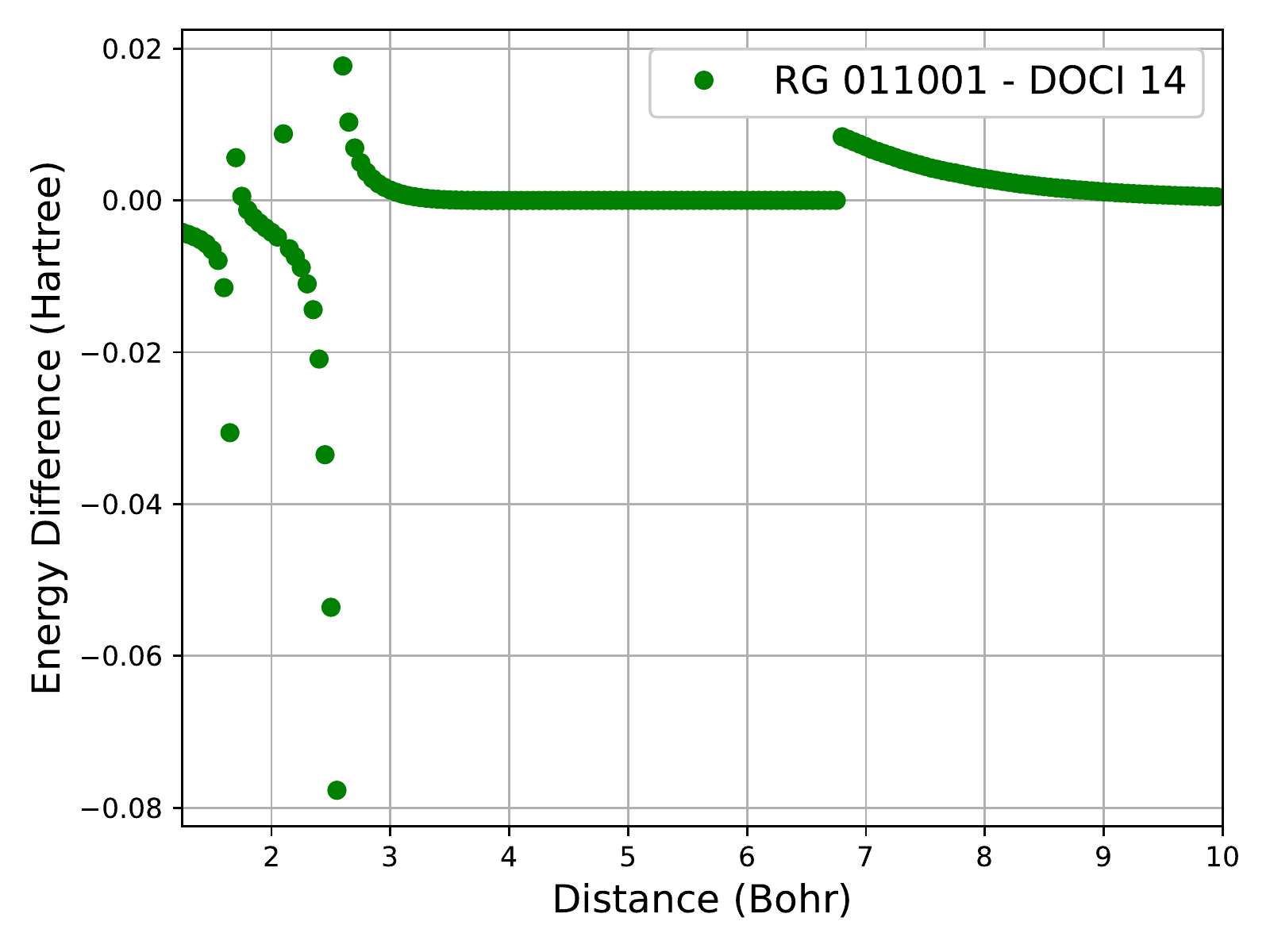} \hfill
		\includegraphics[width=0.24\textwidth]{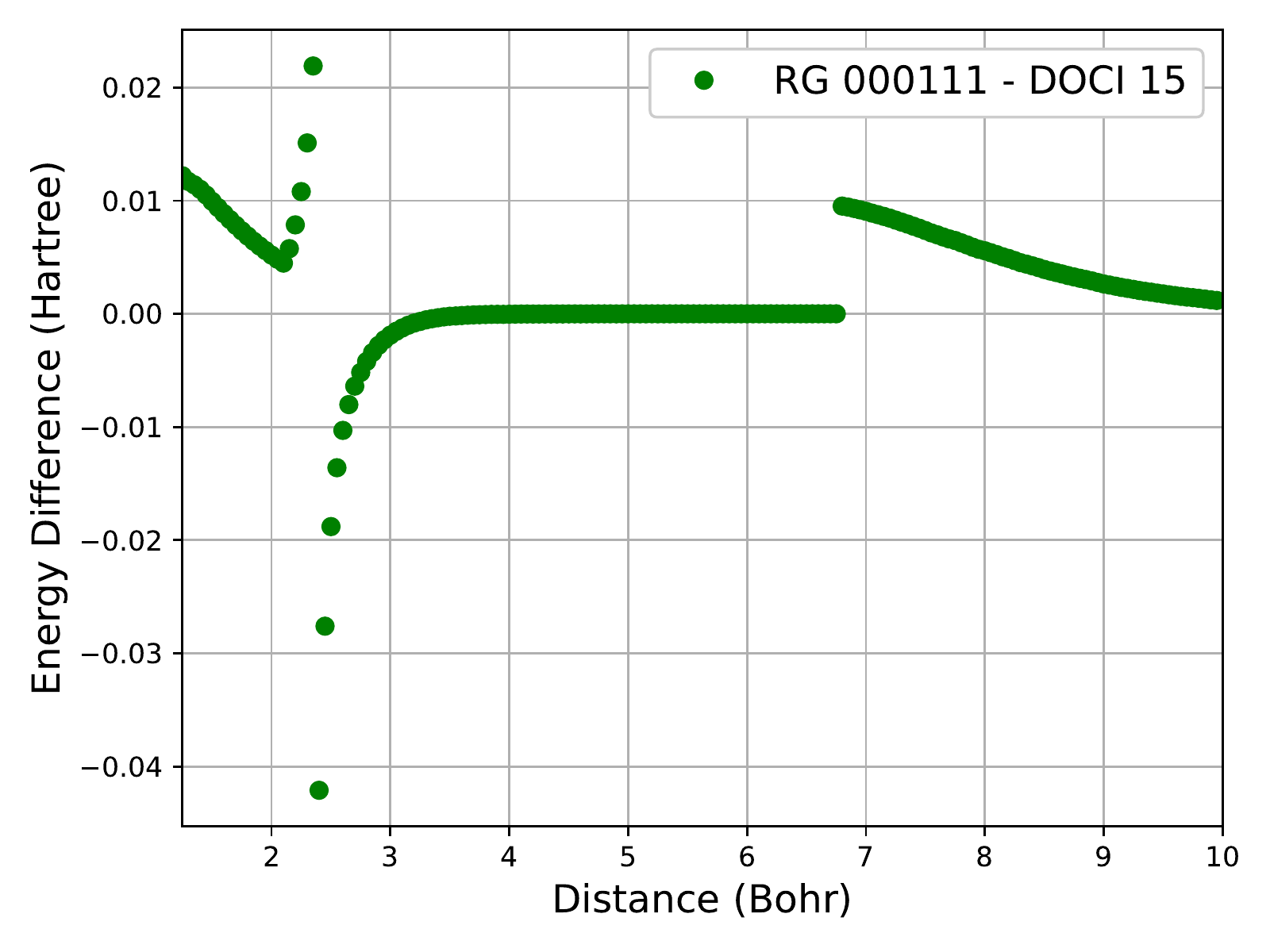}
	\end{subfigure}
	\begin{subfigure}{\textwidth}
		\includegraphics[width=0.24\textwidth]{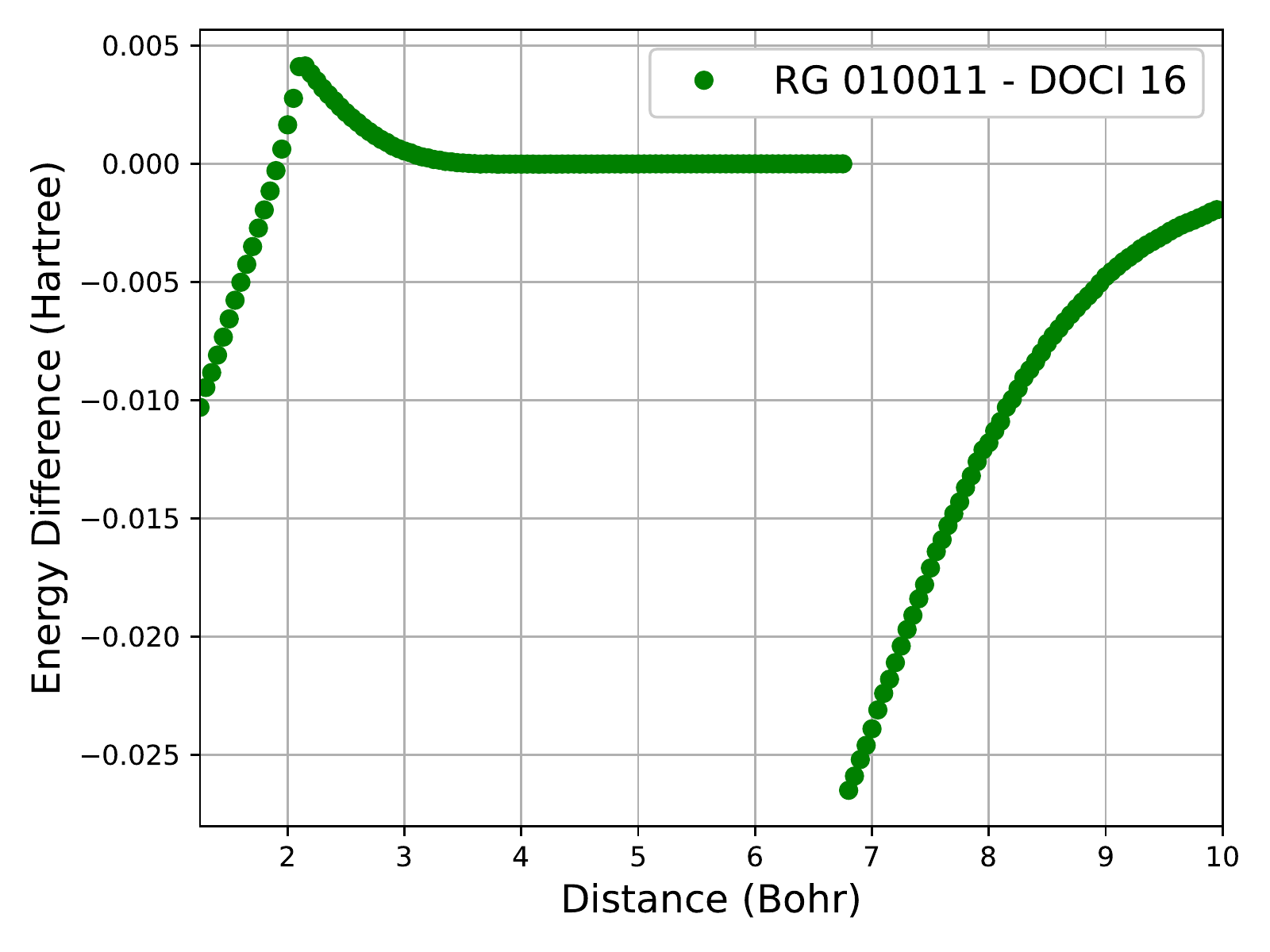} \hfill
		\includegraphics[width=0.24\textwidth]{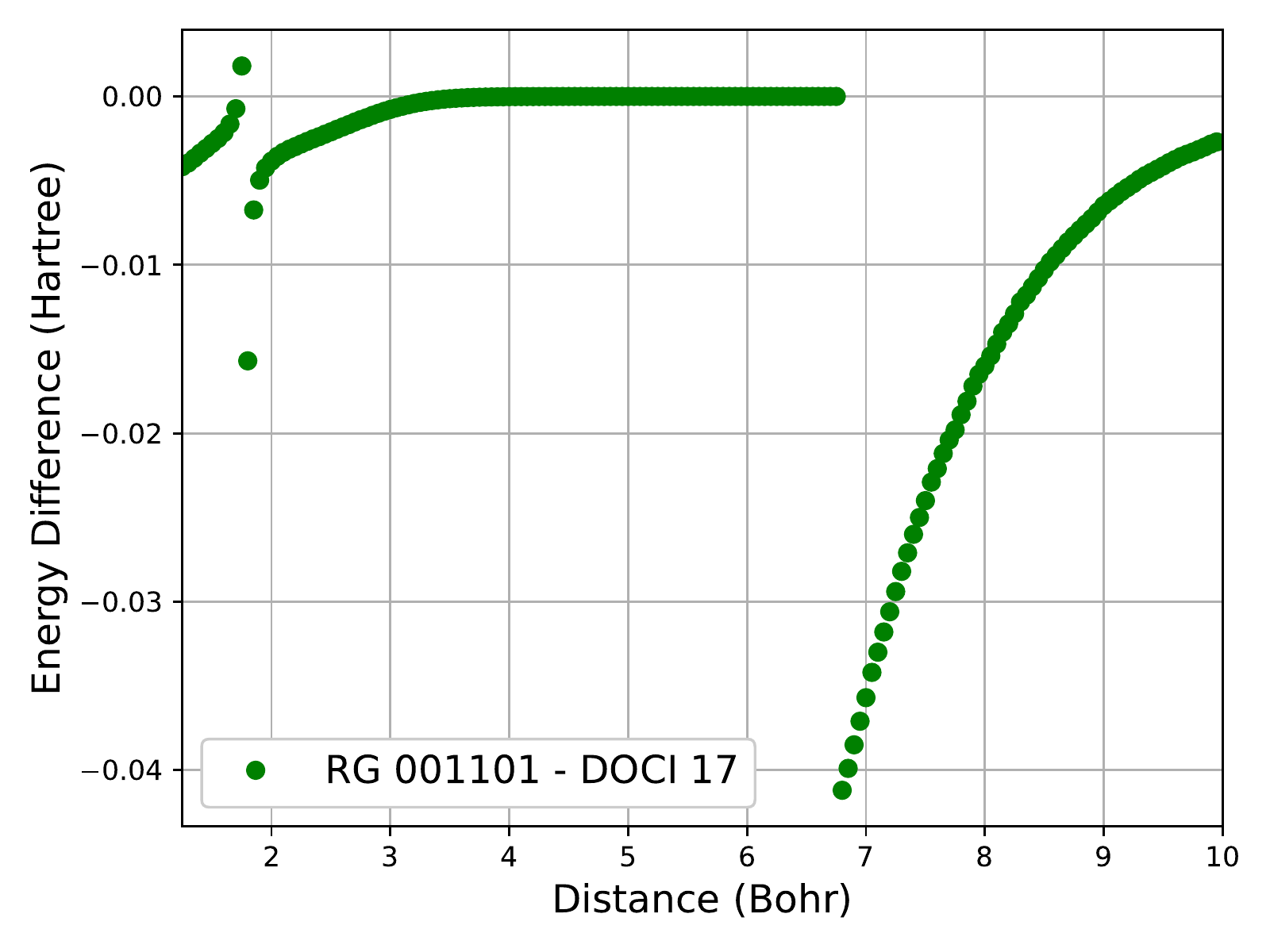} \hfill
		\includegraphics[width=0.24\textwidth]{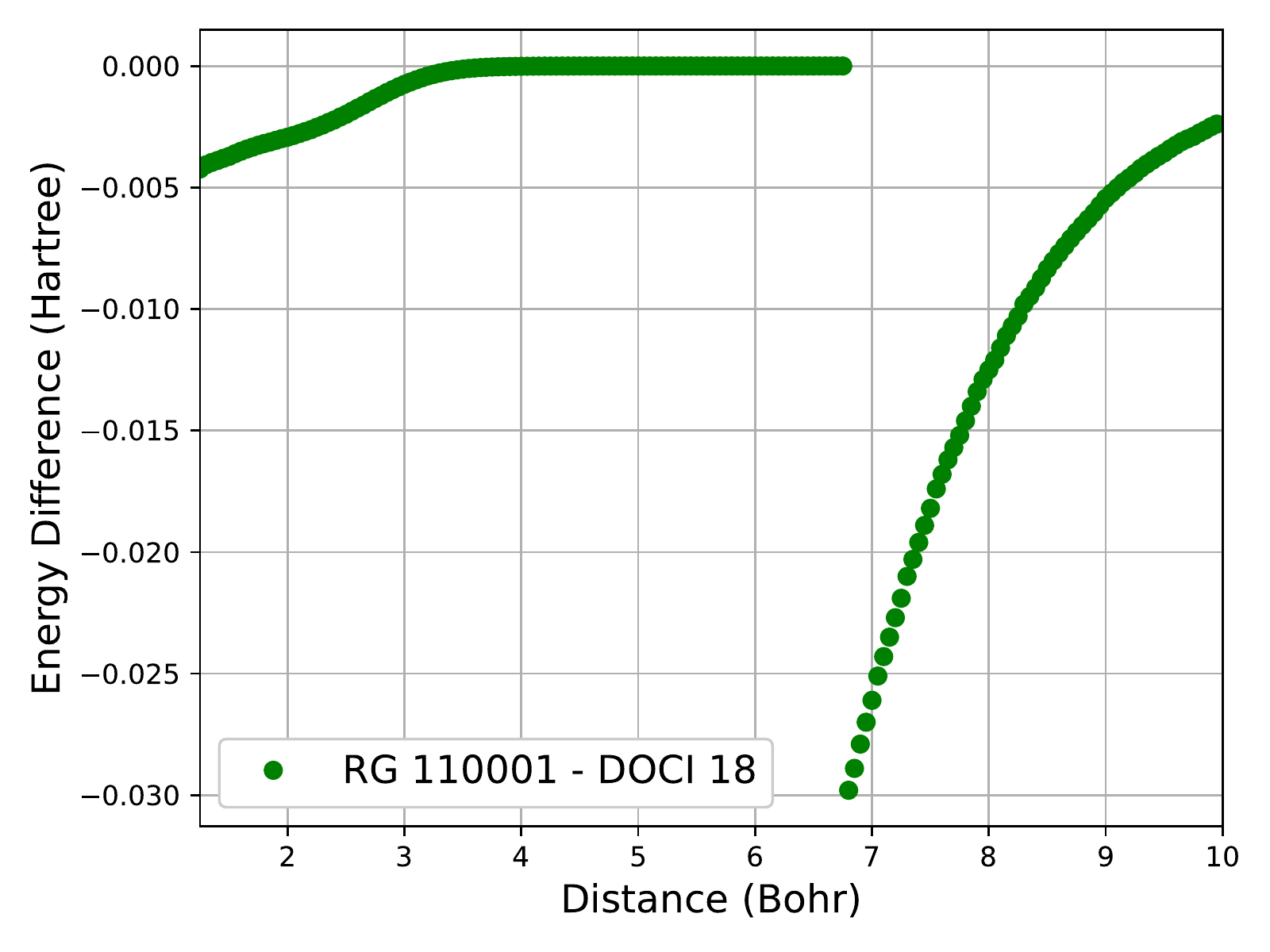} \hfill
		\includegraphics[width=0.24\textwidth]{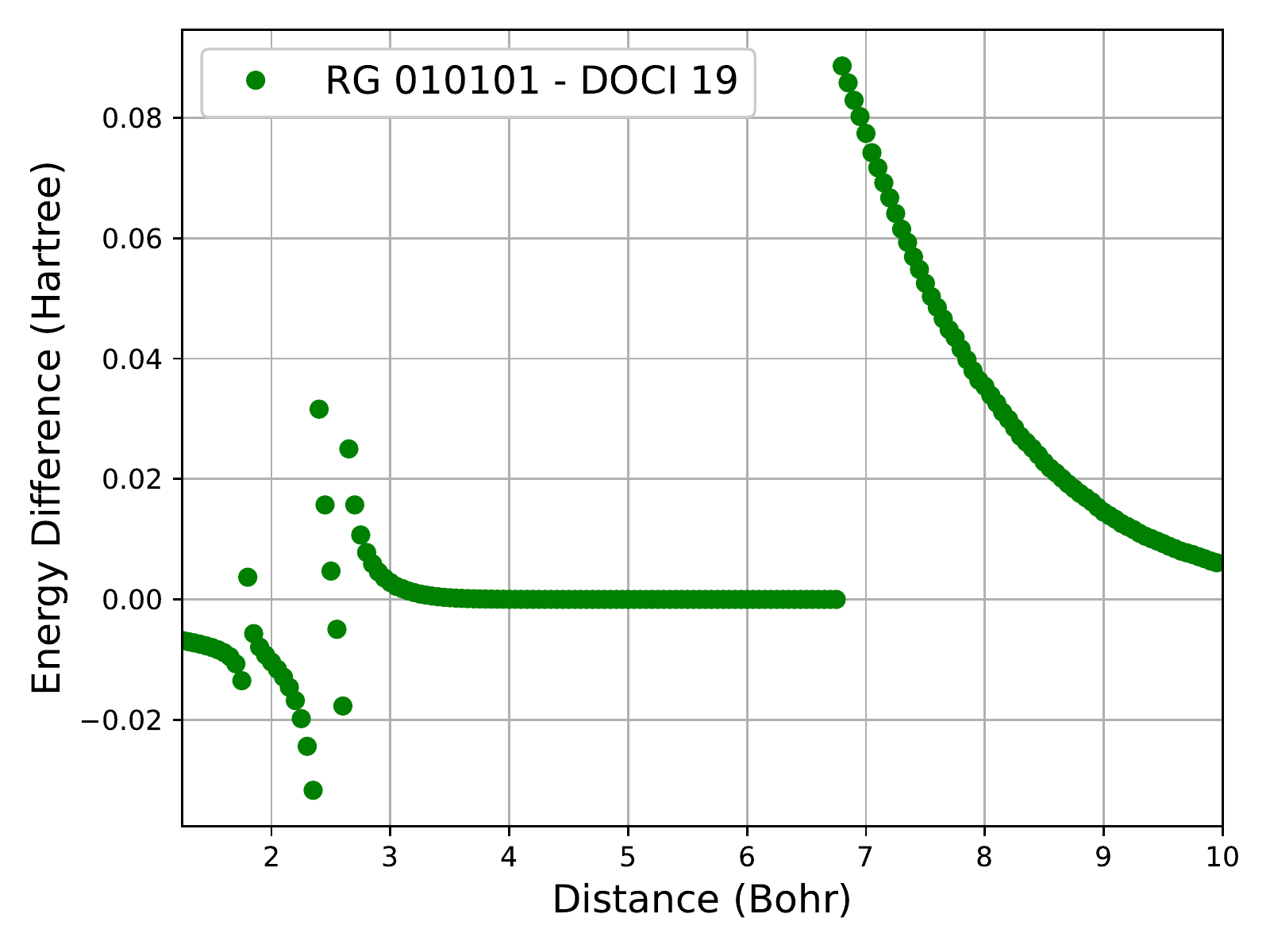}
	\end{subfigure}
	\caption{Energy differences between RG states computed with reduced BCS parameters defining the Hamiltonian corresponding to the optimal 101010 solution and DOCI states. All results computed in the STO-6G basis set in the basis of OO-DOCI orbitals.}
\end{figure}

\bibliography{dists_arxiv}

\bibliographystyle{unsrt}

\end{document}